\documentclass[prx,twocolumn, superscriptaddress,longbibliography]{revtex4-2}% revtext4-1 is ok as well, but prx is actually not defined

\usepackage[]{fontenc}
\usepackage[normalem]{ulem}
\usepackage{graphicx}
\usepackage{bm}
\usepackage{mathtools} 
\usepackage{leftindex}
\usepackage[caption=false]{subfig}
\usepackage{chngcntr} 
\usepackage{multirow}
\usepackage{subeqnarray}
\usepackage{wasysym}

\usepackage[british]{babel}
\usepackage{amsmath}
\usepackage{amssymb}
\usepackage{framed}
\usepackage{soul}
\usepackage{color}
\usepackage{alltt}
\usepackage{pstricks,pst-node,pst-tree,pst-grad,pst-text,graphics}
\usepackage{dcolumn}
\usepackage{url}
\usepackage{cases}
\usepackage{natbib}

\usepackage{ifthen}
\usepackage{afterpage}
\usepackage{comment}
\usepackage{supertabular}
\usepackage{xspace}
\usepackage{setspace}
\usepackage{textcomp}
\usepackage{tikz}
\usepackage{tabularx}
\usepackage[compat=1.1.0]{tikz-feynman}

\usepackage[hidelinks,colorlinks=true,linkcolor=blue,filecolor=blue,urlcolor=blue,citecolor=blue]{hyperref}
\usetikzlibrary{decorations.pathmorphing}
\usetikzlibrary{decorations.markings}
\usetikzlibrary{arrows,shapes,decorations,automata,backgrounds,petri}

\newcommand{\defequal}{\vcentcolon=}
\newcommand{\plaind}{\mathrm{d}}
\newcommand{\dint}[1]{\mathchoice{\!\plaind#1\,}{\!\plaind#1\,}{\!\plaind#1\,}{\!\plaind#1\,}}

\DeclareFontFamily{U}{wncy}{}
\DeclareFontShape{U}{wncy}{m}{n}{<->wncyr10}{}
\DeclareSymbolFont{mcy}{U}{wncy}{m}{n}
\DeclareMathSymbol{\sha}{\mathord}{mcy}{"58}

\usepackage{dsfont}

\newcommand{\gpset}[1]{\mathds{#1}}

\newcommand{\canetset}[1]{{\mathchoice {\hbox{$\sf\textstyle #1\kern-0.4em #1$}}
{\hbox{$\sf\textstyle #1\kern-0.4em #1$}}
{\hbox{$\sf\scriptstyle #1\kern-0.3em #1$}}
{\hbox{$\sf\scriptscriptstyle #1\kern-0.2em #1$}}}}

\newcommand{\Zset}{\gpset{Z}}

\newcommand{\Rset}{\gpset{R}}

\def\nbZ{{\mathchoice {\hbox{$\sf\textstyle Z\kern-0.4em Z$}}
{\hbox{$\sf\textstyle Z\kern-0.4em Z$}}
{\hbox{$\sf\scriptstyle Z\kern-0.3em Z$}}
{\hbox{$\sf\scriptscriptstyle Z\kern-0.2em Z$}}}}

\newcommand{\gpvec}[1]{\mathbf{#1}}

\newcommand{\zerovec}{\gpvec{0}}

\newcommand{\kvec}{\gpvec{k}}

\newcommand{\xvec}{\gpvec{x}}

\newcommand{\ident}{\mathbf{1}}

\renewcommand{\exp}[1]{\mathchoice%
{\mathrm{e}^{#1}}%
{\operatorname{exp}(#1)}
{\operatorname{exp}\left(#1\right)}%
{\operatorname{exp}\left(#1\right)}}

\newcommand{\elabel}[1]{\label{eq:#1}}

\newcommand{\Eref}[1]{\mbox{Eq.~\eqref{eq:#1}}}
\newcommand{\Erefs}[1]{\mbox{Eqs.~\eqref{eq:#1}}}

\newcommand{\flabel}[1]{\label{fig:#1}}

\newcommand{\Fref}[1]{Figure~\ref{fig:#1}}

\usepackage{xr}
\newcommand{\latin}[1]{{\it #1}}
\newcommand{\ie}{\latin{i.e.}\@\xspace}
\newcommand{\eg}{\latin{e.g.}\@\xspace}

\newcommand{\av}[1]{\left \langle #1 \right \rangle}

\renewcommand{\exp}[1]{\mathchoice%
{e^{#1}}%
{\operatorname{exp}(#1)}%
{\operatorname{exp}(#1)}%
{\operatorname{exp}(#1)}}

\newcommand{\gs}{\pi}
\newcommand{\phase}{\theta}

\newcommand{\gstilde}{\tilde{\gs}}

\newcommand{\phivec}{\bm{\phi}}

\newcommand{\heavymode}{\sigma}

\newcommand{\perpsub}[1]{#1_{\perp}}

\newcommand{\parsub}[1]{#1_{\parallel}}

\newcommand{\partialperp}{\perpsub{\partial}}
\newcommand{\partialpar}{\parsub{\partial}}

\newcommand{\nablasquared}{\nabla^2}

\makeatletter
\newcommand{\Gbare}[1]{G_0^{\@ifempty{#1}{ab}{#1}}}
\newcommand{\Corrbare}[1]{C_0^{\@ifempty{#1}{ab}{#1}}}
\newcommand{\responsebare}[1]{\response_{0}^{\@ifempty{#1}{ab}{#1}}}
\newcommand{\Pperp}[1]{P^{\@ifempty{#1}{ab}{#1}}_{1}}
\newcommand{\Ppar}[1]{P^{\@ifempty{#1}{ab}{#1}}_{2}}
\newcommand{\Ftensor}[1]{F^{\@ifempty{#1}{abc}{#1}}}
\makeatother

\newcommand{\response}{\chi}

\makeatletter
\newcommand{\diagram}[1]{\mathcal{I}_{\@ifempty{#1}{}{#1}}}
\makeatother

\newcommand{\IRreg}{\mu}

\newcommand{\vertex}{\Gamma}

\makeatletter
\newcommand{\invpropgs}[1]{\vertex_{\gstilde \gs}^{\@ifempty{#1}{ab}{#1}}}
\newcommand{\vertexnoisegs}[1]{\vertex^{\@ifempty{#1}{ab}{#1}}_{\gstilde \gstilde}}
\newcommand{\cubicvertexgs}[1]{\vertex^{\@ifempty{#1}{abc}{#1}}_{\gstilde \gs \gs}}

\newcommand{\invpropgsren}[1]{\vertex^{\@ifempty{#1}{ab}{#1}}_{R,\gstilde \gs}}
\newcommand{\vertexnoisegsren}[1]{\vertex^{\@ifempty{#1}{ab}{#1}}_{R,\gstilde \gstilde}}
\newcommand{\cubicvertexgsren}[1]{\vertex^{\@ifempty{#1}{abc}{#1}}_{R,\gstilde \gs \gs}}
\makeatother

\newcommand{\dimension}[1]{[#1]}

\newcommand{\noise}{\Gamma}

\newcommand{\lambdatilde}{\tilde{\lambda}}

\newcommand{\lambdaR}{\lambda_R}

\newcommand{\deltaR}{\delta_R}

\newcommand{\imag}{\mathring{\imath}}

\usepackage{color}
\definecolor{darkgreen}{rgb}{0,0.6,0}
\definecolor{darkblue}{rgb}{0,0,0.6}
\definecolor{darkred}{rgb}{0.6,0,0}
\definecolor{darkpurple}{rgb}{0.5,0,0.5}
\hypersetup{
bookmarksopen=true,
pdftitle="",
pdfauthor="", 
pdftoolbar=false, 
pdfstartview={FitH},		
pdfmenubar=true,			
pdfhighlight=/O,			
colorlinks=true,			
urlcolor=darkblue,
citecolor=darkblue,		
linkcolor=darkblue}

\makeatletter
\newcommand{\customlabel}[2]{%
   \protected@write \@auxout {}{\string \newlabel {#1}{{#2}{\thepage}{#2}{#1}{}} }%
   \hypertarget{#1}{\hspace{0pt}}
}
\makeatother

\begin{document}

%% New commands are here

%\newcommand{\titleText}{Kosterlitz Thouless Transition at the Onset of Activity in Malthusian Flocks}
\newcommand{\titleText}{Large Spin-Wave Fluctuations Suppress Activity in Malthusian Flocks}
%through a Berezinskii-Kosterlitz-Thouless type Transition}
%\newcommand{\titleText}{A Berezinskii-Kosterlitz-Thouless Transition in Malthusian Flocks}
\title{\titleText}

\author{Emir Sezik}%
 \email{emir.sezik19@imperial.ac.uk}
\affiliation{%
Department of Mathematics
and Centre for Complexity Science, 
Imperial College London, London SW7 2AZ, United Kingdom}%

\author{Gunnar Pruessner}
\email{g.pruessner@imperial.ac.uk}
\affiliation{%
Department of Mathematics
and Centre for Complexity Science, 
Imperial College London, London SW7 2AZ, United Kingdom}%

\date{\today}        

\begin{abstract}
%\noindent 
Novel phases, beyond long-range order in two dimensions, have continued to be discovered within flocking models, establishing flocking as one of the pivotal paradigms in active matter. However, much of the discussion around ``Malthusian'' (constant density) flocks, an analytically more tractable alternative to the Vicsek model, has centred around the scaling exponents governing the intermediate regime prior to the proliferation of asters, leaving open the question of what other phases the model might display. Here, we study the two-dimensional dynamics of Malthusian flocks and identify a previously unnoticed phase, where the dynamics is that of the equilibrium XY Model. By identifying the symmetries of the model, we derive the effective equations of motion for the Goldstone modes and analyse the spin-wave fluctuations. We identify a novel critical point separating two distinct phases and, using a perturbative RG procedure, determine the RG flows in its vicinity. This allows us to calculate the universal scaling behaviour at the critical point, along with its logarithmic corrections.
%Unlike the equilibrium BKT transition, the novel phase transition presented here is driven only by the interplay between activity and spin-waves. 
%XXX
%%Unlike the equilibrium counterpart, which 
%displays a phase transition in 
%%features
%%effective degrees of freedom, namely vortices, the novel phase transition here is due to the interaction of activity and spin-waves.
The novel phase transition here is due to the interaction of activity and spin-waves, unlike the equilibrium counterpart, which undergoes a phase transition in effective degrees of freedom, namely vortices.
%
%
%Strikingly, these flows are precisely those of the 
Nevertheless, the RG flows are similar to those of the
Berezinskii-Kosterlitz-Thouless transition, and we show that for sufficiently strong noise, the activity becomes irrelevant and the system crosses over to the equilibrium XY universality class.
\end{abstract}

                              
\maketitle

\newcommand{\longtodo}[1]{\todo[inline,size=\tiny]{#1}}

%%%%%%%%%%%
%%  Introduction  %%
%%%%%%%%%%%
%Flocking is an important paradigm in active matter as it is one of the few known mechanisms by which a system can spontaneously break a continuous symmetry and sustain long-range order in two dimensions \cite{toner_long-range_1995, vicsek_novel_8, mahault_quantitative_2019}, circumventing the celebrated Mermin-Wagner theorem \cite{mermin_absence_1966, hohenberg_existence_1967}. 
%

Flocking is an important paradigm in active matter, in its own right, but also because its non-equilibrium nature rules out the Mermin-Wagner theorem \cite{mermin_absence_1966, hohenberg_existence_1967, tasaki_prl}. The Vicsek Model \cite{vicsek_novel_8} famously and strikingly breaks a continuous symmetry, and displays long-range order in two dimensions \cite{toner_long-range_1995, vicsek_novel_8, mahault_quantitative_2019}. 

Although first investigated numerically more than three decades ago \cite{vicsek_novel_8}, to this day flocking systems keep producing surprising phenomena such as  phase separation \cite{miller_chemotaxis, miller_spinodal, miller_phase_separation}, susceptibility to domain-wall formation \cite{small_obstacle_large_flock, granek2026continuous, metastability_discrete_flocks}, and stability against frustration \cite{lardetDisorderedDirectedEmergence2024, lardet_flocking_beyond_one} . At the heart of all this novel physics lies the breaking of detailed balance through directed motion, enabling the emergence of non-trivial phases. 

Numerically, the rich phases of flockers are well understood, yet a comprehensive analytical treatment is still lacking. Indeed, the hydrodynamic descriptions of flocking models have been notoriously hard to analyse, mainly because of the existence of two coupled slow modes, one associated with the velocity of the particles and the other with their density \cite{toner_flocks_10, toner_reanalysis_2012, toner_hydrodynamics_2005}. Even after thirty years, the complete hydrodynamic description and the scaling behaviour of the Vicsek model remain contested \cite{chate_dynamic_2024, maitra_inconvenient_2025, jentsch_new_2024}. 

Motivated by this analytical difficulty, simpler models of flocking, such as so-called Malthusian (constant density) flocks \cite{toner_birth_2012}, can be devised by considering physical scenarios in which the density field becomes a fast variable, \eg through birth and death processes, and can be subsequently integrated out \cite{chen_novel_2020, chen_moving_2020, di_carlo_evidence_2022, toner_birth_2024}. This yields an overall simpler hydrodynamic description involving only one field, though at a notable cost: it is numerically well observed that the model no longer displays true long-range order due to the proliferation of asters \cite{besseMetastabilityConstantdensityFlocks2022, rouzaire_non-reciprocal_2024}. Although Malthusian flocks potentially do not display Vicsek-like flocking, they have found broad applicability in active matter systems. Specifically, 
%recent works have established that 
they were found to belong to the same universality class as the continuum limit of spin systems on a lattice with vision-cone interactions \cite{Loos_XY,nonmutual_torques_ramaswamy, rouzaire_non-reciprocal_2024, dopierala_inescapable_2025, orderinganddefectxy, rouzaireDynamicsO2Excitations2026a}. 
%Moreover, 
%it has been argued \cite{visionconeRG_Sezik} that 
One may argue that this universality class of Malthusian flocks is the unique extension of isotropic Model~A \cite{hohenberg_theory_1977} to active matter systems, thus constituting an important and tractable model in the search of active universality classes. 

\newcommand{\citeAuthor}[1]{\citeauthor{#1} \cite{#1}}

The ordering dynamics \cite{di_carlo_evidence_2022} and the dynamics of the Goldstone modes \cite{chen_moving_2020, chen_novel_2020} of this universality class have been analysed near the upper critical dimension, $d_c=4$, via a perturbative renormalisation group (RG) scheme. For $d=2$, however, an agreed analytical description of the Goldstone modes is still lacking, as few perturbative methods are capable of capturing the physics at the lower critical dimension. Although proliferation of asters was numerically observed to destroy any potential long-range order, an intermediate regime exists in which universal scaling behaviour is observed \cite{besseMetastabilityConstantdensityFlocks2022}, and it is the scaling exponents governing this regime that have become the focal point of contention in the field \cite{maitra_inconvenient_2025, chate_dynamic_2024}. \citeAuthor{chate_dynamic_2024} claim to have identified the \emph{exact} scaling exponents for the Goldstone modes through several hyperscaling relations, whereas \citeAuthor{maitra_inconvenient_2025} maintain that obtaining exact scaling relations is impossible. Although the search for the scaling exponents represents an important research avenue, it may not constitute the full picture of Malthusian flocks. Novel phases have continued to be discovered within flocking models \cite{metastability_discrete_flocks, granek2026continuous, lardet_flocking_beyond_one, lardetDisorderedDirectedEmergence2024,  small_obstacle_large_flock, jentsch2025diversity, solon_tailler_liquid_gas,Boltzmann_hydro_bertin, chate_onset, miller_phase_separation}, demonstrating that even well-studied systems can harbour unexpected behaviour. This invites the question of whether the phase structure of Malthusian flocks has itself been fully charted, and whether previously unnoticed phases may yet lie hidden within the model.

In this Letter, we study the $d=2$ Goldstone modes of Malthusian flocks, or equivalently, spin systems with vision cone interactions, and identify a novel phase, induced by spin-wave fluctuations only, where the dynamics on large scales is that of the equilibrium XY Model. Using field-theoretic RG, we characterise the nature of the transition from this phase to an active, truly non-equilibrium, ``Malthusian phase'', driven by the noise strength.
%and show that for strong enough noise, the system effectively becomes an equilibrium XY-Model on large scales. 
We find that close to the critical point separating the two phases, the RG flows to leading order in the couplings are similar to those of the celebrated Berezinskii-Kosterlitz-Thouless (BKT) phase transition \cite{kosterlitzOrderingMetastabilityPhase1973, Berezinskii1971}. Unlike the equilibrium BKT transition, the novel phase transition presented here is driven by the interplay between activity and spin-waves. This novel phase transition is identified in Ref.~\cite{JentschErzberger} using a complementary, non-perturbative RG framework. Our results suggest that the phase structure of Malthusian flocks is richer than previously appreciated, and invite further investigation into hidden phases across flocking systems more broadly.

%%%%%%%%%%%%%%%%%%%%%%%%%%%%%
%%  Universality Class and its symmetries  %%
%%%%%%%%%%%%%%%%%%%%%%%%%%%%%
\textit{Universality class and its symmetries.~---}~One of the key insights of the Landau-Ginzburg-Wilson paradigm is that symmetries determine universality classes. The universality class of Malthusian flocks and vision-cone models in the continuum limit can be identified by considering the relevant order parameters and their symmetries. In the present setup, the relevant order parameter is a single real coarse-grained vector field, $\phivec(\xvec,t) \in \Rset^d$, representing the net direction of motion for Malthusian flocks or magnetisation for vision-cone models, so that $d$ is the dimension of the space $\phivec(\xvec,t)$ lives on, $\xvec\in\Rset^d$, and the dimension of the spin-space itself. The universality class is further defined by the following \emph{restricted} rotational $O(d)$ symmetry of the order parameter field 
\begin{equation} \label{eq:symmetrymodelA}
    \phivec(\xvec,t) \to \phivec'(\xvec',t) = \bm{R}_d \cdot \phivec(\bm{R}^{-1}_d \cdot \xvec',t),
\end{equation}
where $\bm{R}_d$ is a $d-$dimensional rotation matrix, satisfying $\bm{R}_d^{T} \cdot \bm{R}_d = \ident$. Unlike equilibrium Model~A, rotating only the fields and not the space, thus dropping $\bm{R}^{-1}_d$ in front of $\xvec$ in \Eref{symmetrymodelA} does not leave the equation of motion invariant, because in the present case activity couples space and order parameter. In the context of Malthusian flocks, this coupling translates to motion in the direction of ``magnetisation". 

%%%%%%%%%%%%%%%%%%%%%%%%%%%%%
%%  Effective EoM for the Goldstone Modes  %%
%%%%%%%%%%%%%%%%%%%%%%%%%%%%%
\textit{Effective Equation of Motion for the Goldstone modes.~---}~Here and in what follows, we set $d=2$. In the symmetry-broken phase of the system, the order parameter, $\phivec(\xvec,t)$, can be parameterised by a single Goldstone mode and a heavy mode 
\begin{equation} \label{eq:phaseangledecomp}
    \phivec(\xvec,t) = \left(\phi_0 + \sigma(\xvec,t) \right)\begin{pmatrix}
        \cos(\theta(\xvec,t))\\
        \sin(\theta(\xvec,t))
    \end{pmatrix}
\end{equation}
where $\phi_0$ corresponds to the magnitude of the net magnetisation of the field, $\phase(\xvec,t)$ to the Goldstone (gapless) mode and $\heavymode(\xvec,t)$ to the heavy mode. Within this parameterisation, the Goldstone mode $\theta(\xvec,t)$ is now a compact variable where $\theta(\xvec,t) + 2 \pi m$ is identified with $\theta(\xvec,t)$ for any $m \in \Zset$. Instead of using the parameterisation \Eref{phaseangledecomp} in an effective equation of motion for $\phivec(\xvec,t)$ and integrating out the heavy mode, we posit the equation of motion for the gapless mode, $\phase(\xvec,t)$, based on symmetry considerations as, for the present case, the former approach might yield an equation with some relevant terms missing in two dimensions \cite{PRLsupplement}.

To ascertain the equation of motion for the Goldstone mode, $\theta(\xvec,t)$, we first need to determine its symmetries. Applying the symmetry operation, \Eref{symmetrymodelA}, to the phase-angle parameterisation of the order parameter field, \Eref{phaseangledecomp}, we find that $\theta'(\xvec',t)$, where
\begin{subequations} \label{eq:symmetriesofangle}
\begin{alignat}{3}
    \theta'(\xvec',t) &= \theta(\xvec,t) + \psi, \qquad &\xvec'&=\begin{pmatrix}
        \cos\psi & -\sin \psi \\
        \sin \psi &\cos \psi
    \end{pmatrix} 
    \xvec \label{eq:sorotation} \\
\text{and/or}\qquad&&&\nonumber\\
%\end{alignat}
%and/or
%\begin{alignat}{3}
    \theta'(\xvec',t) &= -\theta(\xvec,t), \qquad &\xvec'&=\begin{pmatrix}
        1 & 0 \\
        0  &-1
    \end{pmatrix} 
    \xvec \label{eq:parityrotation}
\end{alignat}
\end{subequations}
obeys the same Equation of Motion (EoM) as $\theta(\xvec,t)$, where $\psi \in [0, 2\pi)$ is a constant angle. The two independent symmetries correspond, respectively, to the continuous $SO(2)$ part, parameterised by $\psi$, and the discrete $\Zset_2$ part of the $O(2) \cong SO(2) \rtimes \Zset_2$ group. \Eref{sorotation} captures the rotational symmetry of the system whereas \Eref{parityrotation} captures the parity symmetry of the system. Any term we add to the EoM of $\phase(\xvec,t)$ must be consistent with the symmetries in \Eref{symmetriesofangle}, so that allowing for all such terms will result in the most general EoM compatible with the dynamics of the Goldstone modes of this universality class \cite{PRLsupplement}, however \cite{chate_dynamic_2024}.

%\cite[however][\cf \SMref{EffectiveEquationOfMotion}]{chate_dynamic_2024}.

%This is to be contrasted with the starting point of \citeAuthor{chate_dynamic_2024}, who restrict the number of terms added by positing that every \emph{deterministic} term added to the EoM must be a total divergence. In \cite{SuppMatt}, we discuss the limitations of this assumption. 

%In \SMref{EffectiveEquationOfMotion}, 

In \cite{PRLsupplement}, we show that any EoM consistent with the symmetries \Erefs{symmetriesofangle} must be comprised of the derivatives
\begin{subequations} \label{eq:defn_derivatives}
    \begin{align}
        \partialpar &\defequal \begin{pmatrix}
            \cos\theta(\xvec,t), & \sin\theta(\xvec,t)
        \end{pmatrix}
        %\cdot
        \begin{pmatrix}
            \partial_x \\
            \partial_y
        \end{pmatrix},
         \label{eq:partialpar}
        \\
        \partialperp &\defequal  \begin{pmatrix}
            -\sin\theta(\xvec,t), & \cos\theta(\xvec,t)
        \end{pmatrix}
        %\cdot
        \begin{pmatrix}
            \partial_x \\
            \partial_y
        \end{pmatrix}
          \label{eq:partialperp}.
    \end{align}
\end{subequations}
Using these derivatives, we can spell out the Langevin equation for the Goldstone mode up to second order in gradients
\begin{equation} \label{eq:EoMgeneral}
    \dot{\phase} = \lambdatilde \partialpar \phase + D_1 \partialpar^2 \phase + D_2 \partialperp^2 \phase + g_1 (\partialperp \phase)(\partialpar\phase) + \sqrt{2\noise} \xi
\end{equation}
where $\xi(\xvec,t)$ is a unit Gaussian noise with zero mean, satisfying $\langle\xi(\xvec,t) \xi(\xvec',t')\rangle = \delta(\xvec - \xvec') \delta(t-t')$. Terms like 
$\partialperp \theta$, $(\partialperp \theta)^2$, $\partialperp \partialpar \theta$, $\partialpar\partialperp\theta$ and $(\partialpar\theta)^2$ are excluded in \Eref{EoMgeneral} as they are invariant under the parity symmetry, \Eref{parityrotation}, whereas $\dot{\phase}$ picks up a minus sign. In particular $\partialpar\to\partialpar$ and $\partialperp\to-\partialperp$.
%, even though all of these terms are rotationally invariant, \Eref{sorotation}. 
%In \SMref{EffectiveEquationOfMotion}, 

In \cite{PRLsupplement}, by using the definition of the derivatives \Erefs{defn_derivatives}, we show that the effective EoM of $\theta(\xvec,t)$ is
\begin{align}  \label{eq:EoMgeneral_expanded}
    \dot{\theta}  &= D \lambda \left\{\cos\theta \partial_x + \sin\theta \partial_y\right\} \theta + D \nablasquared \theta + \sqrt{2\noise}\xi  \nonumber\\
    &+\tilde{D}\left\{ \cos(2\theta)(\partial_x^2 - \partial_y^2)\theta + 2\sin(2\theta) \partial_x \partial_y \theta\right\} \nonumber  \\
    &+\tilde{g}\left\{ \sin(2\theta)((\partial_x \theta)^2 - (\partial_y \theta)^2) -  2\cos(2\theta) (\partial_x \theta)(\partial_y \theta)\right\}
\end{align}
where we have defined $D = (D_1 + D_2)/2$, $\tilde{D} = (D_1 - D_2)/2$, $\tilde{g} = (D_2 - D_1 - g_1)/2$ and $\lambdatilde = D \lambda$, consistent with the findings of Ref.~\cite{JentschErzberger}. Each curly bracket in \Eref{EoMgeneral_expanded} as a whole is invariant under  \Eref{symmetriesofangle}. As this symmetry is maintained under the RG flow, the ten terms on the right-hand side of \Eref{EoMgeneral_expanded} will remain parameterised by five couplings. 
% In other words, the terms in the curly brackets stay as they are and they don't break apart.

%five couplings of the ten terms on the right hand side of \Eref{} 

%%%%%%%%%%%%%%%%%%%%%%%%%%%%%
%%  Dimensional Analysis and Relevance Arguments %%
%%%%%%%%%%%%%%%%%%%%%%%%%%%%%
\textit{Dimensional analysis and relevance arguments.~---}~To assess the relevance of the nonlinearities in \Eref{EoMgeneral_expanded}, we perform naive power counting. Choosing $\dimension{x} = L$ to be the unit of length, we demand that fluctuations and noise remain invariant under spatial rescaling, i.e. $[\noise] = B$ and $[D] = A$, so that the unit of time becomes $[t]= L^2/A$. We then find that the field has no length dimensions $[\theta] = B^{1/2}/A^{1/2}$ in two dimensions. Requiring the arguments of $\sin$ and $\cos$ to be dimensionless further fixes the dimension of the noise, $\dimension{\noise} = B = A$, whereby the nonlinearities can be found to have the dimensions $[\lambda] = L^{-1}$ and $[\tilde{D}] = [\tilde{g}] = A$. Here, the spatial dimension of each coupling corresponds to how it enters into observables in combination with the wave-vector magnitude $|\kvec|$ rather than signalling how they behave on a ``system size" $L$. For example, $\lambda$ enters as $\lambda/|\kvec|$, forming an overall dimensionless quantity. The field having no dimension is consistent with its compact nature where $\theta(\xvec,t)$ is equivalent to $\theta(\xvec,t) + 2\pi n$ for all $n \in \Zset$.

The naive power counting above rightly suggests that a drift term like $\partial_x\theta$ is always dominant over all other terms \cite{Pruesssner:DriftAlwaysDominates:2004} in the IR as $\kvec \to \zerovec$. However, the trigonometric function multiplying it, suppresses the drift term on the large scale and thus introduces an additional scale-dependence that undermines this argument \cite{zinn-justin_quantum_2002, amitRenormalisationGroupAnalysis1980a}. There is little scope expanding these functions, as $\theta$ is dimensionless. However, in a perturbation theory about a Gaussian $\theta$, thereby precluding topological defects and allowing $\theta$ to have a simple, logarithmic correlator \cite{PRLsupplement}
%\SMref{pertexpd2} 
in two dimension, trigonometric functions of $\theta$ scale according to
\begin{equation} \elabel{anomalous_scaling_exp}
    \av{\exp{\imag n (\theta(\xvec,t) - \theta(\zerovec,t))}}_0 = \exp{-\frac{n^2}{2}\av{(\theta(\xvec,t) - \phase(\zerovec,t))^2}_0} \propto \frac{1}{|\xvec|^{\frac{n^2 \noise}{2\pi D}}} \ .
\end{equation}
For $n=1$ this is identical to 
$\av{\cos (\theta(\xvec,t) - \theta(\zerovec,t))} = 
\av{\cos \theta(\xvec,t) \cos \theta(\zerovec,t)} + 
\av{\sin \theta(\xvec,t) \sin \theta(\zerovec,t)}$, as the imaginary part of \Eref{anomalous_scaling_exp} vanishes. This suggests that $\cos \theta(\xvec,t)$ scales like $|\xvec|^{-\noise/(4\pi D)}$.
The effective engineering dimension of $\lambda$ is therefore not $L^{-1}$, but rather $L^{-1+\noise/(4\pi D)}$ \cite{zinn-justin_quantum_2002}, so that large enough $\noise$ may render $\lambda$ irrelevant. 
\emph{As $\lambda$ is the fingerprint of activity, the spin-waves captured in \Eref{anomalous_scaling_exp} may effectively suppress activity.}
Similarly, $\cos(2\theta)$ and $\sin(2\theta)$ have scaling dimension $L^{\noise/(\pi D)}$ according to \Eref{anomalous_scaling_exp}, rendering $\tilde{D}$ in \Eref{EoMgeneral_expanded} irrelevant instead of marginal. In other words, on the large scale terms like $\tilde{D} \cos(2\theta) \partial_x^2$ vanish in comparison to $D\nabla^2$. Similarly, $\tilde{g}$ becomes irrelevant due to the presence of trigonometric functions multiplying it.
% GP 2 June 2026
% On the one hand, below we will assume that theta is Gaussian and has infinite support, on the other hand, such a theta will almost certainly come from a \phi that has (isolated) singularities, thereby producing ridges in theta. Or, in short: Allowing large theta seems incompatible with its differentiability, given its origin as the angle of \phi. 
% Maybe all we need is differentiability to construct the Gaussian further down. Let's ditch "small" fluctuations:
Henceforth, 
%in the spirit of small fluctuations of $\theta(\xvec,t)$,
we will assume $\theta(\xvec,t)$ to be differentiable everywhere, thereby focussing solely on spin waves.

Dropping therefore $\tilde{D}$ and $\tilde{g}$ from \Eref{EoMgeneral_expanded} we thus arrive at the following effective dynamics 
\begin{equation} \label{eq:model_final}
    \dot{\theta}  = D \lambda \left\{\cos\theta \partial_x + \sin\theta \partial_y\right\} \theta + D \nablasquared \theta + \sqrt{2\noise}\xi.
\end{equation}
Given the effective scaling dimension $L^{-1 + \noise/(4\pi D)}$ of $\lambda$, its \emph{relevance} depends on the ratio $\noise/ D$ of the noise strength and the diffusivity. For $\noise > 4 \pi D$, the nonlinearity $\lambda$ is irrelevant as the spin-waves screen the activity out and the theory is effectively free on large length scales. In this case, $\theta(\xvec,t)$ displays Gaussian fluctuations \Eref{anomalous_scaling_exp}, obtained by considering \Eref{model_final} without any nonlinearity, which is the Edwards-Wilkinson equation.
This defines one phase of our model, which we dub the (high-temperature) ``XY phase''. %Because we have not attempted to include vortices in our treatment, this phase will be trivially vortex-free.

% 15 July 2026: Old version (before Patrick-induced edits)
% This
% defines one phase of our model, which we dub the (vortex-free) ``XY phase''. 
% GP: Patrick seems to oppose the idea of a vortex-free phase. And he is right. When we wrote vortex-free, what we had in mind, was the vortices being free, being unbound. It's the high-temperature XY phase. We don't capture vortices, but they don't change the picture.

On the other hand, for $\noise < 4 \pi D$, the coupling $\lambda$ becomes relevant and the dynamics is driven out of equilibrium with the drift term $\propto\lambda$. This regime defines the other phase displayed by the system, which we dub the ``Malthusian phase''. The two phases are separated by the Gaussian fixed point
\begin{equation} \label{eq:Gaussian_critical_point}
    \noise = 4\pi D, \qquad \lambda = 0.
\end{equation}
% The situation is the same as in std phi4 theory in d>4? Why then RG?
The ensuing renormalisation group (RG) calculation will capture the logarithmic behaviour in our system that is similarly observed in $\varphi^4$-theory \cite{bellac_quantum_1992, amit_field_2005, zinn-justin_quantum_2002} at the upper critical dimension $d_c=4$, where there is likewise (only) a Gaussian fixed point.

%which equally has (only) a Gaussian fixed point.

%%%%%%%%%%%%%%%%%%%%%%%%%%%%%
%% Leading Order flow functions and Phase Diagram %%
%%%%%%%%%%%%%%%%%%%%%%%%%%%%%
\textit{Leading order RG flows and the phase diagram~---}~To understand the behaviour of the model close to the fixed point, \Eref{Gaussian_critical_point}, we perform a perturbative RG procedure on \Eref{model_final} using its associated MSRJD field theory, expanding in small
%about the Gaussian critical point, \Eref{Gaussian_critical_point} where both
\begin{equation}\elabel{def_delta}
    \delta \equiv \frac{\noise}{4 \pi D} - 1
\end{equation}
and $\lambda$. In the present case, $\delta$ plays the role of $\epsilon = 4-d$ in $\varphi^4$-theories, as its sign determines the relevance of the nonlinear coupling $\lambda$ and, consequently, the stability of the Gaussian fixed point \cite{amitRenormalisationGroupAnalysis1980a, zinn-justin_quantum_2002}. In distinction $\varphi^4$-theories, however, $\delta$ is itself allowed to flow under the RG, as it can receive non-trivial corrections through the renormalisation of $D$ or $\noise$, giving rise to a much richer flow diagram.

Assuming renormalisability close to the Gaussian fixed point, we perform the leading order renormalisation in 
%\SMref{rg}. 
\cite{PRLsupplement}. The present perturbative expansion is no longer organised in loops but in $\lambda$ and $\delta$, as at any given order in $\lambda$, a diagram with arbitrarily many loops can be constructed due to the infinite number of vertices generated by the sinusoidal nonlinearities. We find that the flow functions of the couplings $\lambda$ and $\delta$, on an inverse length-scale $\mu\downarrow0$, to leading order are given by
\begin{subequations} \label{eq:flow_functions}
    \begin{align}
            \IRreg\frac{\plaind \lambda_R}{\plaind \IRreg} &= \deltaR \lambdaR \\
            \IRreg\frac{\plaind \delta_R}{\plaind \IRreg} &= \frac{\alpha}{4}\lambdaR^2
    \end{align}
\end{subequations}
where $\alpha$ is a constant defined by the integral $\alpha = \int_{0}^{\infty}\dint{u}\exp{-u-\int_u^{\infty}\frac{ds}{s}\exp{-s}} = 0.62433\ldots$. \Erefs{flow_functions} represent the key finding of the present work. They describe the RG flow close to the Gaussian fixed point, \Eref{Gaussian_critical_point}, separating the XY and the Malthusian phase. Strikingly, the flow functions are similar to those of the BKT phase transition \cite{Chaikin_Lubensky_1995, Berezinskii1971, kosterlitzOrderingMetastabilityPhase1973}. In equilibrium, the BKT transition occurs due to vortex dynamics, whereas in the present case, the transition is a result of the interplay between activity $\lambda$ and the spin-waves. The resemblance to the BKT can be made more explicit by rewriting the flow functions, \Erefs{flow_functions}, in terms of $\lambda'_R = \sqrt{\alpha}/2 \lambdaR$. The resemblance of the RG flows near this critical point to the BKT transition is also uncovered independently in Ref.~\cite{JentschErzberger}. 

Similar to the BKT transition \cite{Chaikin_Lubensky_1995}, the behaviour of the flow functions can be classified into three physically different regions, \Fref{phase_diagram}. Firstly, for $\deltaR >0$ and $|\lambdaR| \leq 2\deltaR/\sqrt{\alpha}$, the RG flow as $\mu\downarrow0$ brings the theory into the line of fixed points where $\lambdaR = 0$. This line of fixed points corresponds to the vortex-free XY phase where the nonlinearity vanishes and the dynamics corresponds to that of the equilibrium XY model. The stability of the line of fixed points implies that for large enough noise, the activity becomes irrelevant and the model becomes asymptotically equilibrium. Secondly, for $\delta_R>0$, so that $\noise>4\pi D$, \Eref{def_delta}, and $|\lambdaR| > 2\deltaR/\sqrt{\alpha}$, the coupling $\lambdaR$ seems irrelevant at first, but it will eventually flow to a region where $\deltaR<0$ where $\lambdaR$ keeps growing as $\mu\to0$, bringing the model to the Malthusian phase. As our results are perturbative, we cannot predict the flow of the couplings deep inside the Malthusian phase where $\lambdaR$ and $\deltaR$ are large. Finally, for $\deltaR<0$, the coupling $\lambdaR$ is relevant and keeps growing under the flow. These two regions where $\lambdaR$ keeps growing, either $\deltaR<0$ or $|\lambdaR| > 2\deltaR/\sqrt{\alpha}$, define the Malthusian phase where the nonlinearity has to be kept in the effective description of the system. Unfortunately, our RG calculation reveals that the fixed point describing the Malthusian phase is not perturbatively accessible and we need to resort to different methods to characterise it. Indeed, non-perturbative RG has been employed in Ref.~\cite{JentschErzberger} to elucidate the physics at the Malthusian phase. The two phases, the vortex-free XY phase and the Malthusian phase, are separated by the half-stable Gaussian critical point $\lambdaR = 0 = \deltaR$, accessible under the RG flow,  $\mu \downarrow 0$, only when $\deltaR = \sqrt{\alpha} |\lambdaR|/2$, \ie the flow along the separatrix of the two phases starting from positive $\deltaR$.  
\begin{figure}
    \centering
    \includegraphics[width=1\linewidth]{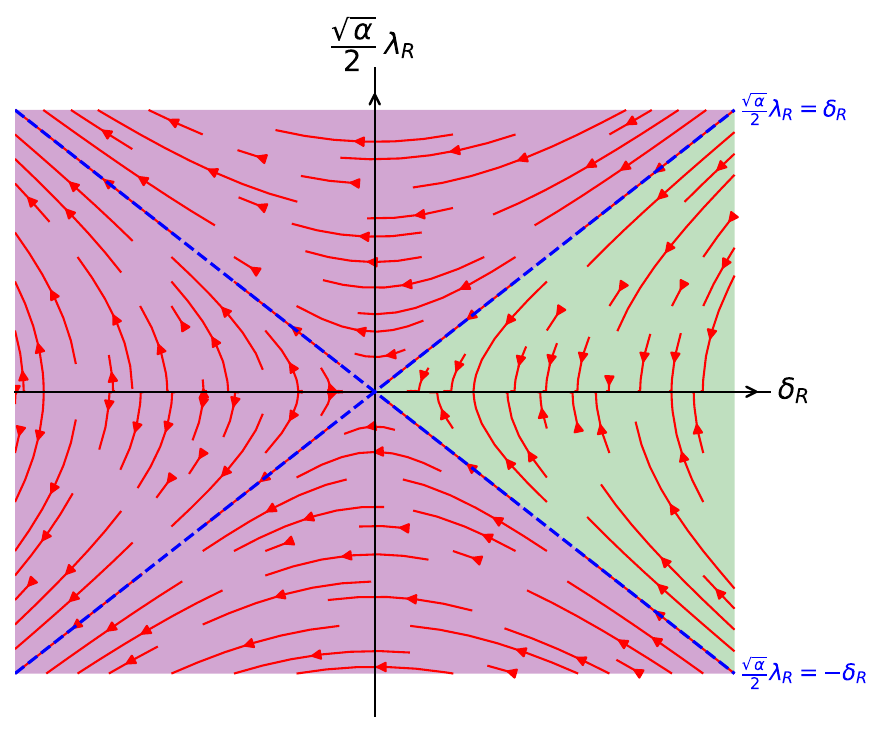}
    \caption{RG flows \Eref{flow_functions} as $\mu\downarrow0$ of the model near the Gaussian critical point $\lambdaR = 0 = \deltaR$. The Malthusian and XY phases are indicated by purple and green, respectively. These phases are separated by a half-stable Gaussian fixed point, that is accessible only from $\deltaR>0$ along the separatrix $\sqrt{\alpha}|\lambdaR|/2  = |\deltaR|$. The XY phase, $\lambdaR = 0$, is stable at large noise strengths $\noise$ and small nonlinearity $\lambda$. }
    \flabel{phase_diagram}
\end{figure}

The logarithmic behaviour of the system at the Gaussian fixed point, \Eref{Gaussian_critical_point}, can be determined at small $\mu$ for $\delta,\lambda$ on the separatrix at $\delta>0$.
%by solving the flow functions in the limit $\mu \downarrow 0$, with the initial condition $\delta = \sqrt{\alpha}|\lambda|/2 $.
%, which ensures that the RG flow leads the system into the Gaussian fixed point. 
In this limit, we find by solving $\plaind \delta/\plaind t=\delta^2$ with $t=\ln(\mu)$,
\begin{equation} \elabel{asymptotic_solutions}
    \delta_R(\mu_0 \ell) = \frac{\sqrt{\alpha}}{2}|\lambda_R(\mu_0\ell)| = 
    -\frac{1}{\ln(\ell)} + \mathcal{O}\left(\frac{1}{\ln^2(\ell)}\right) \ ,
\end{equation}
where $\ell = \mu/\mu_0 <1$ is some scale parameter. Using this to solve the Callan-Symanzik equation for the spin-correlation function we find, 
%\SMref{physicsGFP},
\cite{PRLsupplement},
\begin{equation}
    \av{\cos(\theta(\xvec,t)-\theta(\zerovec, 0))} = \frac{1}{|\xvec|^2 \ln^2(|\xvec|)} \hat{C}\left( \frac{t}{|\xvec|^2 \exp{1/\ln|\xvec|}}\right),
\end{equation}
with $\hat{C}(\bullet)$ a dimensionless scaling function. Both the anomalous scaling exponent $\eta$ and the dynamical scaling exponent $z$ thus acquire logarithmic corrections compared to the values by dimensional analysis, for example $\av{\cos(\theta(\xvec,t)-\theta(\zerovec, t))} \propto 1/|\xvec|^{2+2\delta}$, from \Eref{anomalous_scaling_exp}.

% to the values predicted by dimensional analysis.

% together with the solution of the Callan-Symanzik equation for the spin-correlation function $\av{\cos(\theta(\xvec,t) - \theta(\zerovec, 0))}$, we find that it is given by \SMref{physicsGFP} 

% Using this expression for the renormalised couplings and solving the Callan-Symanzik equation associated with the spin-correlation function $\av{\cos(\theta(\xvec,t) - \theta(\zerovec, 0))}$, we find that it is given by \SMref{physicsGFP} 
% \begin{equation}
%     \av{\cos(\theta(\xvec,t) - \theta(\zerovec, 0))} = \frac{1}{|\xvec|^2 \ln^2(|\xvec|)} \hat{C}\left( \frac{t}{|\xvec|^2 \exp{1/\ln|\xvec|}}\right),
% \end{equation}
% where $\hat{C}(\bullet)$ is a non-dimensional function of its argument. We find that both the anomalous scaling exponent $\eta$ and the dynamical scaling exponent $z$ acquire logarithmic corrections to the values predicted by dimensional analysis.
% }

%%%%%%%%%%%%%%%%%%%%%%%%%%%%%
%%  Conclusion and Outlook  %%
%%%%%%%%%%%%%%%%%%%%%%%%%%%%%
\textit{Conclusions \& outlook.~---}~We have studied the symmetry-broken phase of two-dimensional Malthusian flocks, or, equivalently, spin systems with vision-cone interactions. Using the relevant symmetries of the system, we re-derived the effective description in \Eref{model_final} for the Goldstone modes, consistent with Refs.~\cite{chate_dynamic_2024, maitra_inconvenient_2025}. In contrast to \cite{chate_dynamic_2024}, we retain the full sinusoidal nonlinearities, which we show are necessary to capture the anomalous scaling behaviour of the advective nonlinearity. Using ``enhanced'' dimensional analysis, \Eref{anomalous_scaling_exp}, we have identified a novel Gaussian critical point, \Eref{Gaussian_critical_point}, separating two distinct phases. Furthermore, we have tested the stability of this phase with respect to spin-wave fluctuations via a field-theoretic RG analysis leading to \Eref{flow_functions}, and thus \Eref{asymptotic_solutions}. We have found that the high-temperature XY phase is stable for sufficiently large noise strengths, where the nonlinearity becomes RG irrelevant. For all other parameter values, the couplings flow to a perturbatively inaccessible fixed point, where non-perturbative methods are needed to characterise the system or a different effective description altogether. Strikingly, the flows are similar to those of the BKT phase transition. The present transition is distinct from the equilibrium BKT transition for two main reasons: Firstly, it is driven by the interplay between spin waves and activity and secondly, in the present case the transition signals the onset of activity whereas in equilibrium the transition corresponds to vortex-pairing. 

Our results identify a previously unnoticed phase in active systems with a single rotationally invariant order-parameter field. Although activity appears relevant by naive power-counting, we find that it can become irrelevant under RG, with the dynamics corresponding to that of the equilibrium XY model. An important direction for future work is to understand how topological defects, natural in systems with a compact field variable, modify the present dynamics and phase diagram, which draw solely on spin-waves and activity.
%In our analysis, we have neglected the dynamics of the vortices, arising from the compact nature of the field $\phase(\xvec,t)$, by explicitly restricting our attention to spin-wave configurations. 
Accounting for vortices would introduce an additional axis into the phase diagram, namely, the vortex fugacity, and in the resulting full phase diagram one would expect to recover the standard BKT transition, where vortex pairing occurs, on the $\lambda = 0$ plane. The present results therefore correspond to the vortex-free slice of this larger phase diagram. Understanding the interplay between activity and topological defects, and their effect on the phase diagram, remains an interesting open problem. 
% 15 July 2026
However, as vortices cut off correlations even more drastically than spin waves, their presence will undermine the relevance of activity in the form of $\lambda$ even further, \ie vortices have the same effect on activity as spin waves as captured in \Eref{anomalous_scaling_exp}.
Finally, the Malthusian phase itself deserves further study: our results suggest it is perturbatively inaccessible. In Ref.~\cite{JentschErzberger}, authors employ the non-perturbative RG to determine scaling exponents in the Malthusian phase, thus describing the scaling regime without the asters observed in Refs.~\cite{besseMetastabilityConstantdensityFlocks2022}. More broadly, our results are a reminder that flocking systems can harbour important subtleties that are easily overlooked, motivating further investigation. 

%%%%%%%%%%%%%%%%%%%%%%%%%%%%%
%%  Nota bene  %%
%%%%%%%%%%%%%%%%%%%%%%%%%%%%%
\textit{Note added~---}~ While preparing this manuscript we noticed the work by Grosvenor and Patil~\cite{grosvenor2026renormalizationgroupflowactive} which also examines the present model and discusses the relevance of the active term, though it reaches different conclusions. Specifically, the authors claim that activity becomes relevant for $\Gamma < 2\pi D$ and $D$ does not receive any diverging diagrammatic correction.
%, with both of which we disagree.

%%%%%%%%%%%%%%%%%%%%%%%%%%%%%
%%  Acknowledgements  %%
%%%%%%%%%%%%%%%%%%%%%%%%%%%%%
\textit{Acknowledgements~---}~We thank Patrick Jentsch for insightful discussions and, together with Anna Erzberger, for a careful reading of the manuscript. ES also thanks Kevin Grosvenor and Subodh Patil for interesting discussions, and acknowledges support from the Roth Scholarship at Imperial College London.

\bibliography{Novel_PT_VC}% Produces the bibliography via BibTeX.

@STRING{OUP = {Oxford University Press}}

@STRING{OUPaddressUK = {Oxford, UK}}

@STRING{ZPB = {Z. Phys. B}}

@STRING{CUP = {Cambridge University Press}}

@STRING{CUPaddressEng = {Cambridge, UK}}

@article{janssen_lagrangean_1976,
	title = {On a {Lagrangean} for classical field dynamics and renormalization group calculations of dynamical critical properties},
	volume = {23},
	issn = {0340-224X},
	doi = {10.1007/BF01316547},
	number = {4},
	OLDjournal = {Zeitschrift f r Physik B Condensed Matter and Quanta},
    JOURNAL=ZPB,
	author = {Janssen, Hans-Karl},
	year = {1976},
	pages = {377--380},
}

@article{janssen_renormalized_1977,
  author  = {Janssen, Hans-Karl},
  title   = {Renormalized Field Theory of the Critical Dynamics of {$O(n)$}-Symmetric Systems},
  journal = {Z. Phys. B},
  year    = {1977},
  volume  = {26},
  number  = {2},
  pages   = {187--189},
  doi     = {10.1007/BF01325271},
}

@article{bausch_renormalized_1976,
	title = {Renormalized field theory of critical dynamics},
	volume = {24},
	issn = {0340-224X},
	doi = {10.1007/BF01312880},
	number = {1},
	journal = {Z. Phys. B},
	author = {Bausch, R. and Janssen, H. K. and Wagner, H.},
	year = {1976},
	pages = {113--127},
}

@article{toner_hydrodynamics_2005,
	title = {Hydrodynamics and phases of flocks},
	volume = {318},
	issn = {0003-4916},
	url = {https://www.sciencedirect.com/science/article/pii/S0003491605000540},
	doi = {10.1016/j.aop.2005.04.011},
	number = {1},
	journal = {Ann. Phys.},
	author = {Toner, John and Tu, Yuhai and Ramaswamy, Sriram},
	month = jul,
	year = {2005},
	pages = {170--244},
}

@article{toner_flocks_10,
	title = {Flocks, herds, and schools: {A} quantitative theory of flocking},
	volume = {58},
	url = {https://link.aps.org/doi/10.1103/PhysRevE.58.4828},
	doi = {10.1103/PhysRevE.58.4828},
	number = {4},
	journal = {Phys. Rev. E},
	author = {Toner, John and Tu, Yuhai},
	month = jan,
	year = {1998},
	pages = {4828--4858},
}

@article{vicsek_novel_8,
  title = {Novel Type of Phase Transition in a System of Self-Driven Particles},
  author = {Vicsek, Tam\'as and Czir\'ok, Andr\'as and Ben-Jacob, Eshel and Cohen, Inon and Shochet, Ofer},
  journal = {Phys. Rev. Lett.},
  volume = {75},
  issue = {6},
  pages = {1226--1229},
  numpages = {0},
  year = {1995},
  month = {Aug},
  publisher = {American Physical Society},
  doi = {10.1103/PhysRevLett.75.1226},
  url = {https://link.aps.org/doi/10.1103/PhysRevLett.75.1226}
}

@article{hohenberg_existence_1967,
	title = {Existence of {Long}-{Range} {Order} in {One} and {Two} {Dimensions}},
	volume = {158},
	url = {https://link.aps.org/doi/10.1103/PhysRev.158.383},
	doi = {10.1103/PhysRev.158.383},
	number = {2},
	urldate = {2023-11-13},
	journal = {Phys. Rev.},
	author = {Hohenberg, P. C.},
	month = jun,
	year = {1967},
	pages = {383--386},
}

@article{hohenberg_theory_1977,
	title = {Theory of dynamic critical phenomena},
	volume = {49},
	issn = {0034-6861},
	url = {https://link.aps.org/doi/10.1103/RevModPhys.49.435},
	doi = {10.1103/RevModPhys.49.435},
	number = {3},
	urldate = {2023-11-28},
	journal = {Rev. Mod. Phys.},
	author = {Hohenberg, P. C. and Halperin, B. I.},
	month = jul,
	year = {1977},
	pages = {435--479},
}

@article{mahault_quantitative_2019,
	title = {Quantitative {Assessment} of the {Toner} and {Tu} {Theory} of {Polar} {Flocks}},
	volume = {123},
	url = {https://link.aps.org/doi/10.1103/PhysRevLett.123.218001},
	doi = {10.1103/PhysRevLett.123.218001},
	number = {21},
	urldate = {2024-02-07},
	journal = {Phys. Rev. Lett.},
	author = {Mahault, Benoît and Ginelli, Francesco and Chaté, Hugues},
	month = nov,
	year = {2019},
	pages = {218001},
}

@article{toner_long-range_1995,
	title = {Long-{Range} {Order} in a {Two}-{Dimensional} {Dynamical} {XY} {Model}: {How} {Birds} {Fly} {Together}},
	volume = {75},
	copyright = {http://link.aps.org/licenses/aps-default-license},
	issn = {0031-9007, 1079-7114},
	shorttitle = {Long-{Range} {Order} in a {Two}-{Dimensional} {Dynamical} {XY} {Model}},
	url = {https://link.aps.org/doi/10.1103/PhysRevLett.75.4326},
	doi = {10.1103/PhysRevLett.75.4326},
	number = {23},
	urldate = {2024-05-23},
	journal = {Phys. Rev. Lett.},
	author = {Toner, John and Tu, Yuhai},
	month = dec,
	year = {1995},
	pages = {4326--4329},
}

@article{toner_reanalysis_2012,
	title = {Reanalysis of the hydrodynamic theory of fluid, polar-ordered flocks},
	volume = {86},
	copyright = {http://link.aps.org/licenses/aps-default-license},
	issn = {1539-3755, 1550-2376},
	url = {https://link.aps.org/doi/10.1103/PhysRevE.86.031918},
	doi = {10.1103/PhysRevE.86.031918},
	number = {3},
	urldate = {2024-05-23},
	journal = {Phys. Rev. E},
	author = {Toner, John},
	month = sep,
	year = {2012},
	pages = {031918},
}

@article{toner_birth_2012,
	title = {Birth, {Death}, and {Flight}: {A} {Theory} of {Malthusian} {Flocks}},
	volume = {108},
	copyright = {http://link.aps.org/licenses/aps-default-license},
	issn = {0031-9007, 1079-7114},
	shorttitle = {Birth, {Death}, and {Flight}},
	url = {https://link.aps.org/doi/10.1103/PhysRevLett.108.088102},
	doi = {10.1103/PhysRevLett.108.088102},
	
	number = {8},
	urldate = {2024-05-23},
	journal = {Phys. Rev. Lett.},
	author = {Toner, John},
	month = feb,
	year = {2012},
	pages = {088102},
}

@article{chen_novel_2020,
	title = {A novel nonequilibrium state of matter: a $d=4-\epsilon$ expansion study of {Malthusian} flocks},
	volume = {102},
	issn = {2470-0045, 2470-0053},
	shorttitle = {A novel nonequilibrium state of matter},
	url = {http://arxiv.org/abs/2004.00129},
	doi = {10.1103/PhysRevE.102.022610},
	number = {2},
	urldate = {2024-05-23},
	journal = {Phys. Rev. E},
	author = {Chen, Leiming and Lee, Chiu Fan and Toner, John},
	month = aug,
	year = {2020},
	pages = {022610},
}

@article{chen_moving_2020,
	title = {Moving, {Reproducing}, and {Dying} {Beyond} {Flatland}: {Malthusian} {Flocks} in {Dimensions} d {\textgreater} 2},
	volume = {125},
	issn = {0031-9007, 1079-7114},
	shorttitle = {Moving, {Reproducing}, and {Dying} {Beyond} {Flatland}},
	url = {https://link.aps.org/doi/10.1103/PhysRevLett.125.098003},
	doi = {10.1103/PhysRevLett.125.098003},
	number = {9},
	urldate = {2024-05-23},
	journal = {Phys. Rev. Lett.},
	author = {Chen, Leiming and Lee, Chiu Fan and Toner, John},
	month = aug,
	year = {2020},
	pages = {098003},
}

@article{di_carlo_evidence_2022,
	title = {Evidence of fluctuation-induced first-order phase transition in active matter},
	volume = {24},
	issn = {1367-2630},
	url = {https://iopscience.iop.org/article/10.1088/1367-2630/aca9ed},
	doi = {10.1088/1367-2630/aca9ed},
	
	number = {12},
	urldate = {2024-05-23},
	journal = {New J. Phys.},
	author = {Di Carlo, Luca and Scandolo, Mattia},
	month = dec,
	year = {2022},
	pages = {123032},
}

@article{jentsch_new_2024,
  title = {New Universality Class Describes Vicsek's Flocking Phase in Physical Dimensions},
  author = {Jentsch, Patrick and Lee, Chiu Fan},
  journal = {Phys. Rev. Lett.},
  volume = {133},
  issue = {12},
  pages = {128301},
  numpages = {7},
  year = {2024},
  month = {Sep},
  publisher = {American Physical Society},
  doi = {10.1103/PhysRevLett.133.128301},
  url = {https://link.aps.org/doi/10.1103/PhysRevLett.133.128301}
}

@book{tauber_critical_2014,
	address = {Cambridge},
	title = {Critical {Dynamics}: {A} {Field} {Theory} {Approach} to {Equilibrium} and {Non}-{Equilibrium} {Scaling} {Behavior}},
	isbn = {978-0-521-84223-5},
	shorttitle = {Critical {Dynamics}},
	url = {https://www.cambridge.org/core/books/critical-dynamics/041557627C8F8F36D96084B7617BFD5D},
publisher = {Cambridge University Press},
	urldate = {2024-09-16},
	author = {Täuber, Uwe C.},
	year = {2014},
	doi = {10.1017/CBO9781139046213},
}

@article{mermin_absence_1966,
	title = {Absence of {Ferromagnetism} or {Antiferromagnetism} in {One}- or {Two}-{Dimensional} {Isotropic} {Heisenberg} {Models}},
	volume = {17},
	url = {https://link.aps.org/doi/10.1103/PhysRevLett.17.1133},
	doi = {10.1103/PhysRevLett.17.1133},
	number = {22},
	urldate = {2024-09-16},
	journal = {Phys. Rev. Lett.},
	author = {Mermin, N. D. and Wagner, H.},
	month = nov,
	year = {1966},
	pages = {1133--1136},
}

@incollection{janssen_renormalized_1992,
	title = {On the renormalized field theory of nonlinear critical relaxation},
	isbn = {978-981-02-0938-4},
	url = {https://www.worldscientific.com/doi/10.1142/9789814355872_0007},
	urldate = {2024-09-25},
	booktitle = {From {Phase} {Transitions} to {Chaos}},
	author = {Janssen, H. K.},
	month = apr,
	year = {1992},
	doi = {10.1142/9789814355872_0007},
	pages = {68--91},
}

@article{toner_birth_2024,
  title = {Birth, death, and horizontal flight: Malthusian flocks with an easy plane in three dimensions},
  author = {Toner, John},
  journal = {Phys. Rev. E},
  volume = {110},
  issue = {6},
  pages = {064604},
  numpages = {6},
  year = {2024},
  month = {Dec},
  publisher = {American Physical Society},
  doi = {10.1103/PhysRevE.110.064604},
  url = {https://link.aps.org/doi/10.1103/PhysRevE.110.064604}
}

@article{rouzaire_non-reciprocal_2024,
  title = {Nonreciprocal Interactions Reshape Topological Defect Annihilation},
  author = {Rouzaire, Ylann and Pearce, Daniel J. G. and Pagonabarraga, Ignacio and Levis, Demian},
  journal = {Phys. Rev. Lett.},
  volume = {134},
  issue = {16},
  pages = {167101},
  numpages = {6},
  year = {2025},
  month = {Apr},
  publisher = {American Physical Society},
  doi = {10.1103/PhysRevLett.134.167101},
  url = {https://link.aps.org/doi/10.1103/PhysRevLett.134.167101}
}

@article{dopierala_inescapable_2025,
  title = {Inescapable Anisotropy of Nonreciprocal {XY} Models},
  author = {Dopierala, Dawid and Chat\'e, Hugues and Shi, Xia-qing and Solon, Alexandre},
  journal = {Phys. Rev. Lett.},
  volume = {135},
  issue = {8},
  pages = {088302},
  numpages = {7},
  year = {2025},
  month = {Aug},
  doi = {10.1103/r3dx-7lrd},
  url = {https://link.aps.org/doi/10.1103/r3dx-7lrd}
}

@misc{maitra_inconvenient_2025,
      title={The inconvenient truth about flocks}, 
      author={Leiming Chen and Patrick Jentsch and Chiu Fan Lee and Ananyo Maitra and Sriram Ramaswamy and John Toner},
      year={2025},
      eprint={2503.17064},
      archivePrefix={arXiv},
      primaryClass={cond-mat.soft},
      url={https://arxiv.org/abs/2503.17064}, 
}

@article{chate_dynamic_2024,
	title = {Dynamic {Scaling} of {Two}-{Dimensional} {Polar} {Flocks}},
	volume = {132},
	url = {https://link.aps.org/doi/10.1103/PhysRevLett.132.268302},
	doi = {10.1103/PhysRevLett.132.268302},
	number = {26},
	urldate = {2025-05-13},
	journal = {Phys. Rev. Lett.},
	author = {Chaté, Hugues and Solon, Alexandre},
	month = jun,
	year = {2024},
	pages = {268302},
}

@book{toner_physics_2024,
	title = {The {Physics} of {Flocking}: {Birth}, {Death}, and {Flight} in {Active} {Matter}},
	publisher = {Cambridge University Press},
	author = {Toner, John},
	year = {2024},
}

@article{orderinganddefectxy,
  title = {Ordering and Defect Cloaking in Nonreciprocal Lattice {XY} Models},
  author = {Popli, Pankaj and Maitra, Ananyo and Ramaswamy, Sriram},
  journal = {Phys. Rev. Lett.},
  volume = {135},
  issue = {8},
  pages = {088303},
  numpages = {6},
  year = {2025},
  month = {Aug},
  doi = {10.1103/2yky-45sr},
  url = {https://link.aps.org/doi/10.1103/2yky-45sr}
}

@article{nonmutual_torques_ramaswamy,
  title = {Nonmutual torques and the unimportance of motility for long-range order in two-dimensional flocks},
  author = {Dadhichi, Lokrshi Prawar and Kethapelli, Jitendra and Chajwa, Rahul and Ramaswamy, Sriram and Maitra, Ananyo},
  journal = {Phys. Rev. E},
  volume = {101},
  issue = {5},
  pages = {052601},
  numpages = {13},
  year = {2020},
  month = {May},
  doi = {10.1103/PhysRevE.101.052601},
  url = {https://link.aps.org/doi/10.1103/PhysRevE.101.052601}
}

@BOOK{zinn-justin_quantum_2002,
AUTHOR= "Jean Zinn-Justin",
TITLE= {Quantum Field Theory and Critical Phenomena},
PUBLISHER= OUP,
EDITION="4th",
ADDRESS=OUPaddressUK,
YEAR=2002,
  doi = {10.1093/acprof:oso/9780198509233.001.0001},
  url = {https://cds.cern.ch/record/572813}
}

@BOOK{amit_field_2005,
AUTHOR= "Daniel J. Amit and Victor Mart{\'i}n-Mayor",
TITLE= {Field Theory, the Renormalization Group, and Critical Phenomena},
PUBLISHER= {World Scientific},
ADDRESS={Singapore},
ORIGEDITION="3rd edition",
EDITION="3rd",
YEAR=2005 }

@book{bellac_quantum_1992,
  author    = {Michel {Le Bellac}},
  title     = {Quantum and Statistical Field Theory},
  publisher = {Oxford University Press},
  address   = {Oxford},
  year      = {1991},
  note      = {Translated by G. Barton}
}

@article{neudeckerNonlinearCrystalGrowth1982,
  title = {Nonlinear Crystal Growth near the Roughening-Transition},
  author = {Neudecker, B.},
  year = {1982},
  journal = {Z. Phys. B},
  shortjournal = {Z. Phys. B},
  volume = {49},
  number = {1},
  pages = {57--62},
  issn = {1431-584X},
  doi = {10.1007/BF01312969},
  url = {https://doi.org/10.1007/BF01312969},
  urldate = {2025-01-02}
}

@article{amitRenormalisationGroupAnalysis1980a,
  title = {Renormalisation Group Analysis of the Phase Transition in the {{2D Coulomb}} Gas, {{Sine-Gordon}} Theory and {{XY-model}}},
  author = {Amit, D. J. and Goldschmidt, Y. Y. and Grinstein, S.},
  year = {1980},
  journal = {J. Phys. A},
  shortjournal = {J. Phys. A: Math. Gen.},
  volume = {13},
  number = {2},
  pages = {585},
  issn = {0305-4470},
  doi = {10.1088/0305-4470/13/2/024},
  url = {https://dx.doi.org/10.1088/0305-4470/13/2/024},
  urldate = {2024-12-15}
}

@misc{granek2026continuous,
  title   = {How Continuous Symmetry Stabilizes the Ordered Phase of Polar Flocks},
  author  = {Granek, Omer and Chat\'{e}, Hugues and Kafri, Yariv and Ro, Sunghan and Solon, Alexandre and Tailleur, Julien},
  year    = {2026},
  eprint  = {2602.16865},
  archivePrefix = {arXiv},
  primaryClass  = {cond-mat}
}

@article{metastability_discrete_flocks,
  title = {Metastability of Discrete-Symmetry Flocks},
  author = {Benvegnen, Brieuc and Granek, Omer and Ro, Sunghan and Yaacoby, Ran and Chat\'e, Hugues and Kafri, Yariv and Mukamel, David and Solon, Alexandre and Tailleur, Julien},
  journal = {Phys. Rev. Lett.},
  volume = {131},
  issue = {21},
  pages = {218301},
  numpages = {6},
  year = {2023},
  month = {Nov},
  publisher = {American Physical Society},
  doi = {10.1103/PhysRevLett.131.218301},
  url = {https://link.aps.org/doi/10.1103/PhysRevLett.131.218301}
}

@article{small_obstacle_large_flock,
  title = {Small Obstacle in a Large Polar Flock},
  author = {Codina, Joan and Mahault, Beno\^{\i}t and Chat\'e, Hugues and Dobnikar, Jure and Pagonabarraga, Ignacio and Shi, Xia-qing},
  journal = {Phys. Rev. Lett.},
  volume = {128},
  issue = {21},
  pages = {218001},
  numpages = {6},
  year = {2022},
  month = {May},
  publisher = {American Physical Society},
  doi = {10.1103/PhysRevLett.128.218001},
  url = {https://link.aps.org/doi/10.1103/PhysRevLett.128.218001}
}

@misc{lardetDisorderedDirectedEmergence2024,
      title={Disordered Yet Directed: The Emergence of Polar Flocks with Disordered Interactions}, 
      author={Eloise Lardet and Raphaël Voituriez and Silvia Grigolon and Thibault Bertrand},
      year={2024},
      eprint={2409.10768},
      archivePrefix={arXiv},
      primaryClass={cond-mat.soft},
      doi={https://arxiv.org/abs/2409.10768}, 
}

@article{lardet_flocking_beyond_one,
  title = {Flocking beyond one species: Novel phase coexistence in a generalized two-species Vicsek model},
  author = {Lardet, Eloise and Chen, Letian and Bertrand, Thibault},
  journal = {Phys. Rev. Lett.},
  pages = {},
  year = {2026},
  month = {Feb},
  publisher = {American Physical Society},
  doi = {10.1103/lt6f-yjpd},
  url = {https://link.aps.org/doi/10.1103/lt6f-yjpd}
}

@UNPUBLISHED{rouzaireDynamicsO2Excitations2026a,
      title={Dynamics of {$O(2)$} excitations in a non-reciprocal medium}, 
      author={Ylann Rouzaire and Daniel JG Pearce and Ignacio Pagonabarraga and Demian Levis},
      year={2026},
      eprint={2603.23225},
      archivePrefix={arXiv},
      primaryClass={cond-mat.stat-mech},
      url={https://arxiv.org/abs/2603.23225}, 
}

@article{kosterlitzOrderingMetastabilityPhase1973,
  title = {Ordering, Metastability and Phase Transitions in Two-Dimensional Systems},
  author = {Kosterlitz, J. M. and Thouless, D. J.},
  year = {1973},
  journal = {J. Phys. C: Solid State Phys.},
  shortjournal = {J. Phys. C: Solid State Phys.},
  volume = {6},
  number = {7},
  pages = {1181},
  issn = {0022-3719},
  doi = {10.1088/0022-3719/6/7/010},
  url = {https://doi.org/10.1088/0022-3719/6/7/010},
  urldate = {2026-04-20},
  langid = {english}
}

@article{Berezinskii1971,
  author    = {Berezinskii, V. L.},
  title     = {Destruction of Long-Range Order in One-Dimensional and Two-Dimensional
               Systems Having a Continuous Symmetry Group {I}. Classical Systems},
  journal   = {Sov. Phys. JETP},
  year      = {1971},
  volume    = {32},
  number    = {3},
  pages     = {493--500},
}

@article{MSR,
  title = {Statistical Dynamics of Classical Systems},
  author = {Martin, P. C. and Siggia, E. D. and Rose, H. A.},
  journal = {Phys. Rev. A},
  volume = {8},
  issue = {1},
  pages = {423--437},
  numpages = {0},
  year = {1973},
  month = {Jul},
  doi = {10.1103/PhysRevA.8.423},
  url = {https://link.aps.org/doi/10.1103/PhysRevA.8.423}
}

@article{Dominicis1976TECHNIQUESDR,
  title={TECHNIQUES DE RENORMALISATION DE LA TH{\'E}ORIE DES CHAMPS ET DYNAMIQUE DES PH{\'E}NOM{\`E}NES CRITIQUES},
  author={Cyrano De Dominicis},
  journal={J. Phys. Colloq.},
  year={1976},
  volume={37},
  url={https://api.semanticscholar.org/CorpusID:98380494}
}

@book{gradshteyn2014,
  author    = {Gradshteyn, I. S. and Ryzhik, I. M.},
  editor    = {Zwillinger, Daniel and Moll, Victor Hugo},
  title     = {Table of Integrals, Series, and Products},
  edition   = {8th},
  publisher = {Academic Press},
  address   = {Amsterdam},
  year      = {2014},
  isbn      = {978-0-12-384933-5}
}

@misc{mathematica,
  author    = {{Wolfram Research, Inc.}},
  title     = {Mathematica},
  year      = {2026},
  note      = {Version 13.2.1.0},
  publisher = {Wolfram Research, Inc.},
  address   = {Champaign, IL},
  url       = {https://www.wolfram.com/mathematica}
}

@BOOK{Chaikin_Lubensky_1995,
AUTHOR= "P. M. Chaikin and T. C. Lubensky",
TITLE= {Principles of condensed matter physics},
PUBLISHER= CUP,
ADDRESS= CUPaddressEng,
YEAR=1995 }

@article{besseMetastabilityConstantdensityFlocks2022,
    title = {Metastability of constant-density flocks},
    volume = {129},
    url = {https://link.aps.org/doi/10.1103/PhysRevLett.129.268003},
    doi = {10.1103/PhysRevLett.129.268003},
    number = {26},
    urldate = {2024-02-09},
    journal = {Phys. Rev. Lett.},
    publisher = {American Physical Society},
    author = {Besse, Marc and Chaté, Hugues and Solon, Alexandre},
    month = dec,
    year = {2022},
    pages = {268003},
}

@misc{jentsch2025diversity,
      title={Diversity of critical phenomena in the ordered phase of polar active fluids}, 
      author={Patrick Jentsch and Chiu Fan Lee},
      year={2025},
      eprint={2512.18846},
      archivePrefix={arXiv},
      primaryClass={cond-mat.soft},
      url={https://arxiv.org/abs/2512.18846}, 
}

@article{miller_chemotaxis,
  title = {Following Your Nose: Autochemotaxis and Other Mechanisms for Spinodal Decomposition in Flocks},
  author = {Miller, Maxx and Toner, John},
  journal = {Phys. Rev. Lett.},
  volume = {132},
  issue = {12},
  pages = {128301},
  numpages = {5},
  year = {2024},
  month = {Mar},
  publisher = {American Physical Society},
  doi = {10.1103/PhysRevLett.132.128301},
  url = {https://link.aps.org/doi/10.1103/PhysRevLett.132.128301}
}

@article{miller_spinodal,
  title = {Spinodal decomposition and phase separation in polar active matter},
  author = {Miller, Maxx and Toner, John},
  journal = {Phys. Rev. E},
  volume = {109},
  issue = {3},
  pages = {034606},
  numpages = {24},
  year = {2024},
  month = {Mar},
  publisher = {American Physical Society},
  doi = {10.1103/PhysRevE.109.034606},
  url = {https://link.aps.org/doi/10.1103/PhysRevE.109.034606}
}

@article{miller_phase_separation,
  title = {Phase separation in ordered polar active fluids: A completely different universality class from that of equilibrium fluids},
  author = {Miller, Maxx and Toner, John},
  journal = {Phys. Rev. E},
  volume = {110},
  issue = {5},
  pages = {054607},
  numpages = {21},
  year = {2024},
  month = {Nov},
  publisher = {American Physical Society},
  doi = {10.1103/PhysRevE.110.054607},
  url = {https://link.aps.org/doi/10.1103/PhysRevE.110.054607}
}

@article{Boltzmann_hydro_bertin,
  title = {Boltzmann and hydrodynamic description for self-propelled particles},
  author = {Bertin, Eric and Droz, Michel and Gr\'egoire, Guillaume},
  journal = {Phys. Rev. E},
  volume = {74},
  issue = {2},
  pages = {022101},
  numpages = {4},
  year = {2006},
  month = {Aug},
  publisher = {American Physical Society},
  doi = {10.1103/PhysRevE.74.022101},
  url = {https://link.aps.org/doi/10.1103/PhysRevE.74.022101}
}

@article{chate_onset,
  title = {Onset of Collective and Cohesive Motion},
  author = {Gr\'egoire, Guillaume and Chat\'e, Hugues},
  journal = {Phys. Rev. Lett.},
  volume = {92},
  issue = {2},
  pages = {025702},
  numpages = {4},
  year = {2004},
  month = {Jan},
  publisher = {American Physical Society},
  doi = {10.1103/PhysRevLett.92.025702},
  url = {https://link.aps.org/doi/10.1103/PhysRevLett.92.025702}
}

@article{solon_tailler_liquid_gas,
  title = {From Phase to Microphase Separation in Flocking Models: The Essential Role of Nonequilibrium Fluctuations},
  author = {Solon, Alexandre P. and Chat\'e, Hugues and Tailleur, Julien},
  journal = {Phys. Rev. Lett.},
  volume = {114},
  issue = {6},
  pages = {068101},
  numpages = {5},
  year = {2015},
  month = {Feb},
  publisher = {American Physical Society},
  doi = {10.1103/PhysRevLett.114.068101},
  url = {https://link.aps.org/doi/10.1103/PhysRevLett.114.068101}
}

@article{tasaki_prl,
  title = {Hohenberg-Mermin-Wagner-Type Theorems for Equilibrium Models of Flocking},
  author = {Tasaki, Hal},
  journal = {Phys. Rev. Lett.},
  volume = {125},
  issue = {22},
  pages = {220601},
  numpages = {6},
  year = {2020},
  month = {Nov},
  publisher = {American Physical Society},
  doi = {10.1103/PhysRevLett.125.220601},
  url = {https://link.aps.org/doi/10.1103/PhysRevLett.125.220601}
}

@article{Loos_XY,
  title = {Long-Range Order and Directional Defect Propagation in the Nonreciprocal $\mathit{XY}$ Model with Vision Cone Interactions},
  author = {Loos, Sarah A. M. and Klapp, Sabine H. L. and Martynec, Thomas},
  journal = {Phys. Rev. Lett.},
  volume = {130},
  issue = {19},
  pages = {198301},
  numpages = {6},
  year = {2023},
  month = {May},
  publisher = {American Physical Society},
  doi = {10.1103/PhysRevLett.130.198301},
  url = {https://link.aps.org/doi/10.1103/PhysRevLett.130.198301}
}

@article{Pruesssner:DriftAlwaysDominates:2004,
  title = {Drift Causes Anomalous Exponents in Growth Processes},
  author = {Pruessner, Gunnar},
  journal = {Phys. Rev. Lett.},
  volume = {92},
  issue = {24},
  pages = {246101},
  numpages = {4},
  year = {2004},
  month = {Jun},
  publisher = {American Physical Society},
  doi = {10.1103/PhysRevLett.92.246101},
  url = {https://link.aps.org/doi/10.1103/PhysRevLett.92.246101}
}

@footnote{PRLsupplement,
  note = {See Supplemental Material [url] for technical details,
  which includes
  Refs.~\cite{XXX}.}
}

@incollection{Grinstein:1995,
  author    = {G. Grinstein},
  title     = {Generic scale invariance and self-organized criticality},
  booktitle = {Scale Invariance, Interfaces, and Non-Equilibrium Dynamics},
  editor    = {Alan McKane and Michel Droz and Jean Vannimenus and Dietrich Wolf},
  publisher = {Plenum Press},
  address   = {New York},
  year      = {1995},
  pages     = {261--293},
  note      = {Proceedings of the {NATO} Advanced Study Institute, Cambridge, UK, 20--30 June 1994}
}

@article{KPZ_1986,
  title = {Dynamic Scaling of Growing Interfaces},
  author = {Kardar, Mehran and Parisi, Giorgio and Zhang, Yi-Cheng},
  journal = {Phys. Rev. Lett.},
  volume = {56},
  issue = {9},
  pages = {889--892},
  numpages = {0},
  year = {1986},
  month = {Mar},
  publisher = {American Physical Society},
  doi = {10.1103/PhysRevLett.56.889},
  url = {https://link.aps.org/doi/10.1103/PhysRevLett.56.889}
}

@book{Meakin,
  author    = {Meakin, Paul},
  title     = {Fractals, Scaling and Growth Far from Equilibrium},
  series    = {Cambridge Nonlinear Science Series},
  publisher = {Cambridge University Press},
  address   = {Cambridge},
  year      = {1998},
  isbn      = {9780521452533},
  doi       = {10.1017/CBO9780511189807}
}

@book{PimpinelliVillain,
  author    = {Pimpinelli, Alberto and Villain, Jacques},
  title     = {Physics of Crystal Growth},
  series    = {Collection Al\'ea-Saclay: Monographs and Texts in Statistical Physics},
  publisher = {Cambridge University Press},
  address   = {Cambridge},
  year      = {1998},
  isbn      = {9780521558556},
  doi       = {10.1017/CBO9780511622526}
}

@book{VicsekFractalGrowth,
  author    = {Vicsek, Tam\'as},
  title     = {Fractal Growth Phenomena},
  edition   = {1},
  publisher = {World Scientific},
  address   = {Singapore},
  year      = {1989},
  isbn      = {9789971508302},
  doi       = {10.1142/0511}
}

@book{BarabasiStanley,
  author    = {Barab\'asi, Albert-L\'aszl\'o and Stanley, H.~Eugene},
  title     = {Fractal Concepts in Surface Growth},
  publisher = {Cambridge University Press},
  address   = {Cambridge},
  year      = {1995},
  isbn      = {9780521483186},
  doi       = {10.1017/CBO9780511599798}
}

@misc{grosvenor2026renormalizationgroupflowactive,
      title={On the Renormalization Group Flow of Active Flocks}, 
      author={Kevin T. Grosvenor and Subodh P. Patil},
      year={2026},
      eprint={2606.20552},
      archivePrefix={arXiv},
      primaryClass={cond-mat.soft},
      url={https://arxiv.org/abs/2606.20552}, 
}

@unpublished{JentschErzberger,
  author = {Jentsch, P. and Erzberger, A.},
  title  = {},
  note   = {to be published},
}

\end{document}

%% New commands are here

\newcommand{\titleText}{Supplemental Material for ``Large Spin-Wave Fluctuations Suppress Activity in Malthusian Flocks"}
\title{\titleText}

\author{Emir Sezik}%
 \email{emir.sezik19@imperial.ac.uk}
\affiliation{%
Department of Mathematics
and Centre of Complexity Science, 
Imperial College London, London SW7 2AZ, United Kingdom}%

\author{Gunnar Pruessner}
\email{g.pruessner@imperial.ac.uk}
\affiliation{%
Department of Mathematics
and Centre of Complexity Science, 
Imperial College London, London SW7 2AZ, United Kingdom}%

\date{\today}        

\keywords{Active matter, field theory, phase transitions}
                              
\maketitle

\hrule
\vspace{-3\baselineskip} % adjust as needed
\tableofcontents
\vspace{1em}
\hrule

\newcommand{\longtodo}[1]{\todo[inline,size=\tiny]{#1}}

\section{Effective Equation of Motion}
\seclabel{EffectiveEquationOfMotion}
%\label{sec:Dynamicsd2}
% Do we not dive in too quickly? Don't we need a bit of intro? Something about the physics or the maths, more than just half a sentence.
In this section, we derive the effective equation of motion (EoM) for the Goldstone mode of the universality class encompassing Malthusian flocks and vision-cone models, purely on the basis of symmetry considerations, namely that rotating spins must be accompanied by an equal rotation of the location of these spins in space. For this universality class, there is only one coarse-grained hydrodynamic variable, $\phivec(\posvec,t) \in \Rset^d$, representing the local direction of motion for Malthusian flocks and magnetisation for vision-cone models. The universality class is defined by the following \emph{restricted} rotational $O(d)$ symmetry of the order parameter
\begin{equation} \label{eq:symmetrymodelA}
    \phivec(\posvec,t) \to \phivec'(\posvec',t) = \bm{R}_d \cdot \phivec(\bm{R}^{-1}_d \cdot \posvec',t),
\end{equation}
where $\bm{R}_d$ is a $d-$dimensional rotation matrix, satisfying $\bm{R}_d^{T} \cdot \bm{R}_d = \ident$. In other words, if $\phivec$ with noise $\xivec$ solves the EoM of a vision-cone model, so does $\phivec'$ with noise $\xivec'$.

% pruess 1 June 2026: Emir to read from here

\Eref{symmetrymodelA} is stricter than in traditional ferromagnetic phase transitions. In, say, equilibrium Model~A \cite{hohenberg_theory_1977}, globally rotating the fields alone, without rotating space, leaves the EoM invariant. Its purely spatial $O(d)$-symmetry has no direct bearing on the available excitations (although, of course, dimensionality itself has). The transition is characterised by $O(N)\to O(N-1)$ with $N$ being the dimensionality of the spins, and $d$ entering to characterise the spectral density of harmonic excitations. In the present vision cone model, spin-space and space have to be rotated simultaneously to leave the EoM invariant and $N$ must be equal to $d$, as the spaces of $\posvec\in\Rset^d$ and $\phivec\in\Rset^N$``talk to each other'' in terms like $\lambdaone (\phivec(\posvec,t) \cdot \nabla)\phivec(\posvec,t)$, that project the spatial $\nabla$ against the spin $\phivec$ to implement the preferred direction of a vision cone.

%XXX CONT HERE 

%Unlike equilibrium Model~A, rotating only the fields do not leave the equation of motion invariant as the space and the order parameter field are now coupled together due to the presence of activity. 

Even with the reduced rotational symmetry, \Eref{symmetrymodelA}, the vision cone system is isotropic as there is no preferred direction for the net magnetisation. Near the upper critical dimension $d_c = 4$, the universality class is defined by the Langevin equation
% \gpcomment{1 June 2026. Emir needs to check that my change from $\lambdatwo(\nabla \cdot \phivec(\posvec,t))\phivec(\posvec,t)$ to $\lambdatwo\phivec(\posvec,t)(\nabla \cdot \phivec(\posvec,t))$ is fine. Looking at the expression, I also removed confusing brackets. I left everything in a comment:}
\begin{align} \label{eq:activeModelA}
% Original
    % \dot{\phivec}(\posvec,t) &+\lambdaone (\phivec(\posvec,t) \cdot \nabla)\phivec(\posvec,t) + 
    % \lambdatwo(\nabla \cdot \phivec(\posvec,t))\phivec(\posvec,t) + \lambdathree\nabla(\phivec^2(\posvec,t)) =
% New
    \dot{\phivec}(\posvec,t) &+\lambdaone \phivec(\posvec,t) \cdot \nabla \phivec(\posvec,t) + 
    \lambdatwo \phivec(\posvec,t) \nabla \cdot \phivec(\posvec,t) + \lambdathree\nabla(\phivec(\posvec,t)\cdot\phivec(\posvec,t)) = \nonumber \\
    &-\big\{ r - \Dperp \nabla^2 + \frac{u}{6}\phivec^2(\posvec,t)\big\}\phivec(\posvec,t) -(\Dpar - \Dperp)\nabla(\nabla \cdot \phivec(\posvec,t)) + \sqrt{2
    \noise}\xivec(\posvec,t).
\end{align}
where the terms inside the large curly brackets on the right are the usual equilibrium Model~A terms, $(\Dpar - \Dperp)$ quantifies the anisotropic diffusion and the $\lambda_i$ terms drive the system out of equilibrium.   
% \gpcomment{1 June 2026: Emir, I seem to remember that you had a very clear motivation for this notation ($\Dpar$ and $\Dperp$). Maybe that's worth adding here?}

It is common \cite{toner_physics_2024, toner_flocks_10, toner_hydrodynamics_2005, chen_moving_2020} to derive the EoM for the Goldstone modes in the ordered phase by decomposing the field $\phivec(\posvec,t)$ into a longitudinal (heavy) component in the preferred direction, say component $1$, and $d-1$ transversal components, say,
\begin{subequations} \label{eq:fielddecomp}
\begin{align}
    \phi^{1}(\posvec,t) &=\phi_0 + \heavymode(\posvec,t),  \label{eq:fielddecomp_longitudinal} \\
    \phi^{a}(\posvec,t) &= \gs^a(\posvec,t),   \label{eq:fielddecomp_transverse}
\end{align}
\end{subequations}
where the superscripts $1$ and $a = 2, \dots, d$ denote the vector components of the fields, $\gsvec(\posvec,t)$ is the Goldstone mode and $\heavymode(\posvec,t)$ is the heavy mode associated with the fluctuations about the global, uniform magnetisation with magnitude $\phi_0$. While this decomposition is useful to analyse the system near its upper critical dimension, for the present two-dimensional case, we opt instead for a phase-angle decomposition
\begin{equation} \elabel{phaseangledecomp}
    \phivec(\posvec,t) = \left(\phi_0 + \sigma(\posvec,t) \right)\begin{pmatrix}
        \cos(\phase(\posvec,t))\\
        \sin(\phase(\posvec,t))
    \end{pmatrix}
\end{equation}
where $\phi_0$ corresponds to the magnitude of the local magnetisation, $\phase(\posvec,t)$ to the gapless Goldstone mode and $\sigma(\posvec,t)$ to the heavy mode. In this parameterisation, the Goldstone mode $\phase(\posvec,t)$ is understood to be a compact variable where $\phase(\posvec,t) + 2 \pi m$ is identified with $\phase(\posvec,t)$ for any $m \in \Zset$. This is to say that in the parameterisation \Eref{phaseangledecomp} a vector field $\phivec(\posvec,t)$ differentiable everywhere except for countably many isolated singularities will give rise to a scalar field $\phase(\posvec,t)$ such that
\begin{equation}
    \oint\dint{\posvec}\cdot\nabla\phase(\posvec,t) = 2\pi m
\end{equation}
with $m\in\Zset$.

Instead of using the parameterisation \Eref{phaseangledecomp} ``directly'' in \Eref{activeModelA} and integrating out the heavy mode, we posit the EoM for the gapless mode, $\phase(\posvec,t)$, based on symmetry considerations as we worry that in the present case, the ``direct'' approach might yield erroneous results: \Eref{activeModelA} is the correct description of the dynamics near the upper critical dimension, $d_c = 4$, with infinitely many non-linearities neglected based on irrelevance arguments. In two dimension, we might therefore miss important terms. Instead, we therefore attempt to diligently collect all terms allowed by symmetry arguments only.

%Using it in two dimensions, we might miss important terms. 
%On the other hand, in the latter approach, we are guaranteed to capture all the relevant terms consistent with the symmetry, provided we are diligent. 

As discussed in the main text, the dynamics of $\phase(\posvec,t)$ and that of $\phase'(\posvec',t)$, derived from $\phase(\posvec,t)$ via 
% To ascertain the equation of motion for the Goldstone mode, $\phase(\posvec,t)$, we first need to determine its symmetries. Applying the symmetry operation, \Eref{symmetrymodelA}, to the phase-angle parametrisation of the order parameter, \Eref{phaseangledecomp}, we find that
\begin{subequations} \elabel{symmetriesofangle}
\begin{alignat}{3}
    \phase'(\posvec',t) &= \phase(\posvec,t) + \psi, \qquad &\text{with}\ \posvec'&=\begin{pmatrix}
        \cos\psi & -\sin \psi \\
        \sin \psi &\cos \psi
    \end{pmatrix} \posvec\quad\text{and global}\ \psi \in [0, 2\pi)\elabel{sorotation} \\
    \phase'(\posvec',t) &= -\phase(\posvec,t), \qquad &\text{with}\ \posvec'&=\begin{pmatrix}
        1 & 0 \\
        0  &-1
    \end{pmatrix}  \posvec \elabel{parityrotation}
\end{alignat}
\end{subequations}
are identical.
% The two independent symmetries correspond, respectively, to the continuous $SO(2)$ part, parameterised by $\psi$, and the discrete $\Zset_2$ part of the $O(2) \cong SO(2) \rtimes \Zset_2$ group. \Eref{sorotation} captures the rotational symmetry of the system whereas \Eref{parityrotation} captures the parity symmetry of the system. 
Any term featuring in the EoM for $\phase(\posvec,t)$ must be consistent with the symmetries in \Eref{symmetriesofangle}. 

% We argue that including every such term consistent produces the correct EoM describing the dynamics of the Goldstone modes of this universality class. 
In contrast, \citeauthor{chate_dynamic_2024} \cite{chate_dynamic_2024} restrict the terms in the EoM by also demanding that every \emph{deterministic} term must be a total divergence. 
%\gpcomment{1How about here: While their approach results in non-linearities similar to ours, \cf \Eref{EoMgeneral} and their Eq.~(6), we comment here on the validity of the argument. --- they don't arrive at the same eom as we do, right?} 
They arrive at the same EoM (\cf\ Eq.~(7) in \cite{chate_dynamic_2024} and \Eref{model_final} in our main text), based on a reasoning that we believe does not apply.
The authors appear to argue (``Because the symmetry is spontaneously broken, the deterministic part [\ldots] can only come from differences with the local environment. Mathematically, this means that it must be expressed as the divergence of a rank-2 tensor [\ldots].'') that in the ordered state, fluctuations of $\phase(\posvec,t)$ integrate up like ``differences'', so that the change of the total angle $\partial_t \int_A \ddintx{2}{\pos} \phase(\posvec,t)$ in any volume $A$ is essentially uncorrelated noise. As a result, the zero mode, $\int \ddintx{2}{\pos} \phase(\posvec,t)$, of $\phase(\posvec,t)$ performs ``violent'' Brownian motion without dissipation \cite[][p.~268]{Grinstein:1995}. 

The assumption of such ``telescoping differences'' is clearly valid in the presence of reciprocity, where indeed the forces acting on the degrees of freedom overall cancel each other, say $\int_A \ddintx{2}{\pos} \nabla\cdot\phivec=\oint_{\partial A} \dint{\posvec}\cdot\phivec$. Non-reciprocal terms, like a vision-cone or a KPZ-term \cite{KPZ_1986}, do not have this property. 
%While $\int_A \ddintx{2}{\pos} \nabla\cdot\phivec=\oint_{\partial A} \dint{\posvec}\cdot\phivec$
For example 
$\int_A \ddintx{2}{\pos} (\nabla\phase(\posvec,t))^2$ cannot be expected to be mere, uncorrelated noise, despite the dynamics it gives rise to being rotationally symmetric.
%However, we believe this is not necessarily true. This assumption holds for equilibrium systems where the pairwise forces between the spins are reciprocal, but breaks down for active systems where reciprocity is violated. Active systems can remain rotationally invariant while nevertheless violating \citeauthor{chate_dynamic_2024}'s assumption. As an example, consider the microscopic vision-cone model of \cite{dopierala_inescapable_2025}. One can show that
%\escomment{I am confused here. Why do we talk about $\phivec$ when the relevant quantity is $\phase$?} \gpcomment{Because i think of $\phi$ as being the key degree of freedom, so it's intuitive to see how things telescope or not. But i can see your point, please change the examples accordingly. }
More concretely, in the vision-cone model of \cite{dopierala_inescapable_2025},
the quantity $\sum_{i = 1}^{N}\phase_i(t)$, the microscopic equivalent of $\int \ddintx{2}{\pos} \phase(\posvec,t)$, has a non-vanishing deterministic drift due to the non-reciprocal forces. Nevertheless, it is demonstrated there that the model still falls into the Malthusian universality class, as the coarse-grained magnetisation is shown to obey \Eref{activeModelA} in the continuum limit. 
%It is certainly possible that \citeauthor{chate_dynamic_2024}'s assumption holds at the fixed point of the RG flow; however, this needs to be demonstrated rather than assumed.
%\gpcomment{I think we need to explain: 1) That Chate ends up with a similar eom 2) How this is possible if he makes very different assumptions (it's that he allows for $n\times R$ --- which I honestly don't understand: How can you demand that R telescopes and then make it so that the RHS of $\dot{n}$ doesn't?) 3) We need to say that our treatment after having a similar eom differs, because we do not expand in small theta. 1 are addressed in the comment I added above ("1H ow about here"), 3 to be done below ("3How about here"). I don't know what to do about 2 --- my best bet atm is that their whole approach is inconsistent, but I don't know and I worry that I just don't understand.}
Despite the assumption of reciprocal forces acting on $\phase$, \citeauthor{chate_dynamic_2024} recover \Eref{model_final}, because the terms in \Eref{EoMgeneral_expanded} that are non-reciprocal turn out to be irrelevant. In other words, the sole non-linearity in the EoM is a divergence and in this sense reciprocal,
\begin{equation}\elabel{invariant_is_divergence}
    \partial_\parallel \phase(\posvec,t) 
    = \cos\phase(\posvec,t) \partial_x \phase(\posvec,t) + \sin\phase(\posvec,t) \partial_y \phase(\posvec,t) 
    = \partial_x \sin \phase(\posvec,t) - \partial_y \cos \phase (\posvec,t) = -\nabla\cdot \unitnormalperpvec(\phase(\posvec,t)) \ ,
\end{equation}
anticipating \Erefs{defn_unit_vectors} and \eref{def_partpar}.

%To determine the EoM for the Goldstone mode, we first identify the symmetry-allowed ingredients from which we can construct the deterministic terms in our EoM. 

To determine the EoM for the Goldstone modes, we first identify the general characteristics any deterministic term ought to have. For example, the invariance \Eref{sorotation} of $\phase(\posvec,t)$ under shifts by $\psi$ excludes any term depending solely on the absolute value of $\phase(\posvec,t)$, say polynomial in $\phase(\posvec,t)$.
%that is not accompanied by a derivative. 
% CONT HERE 1 June 2026 19.24pm
% 1) Emir needs to approve l.523 ("\Eref{symmetrymodelA} is stricter than in...") to here. DONE
% 1a) I am not so super happy to take Chate on so fiercly. Of course, what they say makes little sense, but I am not up for a fight with them. DONE
% 2) Emir needs to agree to the tensor notation from here to l.668 ("Introducing further...") DONE
% 3) A little bit of clean-up to remove my additional notes. DONE
% 4) Carry on from l.668 
% 1 June 2026 pruess:
% I think a lot of what follows gets quite a bit messier in vector notation. All of this is much cleaner using tensor (index) notation. I totally opt for that. I would introduce the transformation and note that it is unitary (inverse is transpose) and off you go:
In the following we avoid ambiguity by adopting index (tensor) notation and Einstein summation convention where repeated indices are summed over. In this notation, \Eref{symmetriesofangle} becomes
\newcommand{\rotation}{\mathcal{R}}
\newcommand{\parity}{\mathcal{P}}
\begin{subequations}
\begin{align}
    x'_i&=\rotation_{ij}(\psi)x_j \\
    x'_i &= \parity_{ij} x_j
\end{align}
\end{subequations}
with unitary rotation matrix
\begin{equation}
    \rotation_{ij}(\psi)=
    \begin{pmatrix}
    \cos\psi & -\sin\psi\\
    \sin\psi & \cos\psi
    \end{pmatrix}_{ij}
    =
    \rotation^{-1}_{ji}(\psi)
    \end{equation}
such that $x_i=\rotation^{-1}_{ij}x'_j$, 
and unitary parity matrix
\begin{equation}\elabel{def_parity}
    \parity_{ij}=
    \begin{pmatrix}
    1 & 0\\
    0 & -1
    \end{pmatrix}_{ij}
    =
    \parity^{-1}_{ij}
\end{equation} 
such that $x_i=\parity^{-1}_{ij}x'_j$. 

Focusing on rotations for now, spatial derivatives pick up a Jacobian 
\begin{align}
    \frac{\partial}{\partial{x_i}}&=\partial_i=\rotation_{ji}(\psi)\partial_j' 
\end{align}
so that, say $\nabla \phase(\posvec,t)$ is generally not invariant under rotations \Eref{sorotation}. Suitable projections, however, are. To this end, we introduce
\begin{equation}
\label{eq:defn_unit_vectors}
    \unitnormalvec(\phase(\posvec,t)) = \begin{pmatrix}
        \cos\phase(\posvec,t) \\
        \sin\phase(\posvec,t)
    \end{pmatrix}
    \qquad\text{and}\qquad
    \unitnormalperpvec(\phase(\posvec,t)) = \begin{pmatrix}
        -\sin\phase(\posvec,t) \\
        \phantom{-}\cos\phase(\posvec,t)
    \end{pmatrix} \ ,
\end{equation}
% n transforms like x --- don't think of n transforming. Think of them behaving in a certain way.
which have the useful property $n_i(\phase'(\posvec',t))=\rotation_{ij}(\psi)n_j(\phase(\posvec,t))$ and $n_{\perp i}(\phase'(\posvec',t))=\rotation_{ij}(\psi)n_{\perp j}(\phase(\posvec,t))$, using that $\phase'(\posvec',t)=\phase(\posvec,t)+\psi$. It is then easy to show that 
\begin{equation}\elabel{def_partpar}
\partialpar\defequal
    \unitnormalvec(\phase(\posvec,t))\cdot\nabla = 
    n_i(\phase(\posvec,t))\partial_i = 
    n_i(\phase'(\posvec',t))\partial'_i = 
    \unitnormalvec(\phase'(\posvec',t)))\cdot\nabla' =
    \partialpar'
\end{equation}
and similarly
\begin{equation}\elabel{def_partperp}
\partialperp \defequal
    \unitnormalperpvec(\phase(\posvec,t))\cdot\nabla = 
    n_{\perp i}(\phase(\posvec,t))\partial_i = 
    n_{\perp i}(\phase'(\posvec',t))\partial'_i = 
    \unitnormalperpvec(\phase'(\posvec',t))\cdot\nabla'
    =
    \partialperp'
    \ ,
\end{equation}
so that $\partialpar$ and $\partialperp$ are invariant under the rotation \Eref{sorotation}. Our approach to exhaustively explore the space of all terms allowed under the rotational symmetry will therefore focus on the space of terms written in $\partialpar$ and $\partialperp$.

% "8 June 2026: Read2": Is this really true?
%\Erefs{def_partpar} and \eref{def_partperp} are two orthogonal projections of the $\nabla$-operator. \emph{Every} derivative constructed to be invariant under \Eref{sorotation}, namely by multiplying any Jacobian by its inverse, must be decomposable into $\partialpar$ and $\partialperp$. 
%In \Eref{invariant_is_divergence} this inverse is effectively provided by the function being differentiated on the far right of the expression, \ie by the divergence itself, but it can equivalently be written in terms of $\partialpar$ on the far left. 
%Our approach to exhaustively explore the space of all terms allowed under rotational and parity symmetry can therefore focus on the space of terms written in $\partialpar$ and $\partialperp$.

% \gpcomment{I suppose $\partialperp\theta=\nabla\cdot\unitnormalvec$? Yep.}

We further consider the parity transformation \Eref{parityrotation}, whose matrix representation $\parity$ is introduced in \Eref{def_parity}. The two vectors \Eref{defn_unit_vectors} obey 
\begin{equation}
    \unitnormal_i(\phase'(\posvec',t)) 
    = \unitnormal_i(-\phase(\posvec,t))
    = \parity_{ij} \unitnormal_j(\phase(\posvec,t)) 
\end{equation}
where $x_i'=\parity_{ij}x_j$ and similarly 
\begin{equation}
    \unitnormalperp{i}(\phase'(\posvec',t)) 
    = \unitnormalperp{i}(-\phase(\posvec,t)) \ .
    = -\parity_{ij} \unitnormalperp{j}(\phase(\posvec,t)) 
\end{equation}
Projecting those against $\nabla$, we find
\begin{equation}
    \partial_\parallel 
%    = \unitnormalvec(\phase(\posvec,t))\cdot\nabla 
    = \unitnormal_i(\phase(\posvec,t)) \partial_i 
    = \unitnormal_i(\phase(\posvec,t)) \parity_{ij}\partial'_j
    = \unitnormal_j(\phase'(\posvec',t)) \partial'_j
%    = \unitnormalvec'(\phase(\posvec,t))\cdot\nabla' 
  =   \partial'_\parallel
\end{equation}
remaining unchanged and 
\begin{equation}
    \partial_\perp
%    = \unitnormalperpvec(\phase(\posvec,t))\cdot\nabla 
    = \unitnormalperp{i}(\phase(\posvec,t)) \partial_i 
    = \unitnormalperp{i}(\phase(\posvec,t)) \parity_{ij}\partial'_j
    = -\unitnormal_j(\phase'(\posvec',t)) \partial'_j
%    = \unitnormalvec'(\phase(\posvec,t))\cdot\nabla' 
  =   -\partial'_\perp
\end{equation}
acquiring an overall sign.

Constructing the EoM, we were mislead to think that all terms ought to be invariant under \Erefs{symmetriesofangle}. Rather, every term needs to transform in the same way in order to leave the EoM overall invariant under the transformations. This is of particular importance as the parity transform of $\phase(\posvec,t)$ confers a sign to $\dot{\phase}(\posvec,t)$. The overall sign also determines the parity of the number of $\partialperp$ operators, as $\partialperp\phase$ is overall even, while $\partialperp^2\phase$ is odd.

The deterministic part of the EoM, that is, the right hand side of $\dot{\phase}(\posvec,t)$ apart from the noise, is thus limited by the following constraints
\begin{itemize}
\item Every term must be invariant under the rotations in \Erefs{sorotation}. 
%This excludes simple poynomials $\phase(\posvec,t)$
\item Every term must pick a sign under parity, \Eref{parityrotation}.
%\item Every $\phase(\posvec,t)$ 
\end{itemize}
% "8 June 2026: Reread1"
This does not appear to be very restrictive: While derivatives ought to appear in the form \Erefs{def_partpar} and \eref{def_partperp}, there is, \emph{a priori}, no restriction on order or non-linearity with which $\phase(\posvec,t)$ enters. Invariance under \Eref{sorotation} and negative parity under \Eref{parityrotation} need to be tested for each term individually. While for example $\phase(\posvec,t)=\phase'(\posvec',t)-\psi\ne\phase'(\posvec',t)$ clearly fails \Eref{sorotation}, this is not the case for $\partial_\parallel\phase(\posvec,t)=\partial_\parallel'\phase'(\posvec',t)$ and might not be immediately obvious for $\partial_\parallel\phase(\posvec,t)^3$. 
% The latter fails for $\psi$ still entering --- an overall derivbatives helps with linear terms, but here I would need $\partial_\parallel^3\phase(\posvec,t)^3$.
% \gpcomment{Is it obvious that every derivative \emph{must} feature in the form \Erefs{def_partpar} and \eref{def_partperp}? Let's say I take an arbitrary term and multiply it by a suitable ``backtransformation matrix'', would that always boil down to \Erefs{def_partpar} and \eref{def_partperp}? See  "8 June 2026: Read2". How about providing a suitable inverse Jacobian via the function being differentiated? Well, you still boil it down to $\partial_\parallel$ and $\partial_\perp$ --- it's what happens in \Eref{invariant_is_divergence}.}
We proceed by exploring all terms with an increasing number of derivatives. Because we will find \emph{a posteriori} that terms containing three derivatives or more are irrelevant, we will stop the procedure after two derivatives.

% XXX CONT HERE Mon 8 June 2026 18.05pm
% What needs to happen next:
% Check the above, starting from "8 June 2026: Reread1" and also "8 June 2026: Read2"
% We go through all terms, starting without derivatives and dismissing immediately.
% One, two derivatives follow. 
% Emir says, that \partial_\parallel^2 is a total freaking mess, but we will get here.
% We then demonstrate in passing that three derivatives are irrel.

Terms without a derivative, say $f_0(\phase(\posvec,t)$ in the EoM,  
\begin{equation}
    \partial_t \phase(\posvec,t) = \ldots + f_0(\phase(\posvec,t)) + \ldots
\end{equation}
must be translational invariant, $f_0(\phase) = f_0(\phase+\psi)$, so that $f_0(\phase)$ must be constant. At the same time, the function must be odd in $\phase$ to satisfy the parity symmetry, leaving only $f_0(\phase)\equiv0$.

Terms with a single derivative, say  
\begin{equation}
    \partial_t \phase(\posvec,t) = \ldots + f_1(\phase(\posvec,t))\partialpar g_1(\phase(\posvec,t)) + \ldots = \ldots + f_1(\phase(\posvec,t))g_1'(\phase(\posvec,t))\partialpar\phase(\posvec,t)
\end{equation}
will need to have $f_1(\phase)g_1'(\phase)$ constant for the same reason as $f_0(\phase)$ was constant, but because $\partialpar\phase$ is odd, parity symmetry of the whole term no longer disallows non-zero  $f_1(\phase)g_1'(\phase)$. In other words, $\lambdatilde \partialpar \phase$ is allowed. On the other hand, $\partialperp\phase$ is even, so the constant prefactor in front of it must vanish. 

Terms with two derivatives acting on a function of $\phase$ can always be written in terms of derivatives of this function multiplying terms containing two derivatives, namely
$\partialpar^2\phase$, 
$\partialperp^2\phase$,
$\partialpar\partialperp\phase$,
$\partialperp\partialpar\phase$,
$(\partialpar\phase)^2$,
$(\partialperp\phase)^2$ and
$(\partialpar\phase)(\partialperp\phase)$. Of those, only $\partialpar^2\phase$, $\partialperp^2\phase$ and $(\partialpar\phase)(\partialperp\phase)$ have odd parity. Again, translational invariant pre-factors $f_2(\phase)$ with odd parity must vanish, whereas even prefactors are constant. We thus allow for $D_1\partialpar^2\phase$, $D_2\partialperp^2\phase$ and $g_1(\partialpar\phase)(\partialperp\phase)$. Because $\nabla^2\phase=\partialpar^2\phase+\partialperp^2\phase$ the presence of the first two terms is unsurprising, albeit with anisotropy.

There are terms of higher order in derivatives, but we will find out that they are irrelevant in the RG sense. At this stage, the EoM is therefore 
%
%
% That the terms parameterised by $D_1$ and
%
%
% The only constant that has the  
%
%
% Terms with a single derivative are genbe
%
%
%With these ingredients established, the Langevin equation for the Goldstone modes, to second order in derivatives, is given by
\begin{equation} \elabel{EoMgeneral}
    \dot{\phase}(\posvec,t) = \lambdatilde \partialpar \phase(\posvec,t) + D_1 \partialpar^2 \phase(\posvec,t) + D_2 \partialperp^2 \phase(\posvec,t) + g_1 (\partialperp \phase(\posvec,t))(\partialpar\phase(\posvec,t)) + \sqrt{2\noise} \xi(\posvec,t)
\end{equation}
where $\xi(\posvec,t)$ is a unit Gaussian noise with zero mean, satisfying
\begin{equation} \label{eq:noise}
    \langle\xi(\posvec,t) \xi(\posvec',t')\rangle = \delta(\posvec - \posvec') \delta(t-t').
\end{equation}
%Terms like $(\partialperp \phase)^2$, $\partialperp \partialpar \phase$, $\partialpar\partialperp\phase$ and $(\partialpar\phase)^2$ are excluded in \Eref{EoMgeneral} as they are not invariant under the parity symmetry, \Eref{parityrotation}. 

\Eref{EoMgeneral} can be interpreted and analysed in terms of standard growth equations \cite{Meakin,PimpinelliVillain,VicsekFractalGrowth,BarabasiStanley}, with a drift term parameterised by $\lambdatilde$, diffusion by $D_1$ and $D_2$ and a KPZ-like non-linearity by $g_1$. However, all of these terms contain (non-linear) trigonometric functions within $\partialpar$ and $\partialperp$, \Erefs{def_partpar} and \eref{def_partperp}. In order to make relevance arguments, we need to lay bare these pre-factors.

Using the definitions of $\partialperp$ and $\partialpar$, \Erefs{def_partpar} and \eref{def_partperp}, we can rewrite each \Eref{EoMgeneral} in terms of ``plain derivatives'':
%expressed solely in terms of $\phase$, $\partial_x$ and $\partial_y$
\begin{subequations} 
    \begin{align}
        \partialpar^2 \phase &= \left\{ \cos^2(\phase) \partial_x  + \sin^2(\phase)\partial_y  + 2\sin(\phase) \cos(\phase) \partial_x \partial_y \right\} \phase \nonumber \\
        &\qquad\qquad\qquad + \left\{ \sin \phase \cos\phase ( -(\partial_x \phase)^2 + (\partial_y \phase)^2) + (\cos^2 \phase - \sin^2\phase) (\partial_x \phase)(\partial_y \phase) \right\} \nonumber \\
        &=\frac{1}{2}\nabla^2\phase + \left\{\frac{\cos(2\phase)}{2}(\partial^2_x -\partial^2_y)\phase + \sin(2\phase) \partial_x \partial_y \phase  \right\} \nonumber \\ 
        & \qquad \qquad\qquad -\left\{\frac{\sin(2\phase)}{2} \left[ (\partial_x \phase)^2 -  (\partial_y \phase)^2\right] - \cos(2\phase) (\partial_x \phase)(\partial_y \phase)\right\} \\
        \partialperp^2 \phase &= \left\{ \sin^2(\phase) \partial_x  + \cos^2(\phase)\partial_y  - 2\sin(\phase) \cos(\phase) \partial_x \partial_y \right\} \phase  \nonumber \\
        & \qquad \qquad - \left\{ \sin \phase \cos\phase ( -(\partial_x \phase)^2 + (\partial_y \phase)^2) + (\cos^2 \phase - \sin^2\phase) (\partial_x \phase)(\partial_y \phase) \right\} \nonumber \\
        &=\frac{1}{2}\nabla^2\phase - \left\{\frac{\cos(2\phase)}{2}(\partial^2_x -\partial^2_y)\phase + \sin(2\phase) \partial_x \partial_y \phase  \right\} \nonumber \\
        & \qquad \qquad + \left\{\frac{\sin(2\phase)}{2} \left[ (\partial_x \phase)^2 -  (\partial_y \phase)^2\right] - \cos(2\phase) (\partial_x \phase)(\partial_y \phase)\right\} \\
        (\partialperp \phase)(\partialpar \phase) &= \sin\phase \cos \phase\left[ (\partial_y \phase)^2 - (\partial_x \phase)^2 \right] + (\cos^2\phase - \sin^2\phase)(\partial_x \phase)(\partial_y \phase) \nonumber\\
        &= \frac{\sin(2\phase)}{2}\left[ (\partial_y \phase)^2 - (\partial_x \phase)^2 \right] + \cos(2\phase)(\partial_x \phase)(\partial_y \phase)  \ .
    \end{align}
\end{subequations}
%where we have employed the trigonometric double angle identities, $\cos(2\psi) = \cos^2(\psi) - \sin^{2}(\psi)$ and $\sin(2\psi) = 2 \sin(\psi) \cos(\psi)$ to eliminate any appearance of $\cos^2\psi$, $\sin^2\psi$ and $\sin\psi \cos \psi$. 
With the help of the above we can now rewrite \Eref{EoMgeneral} and collect terms,
%
%
% e can now rewrite \Eref{EoMgeneral} using \Eref{23}
%
% Collecting the terms, it follows that  
\begin{align} \label{eq:EoMgeneral_expanded}
    \dot{\phase}  &= \lambdatilde\left\{\cos\phase \partial_x + \sin\phase \partial_y\right\} \phase + D \nablasquared \phase + \sqrt{2\noise}\xi  \nonumber\\
    &+\tilde{D}\left\{ \cos(2\phase)(\partial_x^2 - \partial_y^2)\phase + 2\sin(2\phase) \partial_x \partial_y \phase\right\} + \tilde{g}\left\{ \sin(2\phase)((\partial_x \phase)^2 - (\partial_y \phase)^2) -  2\cos(2\phase) (\partial_x \phase)(\partial_y \phase)\right\}
\end{align}
where we have defined the isotropic diffusivity $D = (D_1 + D_2)/2$, the anisotropic diffusivity $\tilde{D} = (D_1 - D_2)/2$ and the non-linearity $\tilde{g} = (D_2 - D_1 - g_1)/2$. Each term associated with a distinct coupling constant in \Eref{EoMgeneral_expanded} is invariant under the symmetry operations \Eref{symmetriesofangle}:
\begin{subequations}\elabel{invariant_collections}
    \begin{align}
        \cos(2\phase)(\partial_x^2 - \partial_y^2)\phase + 2\sin(2\phase) \partial_x \partial_y \phase &= T^{ij}_1(\theta) \partial_i \partial_j \theta &\quad \text{with} \quad T_1^{ij}(\theta) &= 
        \begin{pmatrix}
            \cos2\theta & \sin2\theta \\
            \sin 2\theta & -\cos2\theta
        \end{pmatrix}_{ij} \\
        \sin(2\phase)((\partial_x \phase)^2 - (\partial_y \phase)^2) -  2\cos(2\phase) (\partial_x \phase)(\partial_y \phase) &= T^{ij}_2(\theta) (\partial_i\theta) (\partial_j \theta) &\quad \text{with} \quad T_2^{ij}(\theta) &= 
        \begin{pmatrix}
            \sin2\theta & -\cos2\theta \\
            -\cos 2\theta & -\sin2\theta
        \end{pmatrix}_{ij}
    \end{align}
\end{subequations}
where $T_1(\phase)$ and $T_2(\phase)$ are invariant under rotations and have suitable parity,
\begin{subequations}
    \begin{align}
        \parity^{-1}T_1(-\phase)\parity &=\rotation^{-1}(\psi)T_1(\phase+\psi)\rotation(\psi)=T_1(\phase)\\
        -\parity^{-1}T_2(-\phase)\parity &=\rotation^{-1}(\psi)T_2(\phase+\psi)\rotation(\psi)=T_2(\phase) \ .
    \end{align}
\end{subequations}
While \Eref{EoMgeneral_expanded} is derived from \Eref{EoMgeneral} using essentially trigonometric identities, the parametrisation used in \Eref{EoMgeneral_expanded} is therefore invariant under renormalisation, which will preserve the symmetries \Eref{symmetriesofangle}. Was it not for \Eref{invariant_collections} demonstrating that each term in \Eref{EoMgeneral_expanded} was invariant under the symmetries in \Eref{symmetriesofangle}, we would need to allow for many more couplings to parameterise the theory.
%, rather than collecting them into ???

%We assume that RG will preserve the symmetries

Neglecting irrelevant terms involving $\cos2\phase$, $\sin2\phase$, which will be justified below, and defining 
\begin{equation}
\elabel{def_lambdatilde}
\lambdatilde = D \lambda \ ,
\end{equation}
we can rewrite \Eref{EoMgeneral_expanded} as
\begin{equation} \label{eq:model_final}
    \dot{\phase}(\posvec,t) 
     =D \lambda \left\{ \partial_x \left(\sin  \phase(\posvec,t) \right) - \partial_y\left( \cos \phase(\posvec,t)\right)\right\} +  D \nablasquared \phase(\posvec,t) + \sqrt{2\noise}\xi(\posvec,t).
\end{equation}
\Eref{model_final} is the central result of this section and forms the basis for all the subsequent analysis.
% \gpcomment{3How about here:} 
In contrast to \cite{chate_dynamic_2024}, we retain the trigonometric non-linearities rather than making a small angle approximation. %}

% XXX CONT HERE 15 June 2026

\section{Field Theory} \label{sec:pertexpd2}
In this section, we derive the field theory associated with the Langevin \Eref{model_final} and set-up the perturbative expansion by finding the bare propagators and the perturbative vertices.

Using well-established techniques \cite{janssen_lagrangean_1976, bausch_renormalized_1976, janssen_renormalized_1977, janssen_renormalized_1992, tauber_critical_2014, MSR, Dominicis1976TECHNIQUESDR}, the Langevin \Eref{model_final}, can be mapped onto a field theory with action $\action=\action_0+\action_I$ that can be separated into a harmonic part,
\begin{equation}  \label{eq:action_harm}
    \action_0[\phasetilde, \phase] = \int \ddintx{2}{\pos} \int \dint{t} \phasetilde(\posvec,t) \left\{ (\partial_t - D \nablasquared + D\mass^2)\phase(\posvec,t) -  \noise\phasetilde(\posvec,t) \right\} \ ,
\end{equation}
where we have introduced a mass term $\mass^2$ to be discussed below,
and a perturbative part
\begin{equation}  \label{eq:action_pert}
    \action_I[\phasetilde, \phase] = D \lambda\int \ddintx{2}{\pos} \int \dint{t} \left\{ \sin \phase(\posvec,t)\partial_x \phasetilde(\posvec,t) - \cos \phase(\posvec,t)\partial_y \phasetilde(\posvec,t) \right\}
\end{equation}
where the $\partial_{x,y}$ of the Langevin \Eref{model_final} has been moved to act only on the auxiliary field $\phasetilde(\posvec,t)$ through integration by parts. For future reference, our sign convention is such that $\action$ enters into the path integral as $\exp{-\action}$. 
% \gpcomment{1 July 2026: Emir to confirm the below, as well as $\av{\phase\phasetilde}_0$ below.}
Henceforth we use the notation
\begin{subequations}
\begin{align}\elabel{def_av}
\av{\bullet}_0 
&\defequal \int\Dint{\phasetilde}\Dint{\phase}
\ \bullet \exp{-\action_0[\phasetilde, \phase]} \\
\av{\bullet} &\defequal \int\Dint{\phasetilde}\Dint{\phase}
\ \bullet\ \exp{-\action_0[\phasetilde, \phase] -\action_I[\phasetilde, \phase] } = \sum_{n=0}^\infty \av{\bullet \frac{\big(-\action_I[\phasetilde, \phase]\big)^n}{n!}}_0
\ .
\end{align}
\end{subequations}

\Eref{action_harm} immediately yields the bare propagators in real space and direct time, where we will carry out most of our loop calculations,
\begin{subequations}
    \elabel{diagramsharmonic}
\begin{align} 
  \begin{tikzpicture}[baseline = (o.base)]
  \begin{feynman}
    \vertex (o) at (0,0) ;
    \vertex (i) at (1.8,0) ;
    % First curve with moderate bend
    \diagram* {
      (o) -- [anti fermion] (i)
    };
  \end{feynman}
\end{tikzpicture}
&\corresponds G_{0}(\posvec, t) \defequal \frac{\Heaviside(t)}{4\pi D t} \exp{-D m^2 t - \frac{\pos^2}{4Dt}} = \av{\phase(\rvec+\rvec_0,t+t_0)\,\phasetilde(\rvec_0,t_0)}_0 \elabel{bare_propagator} \\ 
%\quad\text{and}\quad
\begin{tikzpicture}[baseline = (o.base)]
  \begin{feynman}
    \vertex (o) at (0,0) ;
    \vertex [above left=of o] (i1) ;
    \vertex [below left=of o] (i2) ;
    % \vertex (i1) at (-1.5,-0.75) {\(b\)};
    % \vertex (i2) at (-1.5, 0.75) {\(a\)};
    % First curve with moderate bend
    \diagram* {
      (o) -- [fermion] (i1),
      (o) -- [fermion] (i2)
    };
  \end{feynman}
\end{tikzpicture}
&\corresponds C_{0}(\posvec,t) \defequal %\frac{2\noise}{\left( -\imag \omega + D (\kvec^2 + \mass^2)\right)\left( \imag \omega + D (\kvec^2 + \mass^2)\right)} \ ,
2 \noise \int_{-\infty}^\infty \dint{t'}\int_{\Rset^2}\ddintx{2}{\pos'} G_{0}(\posvec'+\posvec, t'+t) G_{0}(\posvec', t') = \frac{\noise}{4 \pi D}\int_{D\mass^2 |t|}^\infty\frac{\dint{v}}{v} \exp{-v-\frac{\mass^2\pos^2}{4v}}
\elabel{def_C0}
\end{align}
\end{subequations}
where $\Heaviside(t)$ is the Heaviside step function. 

Typically, it is not possible to evaluate the final $v$ integral in \Eref{def_C0} exactly for arbitrary time $t$. However, we can evaluate it at $t = 0$, corresponding to the equal time correlation function
\begin{align} \label{eq:corr_func_real_space_equal_time}
    C_0(\posvec,0) = \frac{\noise}{4\pi D} \int_{0}^{\infty} \frac{\dint{v}}{v} \exp{-v - \frac{\mass^2\pos^2}{4v}} = \frac{\noise}{2\pi D}K_0\left( \mass \pos \right)
\end{align}
where $K_0(x)$ is the modified Bessel function of the second kind. 

The amputated noise vertex
\begin{equation}
    \begin{tikzpicture}[baseline = (o.base)]
  \begin{feynman}
    \vertex (o) at (0,0) ;
    \vertex [above left=0.6 of o] (i1) ;
    \vertex [below left=0.6 of o] (i2) ;
    % \vertex (i1) at (-1.5,-0.75) {\(b\)};
    % \vertex (i2) at (-1.5, 0.75) {\(a\)};
    % First curve with moderate bend
    \diagram* {
      (o) -- [plain] (i1),
      (o) -- [plain] (i2)
    };
  \end{feynman}
\end{tikzpicture}
\defequal 2 \noise,
\end{equation}
may be treated perturbatively whenever convenient.
For completeness we further state our convention of the Fourier transform, as we will occasionally capitalise on its simplicity, 
\begin{subequations} \elabel{fourierconvention}
\begin{align}
\elabel{fourierconvention_phasetilde}
    \phasetilde(\posvec,t) &= \int \ddbarint{2}{k} \int \dbarint{\omega} e^{\imag(\kvec \cdot \posvec - \omega t)} \phasetilde(\kvec, \omega) \\
    \phase(\posvec,t) &= \int \ddbarint{2}{k} \int \dbarint{\omega} e^{\imag(\kvec \cdot \posvec - \omega t)} \phase(\kvec, \omega) ,
\end{align}
\end{subequations}
where $\dbar^2 k = d^2 k/(2\pi)^2$ and $\dbar \omega = d \omega/ (2\pi)$. Correspondingly, we use $\deltabar(\kvec) \defequal (2\pi)^{2}\delta(\kvec)$ and $\deltabar(\omega) \defequal 2\pi \delta(\omega)$.

The present field theory suffers from both infrared (IR) and ultraviolet (UV) divergences whose regularisation determines its critical behaviour. The IR divergence can be cured by considering the theory in a finite, periodic box of volume $V$, allowing for a well-defined steady state, but complicating loop calculations.
%for the $\kvec = \zerovec$ mode, which performs a random walk 
%, while, however, introducing discrete modes and therefore sums rather than integrals in the calculation of loops. 
We rather consider the thermodynamic limit and introduce as a matter of convenience the additional mass term $\mass^2$ in the action \cite{amitRenormalisationGroupAnalysis1980a}, \Eref{action_harm}
which ensures finite correlation and response functions \Erefs{diagramsharmonic} in the IR.
The addition of the mass however breaks the rotational symmetry \Eref{sorotation}, which can be seen by adding the term explicitly as $-\mass^2\phase$ on the right-hand side of \Eref{model_final} and shifting $\phase$ by $\psi$. Yet, adding the mass constitutes only a ``soft breaking of symmetry''  \cite{amitRenormalisationGroupAnalysis1980a} which disappears at the critical point where the (renormalised) mass vanishes, so that neither the flow functions and nor their roots depend on $\mass$. %Furthermore, once the RG procedure is performed, we take $\mass^2$ to be zero, recovering the original model in \Eref{model_final}. 

The mass should strictly be thought of as a small quantity so that it parallels the effect of placing the system in a finite, yet large box of volume $V$. Henceforth we will be interested in correlations across large distances while maintaining $\pos^2 \ll1/\mass^2$, beyond which correlations are mean-field like. 
% With this modification, the bare propagator and the correlation function in Fourier space now read
% \begin{subequations} \label{eq:IRregulatedbareobservables}
%     \begin{align}
%         G_0(\kvec, \omega) &= \frac{1}{-\imag \omega + D( \kvec^2 + \mass^2)}, \label{eq:Ir_regulated_prop}\\
%         C_0(\kvec,\omega) &= \frac{2\noise}{\omega^2 + D^2(\kvec^2 + \mass^2)^2} \label{eq:Ir_regulated_corr}
%     \end{align}
%     \ .
% \end{subequations}

As for the UV regularisation, which is most conveniently implemented by dimensional regularisation \cite{bellac_quantum_1992}, we introduce similarly a UV cutoff $a$ into the correlator \Eref{def_C0} only,
\begin{equation} \label{eq:corr_func_reg}
C_0(\posvec,t;a) = \frac{\noise}{4\pi D} \int_{D\mass^2 |t|}^\infty\frac{\dint{v}}{v} \exp{-v-\frac{\mass^2(\pos^2+a^2)}{4v}}
\end{equation}
%,
% \begin{equation}
%     \frac{\Heaviside(t)}{4\pi D t} \exp{-\frac{\pos^2 + a^2}{4 D t} - D\mass^2 t} \label{eq:bare_prop_real_space}
% \end{equation}
effectively ``preventing'' distances $\pos$ smaller than $a$ from being probed. For equal times, the regularised correlation function can be expressed in closed form, $C_0(\posvec,t=0;a) = \noise K_0( \mass \sqrt{\pos^2+a^2}) )/(2\pi D)$, \Eref{corr_func_real_space_equal_time}.

To find the interaction vertices, we rewrite the perturbative part of the action \Eref{action_pert} by Taylor expanding $\sin(\phase)$ and $\cos(\phase)$ about a constant angle $\phase_0  = 0$,
\begin{equation}  \label{eq:action_pert_expansion}
    \action_I[\phasetilde, \phase] = D \lambda\sum_{n =  0}^{\infty} (-1)^n\int \ddintx{2}{\pos} \int \dint{t} \left\{ \frac{\phase^{2n+1}(\posvec,t)}{(2n+1)!}\partial_x \phasetilde(\posvec,t) - \frac{\phase^{2n}(\posvec,t)}{(2n)!}\partial_y \phasetilde(\posvec,t) \right\},
\end{equation}
from which we can read off infinitely many perturbative vertices
\begin{equation} \label{eq:pert_vertices}
    \begin{tikzpicture}[baseline = (o.base), scale = 1.5]
  \begin{feynman}
    \vertex (origin) at (0,0);
    \vertex (o) at (-0.4,0);
    \vertex (i1) at (0.4, 0.45);
    \vertex (i2) at (0.4, 0.25);
    \vertex (i3) at (0.4, 0.05);
    \vertex (i4) at (0.4, -0.45);
    \diagram* {
      (origin) -- [plain, edge label' = {\scalebox{0.7}{$\kvec$}}] (o),
      (origin) -- [plain] (i1),
      (origin) -- [plain] (i2),
      (origin) -- [plain] (i3),
      (origin) -- [plain] (i4),
    };
  \end{feynman}
  % Dots between i3 and i4 — larger and spaced to bridge the gap
  \node[scale=1] at (0.3, -0.05) {$\vdots$};
  % Curly brace spanning i1 to i4
  \draw [decorate, decoration={brace, amplitude=5pt, raise=4pt}]
    (0.4, 0.5) -- (0.4, -0.5)
    node [midway, right=10pt, scale =0.8] {$2n+1$};
  % Perpendicular tick mark on the outgoing leg, close to origin
  % The outgoing leg goes in the -x direction, so perpendicular is vertical
  \draw (-0.1, -0.06) -- (-0.1, 0.06);
  \node[scale=0.8] at (-0.1, -0.15) {$x$};
  \filldraw[fill=black, draw=black] (0,0) circle (0.03);
\end{tikzpicture}
= \imag (-1)^n D \lambda  k_x,
\qquad
\begin{tikzpicture}[baseline = (o.base), scale = 1.5]
  \begin{feynman}
    \vertex (origin) at (0,0);
    \vertex (o) at (-0.4,0);
    \vertex (i1) at (0.4, 0.45);
    \vertex (i2) at (0.4, 0.25);
    \vertex (i3) at (0.4, 0.05);
    \vertex (i4) at (0.4, -0.45);
    \diagram* {
      (origin) -- [plain, edge label' = {\scalebox{0.7}{$\kvec$}}] (o),
      (origin) -- [plain] (i1),
      (origin) -- [plain] (i2),
      (origin) -- [plain] (i3),
      (origin) -- [plain] (i4),
    };
  \end{feynman}
  % Dots between i3 and i4 — larger and spaced to bridge the gap
  \node[scale=1] at (0.3, -0.05) {$\vdots$};
  % Curly brace spanning i1 to i4
  \draw [decorate, decoration={brace, amplitude=5pt, raise=4pt}]
    (0.4, 0.5) -- (0.4, -0.5)
    node [midway, right=10pt, scale =0.8] {$2n$};
  % Perpendicular tick mark on the outgoing leg, close to origin
  % The outgoing leg goes in the -x direction, so perpendicular is vertical
  \draw (-0.1, -0.06) -- (-0.1, 0.06);
  \node[scale=0.8] at (-0.1, -0.15) {$y$};
\filldraw[fill=black, draw=black] (0,0) circle (0.03);
\end{tikzpicture}
= -\imag (-1)^n D \lambda k_y
\end{equation}
where the dash on the lines emerging towards the left denote multiplication by $k_{x}$ or $k_y$ respectively, corresponding to a derivative in real space with respect to $x$ or $y$, as indicated below the dashes. The components $k_{x}$ and $k_y$ refer to the $\kvec=(k_x,k_y)$ flowing through a propagator that may be attached, \ie $\kvec$ parameterises a physical field $\phase(\kvec,\omega)$ rather than a response field $\phasetilde$. Setting $\phase_0 = 0$ represents a convenient choice, where we measure angle $\phase(\xvec,t)$ with respect to the $x-$axis. Expanding the action about a uniform configuration induces a preferred direction as we discuss below.

% components carried by them, as indicated by $\kvec$ to parameterise a physical field $\phase(\kvec,\omega)$, corresponding to a gradient in real space with $x$ or $y$, as indicated below the dashes. 
%where the dash on the lines emerging towards the left denote multiplication by the $k_{x}$ or $k_y$ components carried by them, as indicated by $\kvec$ to parameterise a physical field $\phase(\kvec,\omega)$, corresponding to a gradient in real space with $x$ or $y$, as indicated below the dashes. 
%The signs above are chosen to reflect the parameterisation by a $\phase$-field attached to them vertices on the left.
%below the dashes referring to the vector component of the outgoing wave-vector. 

\section{Renormalisation Group Procedure} \label{sec:rg}
In this section, we perform the RG procedure to lowest order in the couplings. First, we perform a detailed dimensional analysis to identify the relevant and irrelevant operators which we use to justify the exclusion of certain terms, \Eref{EoMgeneral_expanded}, in the effective description of the Goldstone modes. The dimensional analysis further allows us to identify the Gaussian critical point about which we can perform the perturbative RG scheme. Finally, performing the renormalisation to leading order in the couplings, we derive the flow functions around this critical point and discuss the different phases displayed by the model.

%XXX CONT HERE 22 June 2026 17.11pm

\subsection{Dimensional Analysis and Relevance Arguments}\seclabel{DimAna}
In this section, we perform the usual dimensional analysis, to determine the relevance of the couplings in \Erefs{EoMgeneral_expanded}. Choosing $\dimension{x} = L$ to be the unit of length, we demand spatial interaction $D$ and noise amplitude $\noise$ remain invariant under spatial rescaling, \ie $[\noise] = B$ and $[D] = A$, so that the unit of time $[t]= L^2/A$ and that of mass becomes  $\dimension{\mass} = L^{-1}$. As a result the field $\phase$ has no length dimensions $[\phase] = B^{1/2}/A^{1/2}$. Requiring the arguments of $\sin$ and $\cos$ to be dimensionless further fixes the dimension of the noise, $\dimension{\noise} = B = A$, whereby the drift term, parameterised by $\lambdatilde$ in \Eref{EoMgeneral_expanded} has, as expected, the dimension of a velocity $[\lambdatilde]=L/T$. It follows that $[\lambda] = L^{-1}$, \Eref{def_lambdatilde}, as expected from an advective term, and $[\tilde{D}] = [\tilde{g}] = A$, so that $\tilde{D}$ and $\tilde{g}$ are marginal. 

%. Using these results, we further find that the couplings $\tilde{D}$ and $\tilde{g}$ are marginal, i.e $[\tilde{D}] = [\tilde{g}] = A$. 

The naive power counting above suggests that $\lambda$ is always relevant and, consequently, dominates over the diffusion and the marginal non-linearities parameterised by $\tilde{D}$ and $\tilde{g}$ \cite{Pruesssner:DriftAlwaysDominates:2004}. However, given that the drift-term is not a mere single derivative, but instead is embellished by a trigonometric ``modulation'', we will need to determine the \emph{effective} spatial scaling of $\lambda$.
%, for the present case, naive power counting is not sufficient to determine a term's relevance as it does not account for the anomalous scaling exhibited by some of the terms — unlike standard field theories where such scaling is absent at the mean-field level. 
Specifically, due to the logarithmic growth of correlations between the fields, we find that terms like 
\begin{equation} \label{eq:anomalous_scaling_exp}
    \av{\exp{\pm\imag n (\phase(\posvec,t) - \phase(\zerovec,t))}}_0 = \exp{-\frac{n^2}{2}\av{(\phase(\posvec,t) - \phase(\zerovec,t))^2}_0} = \exp{-n^2C_0(\zerovec, 0;a) + n^2C_0(\posvec, 0;a)} \propto \frac{1}{|\posvec|^{\frac{n^2 \noise}{2\pi D}}}, \qquad \pos^2 \ll 1/m^2\ ,
\end{equation}
acquire anomalous scaling, $\propto r^{-n^2\noise/(2\pi D)}$ above, already at the level of the free theory based on the harmonic action, $\action_0$, even though their naive engineering dimension is $L^0$. The scaling stated in \Eref{anomalous_scaling_exp} is based on the free correlator \Eref{corr_func_real_space_equal_time} and its regularised version \eref{corr_func_reg}, using \cite{gradshteyn2014}
\begin{equation} \label{eq:asymptotic_expansion}
    K_0\left(\mass \sqrt{\pos^2 + a^2}\right) = -\ln\left(\curlyC \mass \sqrt{\pos^2 + a^2} \right) + \mathcal{O}(\mass^2\pos^2),
    , \qquad a^2,\pos^2 \ll 1/m^2 \ ,
\end{equation}
where $\curlyC = e^\gamma/2$ and $\gamma$ is the Euler-Mascheroni constant.

As $\exp{\imag n (\theta(\xvec,t)-\theta(\nullvec,t))}=\exp{\imag n \theta(\xvec,t)}\exp{-\imag n \theta(\nullvec,t)}$, \Eref{anomalous_scaling_exp} suggests that $\exp{\pm\imag n \theta(\xvec,t)}$, and thus $\cos (n\theta(\xvec,t))$ and $\sin (n\theta(\xvec,t))$, scale like $|\xvec|^{-n^2\noise/(4\pi D)}$, which must be accounted for when assessing the scaling and thus the relevance of 
operators in the theory \cite{zinn-justin_quantum_2002}. Since $\lambda$ is multiplied by $\cos\phase$ or $\sin \phase $ in the equation of motion, this anomalous scaling modifies its effective engineering dimension, which is no longer $L^{-1}$ but instead
\begin{equation} \label{eq:dimension_lambda}
    \dimension{\lambda} = L^{-1} \left[\frac{1}{\sin(\phase(\posvec,t)} \right] = L^{\frac{\noise}{4\pi D} - 1}
\end{equation}
Based on the same dimensional argument, the couplings $\tilde{D}$ and $\tilde{g}$ dropped from \Eref{EoMgeneral_expanded} in \Eref{model_final} have subleading effective dimensions $\dimension{\tilde{D}} = \dimension{\tilde{g}} = A L^{\frac{4 \noise}{4 \pi D}}$, indicating that they are RG-irrelevant and do not contribute to the effective description of the Goldstone modes.

Higher order terms, containing higher derivatives, are irrelevant compared to the diffusive coupling $D$. For example, a generic coupling $u$ entering on the right-hand side of \Eref{EoMgeneral} as $u\partialpar^3\phase$ has engineering dimension $[u]=L^3/T=AL$. If such a term is preceded by a trigonometric function of $\phase$, it is only ever more irrelevant.

The relevance of the coupling $\lambda$ depends, according to \Eref{dimension_lambda}, on the ratio $\noise/ D$. When $\noise > 4 \pi D$, the coupling $\lambda$ is irrelevant and the theory is effectively free on large length scales. This defines one phase of our model which we call the free ``XY" phase. When $\noise < 4 \pi D$, the coupling $\lambda$ becomes relevant and the dynamics is driven out of equilibrium with the advective term. This regime defines the other phase displayed by the system, which we call the ``Malthusian" phase. The two phases are separated by a \emph{Gaussian} critical point, where the non-linearity vanishes,
\begin{equation} \label{eq:Gaussian_critical_point}
    \noise = 4\pi D, \qquad\lambda = 0.
\end{equation}
In the following, we perform a perturbative RG procedure  to characterise the physics around the Gaussian critical point.

% XXX CONT HERE 24 June 2026 14.41pm

\subsection{Leading Order Renormalisation} \label{sec:onelooprenormd2}
Perturbative RG procedures can be applied in the vicinity of a fixed point where a theory is fully solvable and the non-linear couplings are small. For example, in the case of $O(N)$-symmetric models with a $\lambda\phivec^4$ interaction, the expansion is performed about the upper critical dimension, $d_c = 4$, where the fixed point is Gaussian, $\lambda=0$. Using dimensional regularisation such that $\epsilon = d_c-d$ is small, the expansion parameter $\lambda$ is found to be of order $\epsilon$ itself.

%The expansion parameter is effectively  and consequently $\lambda\in\mathcal{O}(\epsilon)$.

%,  with the fixed point being Gaussian $\lambda = 0$ and, consequently, $\epsilon = 4-d$ and $\lambda$ serving as the expansion parameters. 
In the present case, the perturbative RG procedure has to be carried out close to the Gaussian critical point in \Eref{Gaussian_critical_point} with the expansion parameters $\lambda$ as well as \cite{amitRenormalisationGroupAnalysis1980a}
\begin{equation} \label{eq:defn_delta}
    \delta \defequal \frac{\noise}{4\pi D} - 1.
\end{equation}
Here, $\delta$ plays the role of $\epsilon = 4-d$ in $\phivec^4$ theories, as its sign determines the relevance of the non-linear coupling $\lambda$ and, consequently, the stability of the Gaussian fixed point \cite{amitRenormalisationGroupAnalysis1980a}. Unlike in $\phivec^4$ theories, however, $\delta$ is itself now allowed to flow under the RG, as it can receive non-trivial corrections through the renormalisation of $D$ or $\noise$, giving rise to a much richer flow diagram.

Assuming the theory is renormalisable close to the Gaussian critical point, \Eref{Gaussian_critical_point}, we introduce the following $Z$-factors to absorb the logarithmic UV divergences that appear in the perturbative expansion and the renormalised quantities, indicated by the subscript $R$: Firstly the field renormalisation,
\begin{equation} \elabel{def_field_renormalisation}
        \phase_{R}(\posvec,t) = \Zfactor{\phase} \phase(\posvec,t), \qquad \phasetilde_{R}(\posvec,t) = \Zfactor{\phasetilde} \phasetilde(\posvec,t), 
\end{equation}
and secondly the renormalisation of the couplings,
\begin{subequations} \elabel{Zfactorsdefn}
    \begin{gather}
        \mass^2_{R} = \mu^{-2} Z_{\mass^2} \mass^2 , ~~D_{R} = \Zfactor{D} D, ~~ \lambdaR = a \Zfactor{\lambda} \lambda \elabel{Zfactorsdefn_propagators}\\
        \noiseR = \Zfactor{\noise} \noise \ ,
    \end{gather}
\end{subequations}
where the renormalised coupling $\lambdaR$ has been rendered dimensionless by the multiplication of $\lambda$ by the UV regulator $a$. The action, \Erefs{action_harm} and \eref{action_pert}, in terms of renormalised quantities then reads
\begin{align}\label{eq:ren_action}
    \action = \int\ddintx{2}{x} \int \dint{t} &\bigg\{ \frac{1}{\Zfactor{\phasetilde}\Zfactor{\phase}} \phasetildeR\left( \partial_t - \frac{D_R}{\Zfactor{D} }\nabla^2 + \frac{D_R \mass_R^2 \mu^2}{Z_{D} \Zfactor{\mass^2}}\right)\phaseR - \frac{\noiseR}{\Zfactor{\phasetilde}^2 \Zfactor{\noise}} \phasetildeR^2 \nonumber \\
    &+ \frac{D_R \lambdaR}{a \Zfactor{D} \Zfactor{\lambda} \Zfactor{\phasetilde}} \left[ \sin\left(\frac{\phaseR}{\Zfactor{\phase}} \right) \partial_x \phasetildeR - \cos\left(\frac{\phaseR}{\Zfactor{\phase}} \right) \partial_y \phasetildeR\right]\bigg\} \ .
\end{align}
Using  \Eref{Zfactorsdefn}, the renormalised counterpart of $\delta$, \Eref{defn_delta},  can be defined
\begin{equation}  \label{eq:defn_deltaR}
    \delta_R \defequal \frac{\noiseR}{4 \pi D_R } - 1 = \frac{\Zfactor{\noise}}{\Zfactor{D}}  \frac{\noise}{4 \pi D} - 1.
\end{equation}

The regularisation of the perturbative expansion through the introduction of the $Z$-factors is performed most conveniently by considering the vertex functions associated with various correlation functions \cite{tauber_critical_2014, bellac_quantum_1992, amit_field_2005}. For the present model, we need to consider only the inverse propagator
\begin{subequations}
    \begin{equation}
        \invpropphase(\kvec,\omega) = \frac{1}{G(\kvec,\omega)}
    \end{equation}
and the noise vertex
    \begin{equation}
        \vertexnoisephase(\kvec,\omega) = \frac{C(\kvec,\omega)}{G(\kvec,\omega) G(-\kvec, -\omega)}
    \end{equation}
\end{subequations}
where $G(\kvec,\omega)$ is the full propagator and $C(\kvec,\omega)$ is the full correlation function. The elimination of the divergences is performed at an arbitrary inverse length scale $\IRreg$. We choose the normalisation point (NP) of the renormalisation scheme at a finite $\mass_R^2$ and demand the following conditions from the renormalised vertices to leading order in the couplings $\delta$ and $\lambda$
%$u \in \{\delta, \lambda\}$ \cite{amitRenormalisationGroupAnalysis1980a}
\begin{subequations} \label{eq:ren_conds}
    \begin{align}
        \left.\invpropphaseren(\kvec,\omega) \right|_{\NP} &= -\imag \omega + D_R k^2 + D_R \mass^2_R\IRreg^2 - \imag \IRreg k_x (\curlyC D_R \lambdaR)\label{eq:ren_cond_invprop}\\
        \left. \vertexnoisephaseren(\kvec,\omega) \right|_{\NP} &= 2 \noise_R \label{eq:ren_cond_noise}
    \end{align}
\end{subequations}
where the last term in \Eref{ren_cond_invprop} derives from the $n=0$ term in the expansion of the non-linearity in \Eref{action_pert}. The renormalisation conditions above are imposed to ensure a finite UV behaviour. To this end, it suffices to extract the divergent parts of the diagrams in the limit $ma \to 0$. This can be thought of as the equivalent of the minimal subtraction scheme \cite{tauber_critical_2014, bellac_quantum_1992} where the $Z$-factors contain only the factors that diverge as $\epsilon \to 0$. The renormalised vertices can be written in terms of the renormalised fields as 
\begin{subequations} \label{eq:Renvertexfuncdefn}
\begin{align}
  \invpropphaseren(\kvec, \omega) &= \frac{1}{G_{R}(\kvec,\omega)} = (\Zfactor{\phase} \Zfactor{\phasetilde})^{-1} \invpropphase(\kvec,\omega), \label{eq:invprop_ren_to_bare} \\
   \vertexnoisephaseren(\kvec,\omega) &= \frac{C_R(\kvec,\omega)}{G_R(\kvec,\omega) G_R(-\kvec, -\omega)} = \Zfactor{\phasetilde}^{-2} \vertexnoisephase(\kvec,\omega), \label{eq:noise_ren_to_bare}
\end{align}
\end{subequations}
where we have used the definitions of the $Z$-factors, \Eref{def_field_renormalisation}, to relate the renormalised vertices to their bare counterparts. Even though there are infinitely many perturbative vertices proportional to $\lambda$, \Erefs{action_pert_expansion} and \eref{pert_vertices}, they all renormalise identically as to preserve the overall form of the non-linearity, \Eref{action_pert} --- a property that derives from the underlying symmetry of the model, \Eref{sorotation}. Since the $n=0$ term in \Eref{action_pert_expansion} contributes a drift-like term to the bare propagator, focusing on its renormalisation will be sufficient to determine that of the full nonlinearity. 

Certain $Z$-factors are determined exactly by using the symmetries of the model. Firstly, the rotational symmetry \Eref{sorotation} evaluated at $\psi = 2\pi$ implies that the action is invariant under shifts of $\phase(\xvec,t)$ by $2\pi$. Since this shift is a linear transformation of the field, the renormalised action must be invariant under shifts of $2\pi$ as well, \ie the action associated with $\phaseR'(\xvec,t) = \phaseR(\xvec,t) + 2\pi$ is the same as the action for $\phaseR(\xvec,t)$. Imposing this symmetry on the renormalised action in \Eref{ren_action}, we find
\begin{equation} \label{eq:Zphase}
    \Zfactor{\phase} = 1
\end{equation}
to all orders in the perturbative expansion. As far as the mass term in the bilinear part is concerned, we allow this to stay as it amounts merely to a ``soft breaking of symmetry'' \cite{amitRenormalisationGroupAnalysis1980a}. 

Secondly, the non-linearities in the second line of  \Eref{ren_action} multiply a spatial derivative of the auxiliary field. As these auxiliary fields form the outgoing legs in any diagram involving a non-linearity, all such diagrams acquire at least one factor of $k_{x,y}$ in each outgoing leg, so that any such perturbative correction vanishes at $k_x=0=k_y$, reducing vertex functions to their bare values,
\begin{subequations} \label{eq:zero_mode_vertices}
\begin{align}
    \invpropphase(\zerovec, \omega) &= -\imag \omega + D \mass^2 \\
    \vertexnoisephase(\zerovec, \omega) &= 2\noise
\end{align}
\end{subequations}
to all orders. Evaluating the vertices in \Eref{zero_mode_vertices} at the NP and using \Erefs{ren_conds}, \eref{Renvertexfuncdefn} and \eref{Zphase}, it follows that 
\begin{equation} \label{eq:otherexactZ}
    \Zfactor{\phasetilde} = \Zfactor{\noise} = \Zfactor{D}\Zfactor{\mass^2} = 1.
\end{equation}
As a consequence, we have the following useful property
\begin{equation} \label{eq:nice_property}
    D_R \mass_R^2\mu^2 = D \mass^2.
\end{equation}
Thus, the only non-trivial $Z$-factors remaining to be determined are $\Zfactor{\lambda}$ and $\Zfactor{D}$.

To determine them, we consider the leading-order diagrammatic contributions to the inverse propagator. To $2^{\rm nd}$ order in the coupling $\lambda$, the inverse propagator is given by 
\begin{equation} \label{eq:invprop_diagrams}
    \invpropphase(\kvec,\omega) = -\imag \omega + Dk^2 + D \mass^2 - \left\{ \normaldaisies +\normaloddsails + \normalevensails + \normalevenmoths + \normaloddmoths  \right\}
\end{equation}
where each diagram inside the curly bracket contains infinitely many diagrams and the overall minus sign may be thought of as a consequence of Dyson summation. We will discuss in the following each of the diagrams in the curly bracket of \Eref{invprop_diagrams}, on the basis of \Erefs{action_harm}, \eref{action_pert} or \eref{action_pert_expansion}. 

As a matter of convenience, we use a mixed parameterisation in terms of real space for all internal fields and thus for the loops, and in terms of Fourier space for all external fields, so that derivative operators become factors of $k_x$ and $k_y$ as envisaged above, \Eref{pert_vertices}. Carrying out the diagrammatics in real space means that products become convolutions. As
\begin{equation}
\av{\phase(\kvec,\omega)\phasetilde(\kvec',\omega')}_0 = \frac{\deltabar(\kvec+\kvec')\deltabar(\omega+\omega')}{-\imag\omega + D(k^2+m^2)}
\end{equation}
% \gpcomment{1 July 2026: Emir to confirm eqn and sentence above and following sentence:}
from \Eref{bare_propagator}, it follows that
\begin{subequations}
\begin{align}
\av{\phase(\kvec,\omega)\phasetilde(\posvec',t')}_0&=\frac{\exp{\imag(\omega t'-\kvec\cdot\posvec')}}{-\imag\omega + D(k^2+m^2)}
= G(\kvec,\omega)\exp{\imag(\omega t'-\kvec\cdot\posvec')} \\
\av{\phase(\posvec,t)\phasetilde(\kvec',\omega')}_0 & =\frac{\exp{\imag(\omega' t-\kvec'\cdot\posvec)}}{\imag\omega' + D(k'^2+m^2)} = G(-\kvec',-\omega')\exp{\imag(\omega' t-\kvec'\cdot\posvec)}
\end{align}
\end{subequations}
from \Eref{fourierconvention_phasetilde}, with the exponentials entering into the loop integral whenever an external bare propagator connects to an internal field.

%\gpcomment{1) Should we comment on mixing real space and $k$-space (in the loop versus outside)? I think this isn't an issue, as long as every field is ever only evaluated in $x$ or $k$. Within a vertex, I suppose you get the $\delta(k)$ only after taking all fields into $k$. Maybe this needs commenting. 2) I wonder about there not being a k-label on some of the dashes. --> Emir to move the k-labels all the way to the left, to indicate that these are the flowing through the legs, ie the k of the left (but amputated) physical field.} 

The first diagram contains the following ``daisy" diagrams
\begin{align} \label{eq:daisy_diagrams}
    \normaldaisies &\defequal 
\begin{tikzpicture}[baseline = (origin.base), scale = 1.5]
  \begin{feynman}
    \vertex (origin) at (0,0);
    \vertex (o) at (-0.4,0);
    \vertex (i) at (0.4, 0);
    \diagram* {
      (origin) -- [plain, edge node={node[pos=0.9,auto,swap]{\scalebox{0.7}{$\kvec$}}}] (o),
      (origin) -- [plain] (i),
    };
  \end{feynman}
  % Filled black square at the origin
  % The outgoing leg goes in the -x direction, so perpendicular is vertical
  \draw (-0.1, -0.06) -- (-0.1, 0.06);
  \node[scale=0.8] at (-0.1, -0.15) {$x$};
\filldraw[fill=black, draw=black] (0,0) circle (0.03);
\end{tikzpicture}
+ 
\begin{tikzpicture}[baseline = (origin.base), scale = 1.5]
  \begin{feynman}
    \vertex (origin) at (0,0);
    \vertex (o) at (-0.4,0);
    \vertex (i) at (0.4, 0);
    \vertex (l1) at (0.6, 0.4);
    \vertex (l2) at (0.6, -0.4);
    \diagram* {
      (origin) -- [plain, edge node={node[pos=0.9,auto,swap]{\scalebox{0.7}{$\kvec$}}}] (o),
      (origin) -- [plain] (i),
       (origin) -- [anti fermion, bend left = 30, arrow size = 0.7pt] (l1), 
      (origin) -- [anti fermion, bend right = 30,  arrow size = 0.7pt] (l1),
       };
  \end{feynman}
  % Filled black square at the origin
  % The outgoing leg goes in the -x direction, so perpendicular is vertical
  \draw (-0.1, -0.06) -- (-0.1, 0.06);
  \node[scale=0.8] at (-0.1, -0.15) {$x$};
\filldraw[fill=black, draw=black] (0,0) circle (0.03);
\end{tikzpicture}
+
\begin{tikzpicture}[baseline = (origin.base), scale = 1.5]
  \begin{feynman}
    \vertex (origin) at (0,0);
    \vertex (o) at (-0.4,0);
    \vertex (i) at (0.4, 0);
    \vertex (l1) at (0.6, 0.4);
    \vertex (l2) at (0.6, -0.4);
    \diagram* {
      (origin) -- [plain, edge node={node[pos=0.9,auto,swap]{\scalebox{0.7}{$\kvec$}}}] (o),
      (origin) -- [plain] (i),
       (origin) -- [anti fermion, bend left = 30, arrow size = 0.7pt] (l1), 
      (origin) -- [anti fermion, bend right = 30,  arrow size = 0.7pt] (l1),
      (origin) -- [anti fermion, bend left = 30, arrow size = 0.7pt] (l2), 
      (origin) -- [anti fermion, bend right = 30,  arrow size = 0.7pt] (l2), 
    };
  \end{feynman}
  % Filled black square at the origin
  % The outgoing leg goes in the -x direction, so perpendicular is vertical
  \draw (-0.1, -0.06) -- (-0.1, 0.06);
  \node[scale=0.8] at (-0.1, -0.15) {$x$};
\filldraw[fill=black, draw=black] (0,0) circle (0.03);
\end{tikzpicture}
+
\begin{tikzpicture}[baseline = (origin.base), scale = 1.5]
  \begin{feynman}
    \vertex (origin) at (0,0);
    \vertex (o) at (-0.4,0);
    \vertex (i) at (0.4, 0);
    \vertex (l1) at (0.6, 0.4);
    \vertex (l2) at (0.6, -0.4);
    \vertex (l3) at (0.3, 0.7);
    \diagram* {
      (origin) -- [plain, edge node={node[pos=0.9,auto,swap]{\scalebox{0.7}{$\kvec$}}}] (o),
      (origin) -- [plain] (i),
       (origin) -- [anti fermion, bend left = 20, arrow size = 0.7pt] (l1), 
      (origin) -- [anti fermion, bend right = 20,  arrow size = 0.7pt] (l1),
      (origin) -- [anti fermion, bend left = 20, arrow size = 0.7pt] (l2), 
      (origin) -- [anti fermion, bend right = 20,  arrow size = 0.7pt] (l2), 
      (origin) -- [anti fermion, bend left = 20, arrow size = 0.7pt] (l3), 
      (origin) -- [anti fermion, bend right = 20,  arrow size = 0.7pt] (l3), 
    };
  \end{feynman}
  % Filled black square at the origin
  % The outgoing leg goes in the -x direction, so perpendicular is vertical
  \draw (-0.1, -0.06) -- (-0.1, 0.06);
  \node[scale=0.8] at (-0.1, -0.15) {$x$};
\filldraw[fill=black, draw=black] (0,0) circle (0.03);
\end{tikzpicture}
+ 
\begin{tikzpicture}[baseline = (origin.base), scale = 1.5]
  \begin{feynman}
    \vertex (origin) at (0,0);
    \vertex (o) at (-0.4,0);
    \vertex (i) at (0.4, 0);
    \vertex (l1) at (0.6, 0.4);
    \vertex (l2) at (0.6, -0.4);
    \vertex (l3) at (0.3, 0.7);
    \vertex (l4) at (0.3, -0.7);
    \diagram* {
      (origin) -- [plain, edge node={node[pos=0.9,auto,swap]{\scalebox{0.7}{$\kvec$}}}] (o),
      (origin) -- [plain] (i),
       (origin) -- [anti fermion, bend left = 20, arrow size = 0.7pt] (l1), 
      (origin) -- [anti fermion, bend right = 20,  arrow size = 0.7pt] (l1),
      (origin) -- [anti fermion, bend left = 20, arrow size = 0.7pt] (l2), 
      (origin) -- [anti fermion, bend right = 20,  arrow size = 0.7pt] (l2), 
      (origin) -- [anti fermion, bend left = 20, arrow size = 0.7pt] (l3), 
      (origin) -- [anti fermion, bend right = 20,  arrow size = 0.7pt] (l3), 
      (origin) -- [anti fermion, bend left = 20, arrow size = 0.7pt] (l4), 
      (origin) -- [anti fermion, bend right = 20,  arrow size = 0.7pt] (l4), 
    };
  \end{feynman}
  % Filled black square at the origin
  % The outgoing leg goes in the -x direction, so perpendicular is vertical
  \draw (-0.1, -0.06) -- (-0.1, 0.06);
  \node[scale=0.8] at (-0.1, -0.15) {$x$};
\filldraw[fill=black, draw=black] (0,0) circle (0.03);
\end{tikzpicture}
+ \dots = \imag k_x \diagram{\smalldaisies},
\end{align}
where an increasing number of petals dress an inverse bare propagator, given by the $n=0$ term in \Eref{action_pert_expansion}. Each petal contributes the loop integral $2\noise\int\ddintx{2}{\pos'}\dint{t'}G_0(\posvec',t')^2=C_0(\zerovec,0;a)$. After some accounting of the symmetry factors, the summation can be carried out to give
\begin{equation} \label{eq:sum_of_daisies}
    \diagram{\smalldaisies} = D \lambda \sum_{n = 0}^{\infty} \frac{1}{n!} \left(-\frac{C_0(\zerovec,0;a)}{2}\right)^{n } = D \lambda\exp{-\frac{1}{2} C_0(\zerovec, 0;a)}.
\end{equation}
% Emir remarks that a log-divergent C_0 is now well contained.
%where the infinitely many daisies dress the tree-level contribution to the inverse propagator, given by the $n=0$ term in \Eref{action_pert_expansion}. The $k_x$ independent part of the diagram, defined as,
% \begin{equation}
%     \normaldaisies \defequal \imag k_x \diagram{\smalldaisies},
% \end{equation}
% can be calculated by considering the sum of the infinitely many daisy diagrams
% \begin{equation} \label{eq:sum_of_daisies}
%     \diagram{\smalldaisies} = D \lambda \sum_{n = 0}^{\infty} \frac{1}{n!} \left(-\frac{C_0(\zerovec,0;a)}{2}\right)^{n } = D \lambda\exp{-\frac{1}{2} C_0(\zerovec, 0;a)}.
% \end{equation}
There is no contribution of this form proportional to $k_y$, because of the even number of internal $\phase$-fields this term comes with, \Eref{action_pert_expansion}. As a result, beyond tree level the inverse propagator will not be isotropic, \ie it will not be invariant under rotations of $\kvec$ alone, which is due to $\phase(\posvec,t)$ being an angle relative to the $x$-axis, \Eref{defn_unit_vectors}. Had we chosen to measure $\phase$ relative to another direction, $\nvec_0$, the daisy diagrams would feature in the form $\nvec_0\cdot\kvec$ rather than $k_x$ as we find it in \Eref{daisy_diagrams}. This anisotropy is not the result of a small-angle expansion, but that of expanding the action about a uniform configuration $\phase(\rvec,t)\equiv\phase_0$. %\escomment{but of the uniform configuration, $\theta_0$, about which we choose to expand the action.}

%, but rather under the full \Eref{sorotation}. The Ward identity resulting from \Eref{sorotation} does not guarantee the isotropy of any vertex alone, but imposes relations between different vertices due to the shift in the fields. 

Apart from the $k_x$ prefactor on the right-hand side of \Eref{daisy_diagrams}, the diagram or the set it expands to does not carry any wave-vector dependence, so that at the current stage, we do not have any corrections to $D$.

%Therefore, it renormalises only the non-linearity $\lambda$. At this order, $D$ does not receive any corrections. 
The UV divergent part of \Eref{sum_of_daisies} as $\mass a\to 0$ can be extracted by using \Erefs{corr_func_real_space_equal_time} and \eref{asymptotic_expansion}
\begin{equation} \label{eq:sum_of_daisies_divergent}
    \diagram{\smalldaisies} = D \lambda\exp{\frac{\noise}{4\pi D} \ln(\curlyC ma)}  = D \lambda (\curlyC ma)^{\frac{\noise}{4\pi D}} =  D \lambda\curlyC m a \left(1 + \delta \ln(\curlyC m a)\right)  + \mathcal{O}(\delta^2) \ .
\end{equation}
%where we have neglected terms of order $\delta^2$ as we are only interested in the leading order renormalisation \cite{amitRenormalisationGroupAnalysis1980a, neudeckerNonlinearCrystalGrowth1982}.
There are no further diagrams to consider to first order in $\lambda$, \Eref{action_pert_expansion}, because the even powers of $\phase$ together a single external $\phase$ cannot be combined into a diagram.

We proceed to second order in $\lambda$ which gives rise to a non-trivial correction to the diffusivity. The first set of second-order diagrams we discuss are the second and third term on the right of \Eref{invprop_diagrams}. These diagrams are what we call the ``sail'' diagrams
\begin{subequations} \label{eq:sail_diagrams}
    \begin{align}
    \normaloddsails &\defequal 
    \begin{tikzpicture}[baseline = (origin.base), scale = 1.5]
  \begin{feynman}[every fermion={arrow size=1pt}, every anti fermion ={arrow size=1pt} ]
    \vertex (origin) at (0,0);
    \vertex (sailtip) at (0.15,0.7);
    \vertex (o) at (-0.4,0);
    \vertex (i) at (0.4, 0);
    \vertex (l1) at (0.3, 0.3);
    \diagram* {
      (origin) -- [plain, edge node={node[pos=0.9,auto,swap]{\scalebox{0.7}{$\kvec$}}}] (o),
      (origin) -- [anti fermion, postaction={decorate},
         decoration={markings,
           mark=at position 0.7 with {
             \draw[line width=0.6pt] (0,-2.5pt) -- (0,2.5pt);
           }
         }
        ] (sailtip),
      (origin) -- [plain] (i),
      (l1) -- [fermion] (sailtip),
      (l1) -- [fermion] (origin),
    };
  \end{feynman}
  % Filled black square at the origin
  \filldraw[fill=black, draw=black] (-0.075, -0.075) rectangle (0.075, 0.075);
  \filldraw[fill=black, draw=black] (0.15-0.075,0.7-0.075) rectangle (0.15+0.075,0.7+0.075);
  % Perpendicular tick mark on the outgoing leg, close to origin
  % The outgoing leg goes in the -x direction, so perpendicular is vertical
  \draw (-0.20, -0.06) -- (-0.20, 0.06);
  \node[scale=0.8] at (-0.2, -0.15) {$x$};
  % \draw (-0.06, 0.5) -- (0.06, 0.5);
  \node[scale=0.8] at (-0.05, 0.52) {$x$};
\end{tikzpicture}
+
\begin{tikzpicture}[baseline = (origin.base), scale = 1.5]
  \begin{feynman}[every fermion={arrow size=1pt}, every anti fermion ={arrow size=1pt} ]
    \vertex (origin) at (0,0);
    \vertex (sailtip) at (0.15,0.7);
    \vertex (o) at (-0.4,0);
    \vertex (i) at (0.4, 0);
    \vertex (l1) at (0.3, 0.3);
    \vertex (l2) at ($(l1) + (0.3,0)$);
    \vertex (l3) at ($(l2) + (0.3,0)$);
    \diagram* {
      (origin) -- [plain, edge node={node[pos=0.9,auto,swap]{\scalebox{0.7}{$\kvec$}}}] (o),
     (origin) -- [anti fermion, postaction={decorate},
         decoration={markings,
           mark=at position 0.7 with {
             \draw[line width=0.6pt] (0,-2.5pt) -- (0,2.5pt);
           }
         }
        ] (sailtip),
      (origin) -- [plain] (i),
      (origin) -- [plain] (i),
      (l1) -- [fermion] (sailtip),
      (l1) -- [fermion] (origin),
      (l2) -- [fermion] (sailtip),
      (l2) -- [fermion] (origin),
      (l3) -- [fermion] (sailtip),
      (l3) -- [fermion] (origin),
    };
  \end{feynman}
  % Filled black square at the origin
  \filldraw[fill=black, draw=black] (-0.075, -0.075) rectangle (0.075, 0.075);
  \filldraw[fill=black, draw=black] (0.15-0.075,0.7-0.075) rectangle (0.15+0.075,0.7+0.075);
  % Perpendicular tick mark on the outgoing leg, close to origin
  % The outgoing leg goes in the -x direction, so perpendicular is vertical
  \draw (-0.20, -0.06) -- (-0.20, 0.06);
  \node[scale=0.8] at (-0.2, -0.15) {$x$};
  % \draw (-0.06, 0.5) -- (0.06, 0.5);
  \node[scale=0.8] at (-0.05, 0.52) {$x$};
\end{tikzpicture}
+
\begin{tikzpicture}[baseline = (origin.base), scale = 1.5]
  \begin{feynman}[every fermion={arrow size=1pt}, every anti fermion ={arrow size=1pt} ]
    \vertex (origin) at (0,0);
    \vertex (sailtip) at (0.15,0.7);
    \vertex (o) at (-0.4,0);
    \vertex (i) at (0.4, 0);
    \vertex (l1) at (0.3, 0.3);
    \vertex (l2) at ($(l1) + (0.3,0)$);
    \vertex (l3) at ($(l2) + (0.3,0)$);
    \vertex (l4) at ($(l3) + (0.3,0)$); 
    \vertex (l5) at ($(l4) + (0.3,0)$); 
    \diagram* {
      (origin) -- [plain, edge node={node[pos=0.9,auto,swap]{\scalebox{0.7}{$\kvec$}}}] (o),
      (origin) -- [anti fermion, postaction={decorate},
         decoration={markings,
           mark=at position 0.7 with {
             \draw[line width=0.6pt] (0,-2.5pt) -- (0,2.5pt);
           }
         }
        ] (sailtip),
      (origin) -- [plain] (i),
      (origin) -- [plain] (i),
      (l1) -- [fermion] (sailtip),
      (l1) -- [fermion] (origin),
      (l2) -- [fermion] (sailtip),
      (l2) -- [fermion] (origin),
      (l3) -- [fermion] (sailtip),
      (l3) -- [fermion] (origin),
      (l4) -- [fermion] (sailtip),
      (l4) -- [fermion] (origin),
      (l5) -- [fermion] (sailtip),
      (l5) -- [fermion] (origin),
    };
  \end{feynman}
  % Filled black square at the origin
  \filldraw[fill=black, draw=black] (-0.075, -0.075) rectangle (0.075, 0.075);
  \filldraw[fill=black, draw=black] (0.15-0.075,0.7-0.075) rectangle (0.15+0.075,0.7+0.075);
  % Perpendicular tick mark on the outgoing leg, close to origin
  % The outgoing leg goes in the -x direction, so perpendicular is vertical
  \draw (-0.20, -0.06) -- (-0.20, 0.06);
  \node[scale=0.8] at (-0.2, -0.15) {$x$};
  % \draw (-0.06, 0.5) -- (0.06, 0.5);
  \node[scale=0.8] at (-0.05, 0.52) {$x$};
\end{tikzpicture} \label{eq:odd_sail_diagrams}
+ \dots
\\
\normalevensails &\defequal
\begin{tikzpicture}[baseline = (origin.base), scale = 1.5]
  \begin{feynman}[every fermion={arrow size=1pt}, every anti fermion ={arrow size=1pt} ]
    \vertex (origin) at (0,0);
    \vertex (sailtip) at (0.15,0.7);
    \vertex (o) at (-0.4,0);
    \vertex (i) at (0.4, 0);
    \diagram* {
      (origin) -- [plain, edge node={node[pos=0.9,auto,swap]{\scalebox{0.7}{$\kvec$}}}] (o),
      (origin) -- [anti fermion, postaction={decorate},
         decoration={markings,
           mark=at position 0.7 with {
             \draw[line width=0.6pt] (0,-2.5pt) -- (0,2.5pt);
           }
         }
        ] (sailtip),
      (origin) -- [plain] (i),
      (origin) -- [plain] (i),
    };
  \end{feynman}
  % Filled black square at the origin
  \filldraw[fill=black, draw=black] (-0.075, -0.075) rectangle (0.075, 0.075);
  \filldraw[fill=black, draw=black] (0.15-0.075,0.7-0.075) rectangle (0.15+0.075,0.7+0.075);
  % Perpendicular tick mark on the outgoing leg, close to origin
  % The outgoing leg goes in the -x direction, so perpendicular is vertical
  \draw (-0.20, -0.06) -- (-0.20, 0.06);
  \node[scale=0.8] at (-0.2, -0.15) {$y$};
  % \draw (-0.06, 0.5) -- (0.06, 0.5);
  \node[scale=0.8] at (-0.05, 0.52) {$y$};
\end{tikzpicture}
+
\begin{tikzpicture}[baseline = (origin.base), scale = 1.5]
  \begin{feynman}[every fermion={arrow size=1pt}, every anti fermion ={arrow size=1pt} ]
    \vertex (origin) at (0,0);
    \vertex (sailtip) at (0.15,0.7);
    \vertex (o) at (-0.4,0);
    \vertex (i) at (0.4, 0);
    \vertex (l1) at (0.3, 0.3);
    \vertex (l2) at ($(l1) + (0.3,0)$);
    \diagram* {
      (origin) -- [plain, edge node={node[pos=0.9,auto,swap]{\scalebox{0.7}{$\kvec$}}}] (o),
      (origin) -- [anti fermion, postaction={decorate},
         decoration={markings,
           mark=at position 0.7 with {
             \draw[line width=0.6pt] (0,-2.5pt) -- (0,2.5pt);
           }
         }
        ] (sailtip),
      (origin) -- [plain] (i),
      (origin) -- [plain] (i),
      (l1) -- [fermion] (sailtip),
      (l1) -- [fermion] (origin),
      (l2) -- [fermion] (sailtip),
      (l2) -- [fermion] (origin),
    };
  \end{feynman}
  % Filled black square at the origin
  \filldraw[fill=black, draw=black] (-0.075, -0.075) rectangle (0.075, 0.075);
  \filldraw[fill=black, draw=black] (0.15-0.075,0.7-0.075) rectangle (0.15+0.075,0.7+0.075);
  % Perpendicular tick mark on the outgoing leg, close to origin
  % The outgoing leg goes in the -x direction, so perpendicular is vertical
  \draw (-0.20, -0.06) -- (-0.20, 0.06);
  \node[scale=0.8] at (-0.2, -0.15) {$y$};
  % \draw (-0.06, 0.5) -- (0.06, 0.5);
  \node[scale=0.8] at (-0.05, 0.52) {$y$};
\end{tikzpicture}
+
\begin{tikzpicture}[baseline = (origin.base), scale = 1.5]
  \begin{feynman}[every fermion={arrow size=1pt}, every anti fermion ={arrow size=1pt} ]
    \vertex (origin) at (0,0);
    \vertex (sailtip) at (0.15,0.7);
    \vertex (o) at (-0.4,0);
    \vertex (i) at (0.4, 0);
    \vertex (l1) at (0.3, 0.3);
    \vertex (l2) at ($(l1) + (0.3,0)$);
    \vertex (l3) at ($(l2) + (0.3,0)$);
    \vertex (l4) at ($(l3) + (0.3,0)$); 
    \vertex (l5) at ($(l4) + (0.3,0)$); 
    \diagram* {
      (origin) -- [plain, edge node={node[pos=0.9,auto,swap]{\scalebox{0.7}{$\kvec$}}}] (o),
      (origin) -- [anti fermion, postaction={decorate},
         decoration={markings,
           mark=at position 0.7 with {
             \draw[line width=0.6pt] (0,-2.5pt) -- (0,2.5pt);
           }
         }
        ] (sailtip),
      (origin) -- [plain] (i),
      (origin) -- [plain] (i),
      (l1) -- [fermion] (sailtip),
      (l1) -- [fermion] (origin),
      (l2) -- [fermion] (sailtip),
      (l2) -- [fermion] (origin),
      (l3) -- [fermion] (sailtip),
      (l3) -- [fermion] (origin),
      (l4) -- [fermion] (sailtip),
      (l4) -- [fermion] (origin),
    };
  \end{feynman}
  % Filled black square at the origin
  \filldraw[fill=black, draw=black] (-0.075, -0.075) rectangle (0.075, 0.075);
  \filldraw[fill=black, draw=black] (0.15-0.075,0.7-0.075) rectangle (0.15+0.075,0.7+0.075);
  % Perpendicular tick mark on the outgoing leg, close to origin
  % The outgoing leg goes in the -x direction, so perpendicular is vertical
  \draw (-0.20, -0.06) -- (-0.20, 0.06);
  \node[scale=0.8] at (-0.2, -0.15) {$y$};
  % \draw (-0.06, 0.5) -- (0.06, 0.5);
  \node[scale=0.8] at (-0.05, 0.52) {$y$};
\end{tikzpicture}
+ \dots\label{eq:even_sail_diagrams}
    \end{align}
\end{subequations}
where the perturbative vertices, shown as filled squares, are dressed with the daisy diagrams given in \Eref{daisy_diagrams}. The presence of two such vertices reflects the second order in $\lambda$. %As with the daisy diagrams, 
As two internal (differentiated) $\phasetilde$ are available, overall an even number of internal $\phase$-fields are needed. However, one internal differentiated $\phasetilde$-field must connect to an internal $\phase$-field, which will be integrated over. In the diagrams above, any such integral vanishes on the basis of odd parity under reflection of the integration domain, as the bare propagator and the correlator are even \Eref{diagramsharmonic} in the dummy variable. To see this more clearly, we identify the ``offending loop'' as 
\begin{equation}
    \int\ddintx{2}{\pos'}\dint{t'} 
    C^n_0(\posvec',t') \partial_{x'} G_0(\posvec',t') = 0 = \int\ddintx{2}{\pos'}\dint{t'} 
    C^n_0(\posvec',t') \partial_{y'} G_0(\posvec',t')
\end{equation}
for $n=1,2,\ldots$. In summary,
%in space. This becomes most obvious when considering any bare 
%$\kvec$-space or by 
%
%Will get down two $\phasetilde$. Will need one $\phasetilde$ to connect to external $\phase$. So, will have one external $\phasetilde$ and one internal $\phasetilde$. That's even. The noise only provides even, so will need even number of $\phase$.
%
%Need even number of $\phase$ to connect to $\noise$ and one more $\phase$ and 
%
%
%Each carries a derivative, 
%
%
%
%
%The sail diagrams carry no $\kvec$ dependence beyond a single prefactor arising from the perturbative vertex connecting the ingoing and outgoing legs. However, because of this, they vanish identically: the derivative acting on the bare propagator connecting the two vertices renders the loop integrand odd under the parity transformation $\mathbf{k} \to -\mathbf{k}$, so that upon integration over the loop momentum, 
\begin{equation} \label{eq:sum_of_sails_divergent}
    \normaloddsails = \normalevensails = 0 \ .
\end{equation}

A second set of diagrams can be found to second order in $\lambda$, shown as the rightmost diagrams in \Eref{invprop_diagrams}, which we may call the ``moth'' (or ``Gibson Flying V'') diagrams,
\begin{subequations} \label{eq:moth_diagrams}
    \begin{align}
        \normalevenmoths &\defequal 
        \begin{tikzpicture}[baseline = (origin.base), scale = 1.5]
  \begin{feynman}
    \vertex (origin) at (0,0);
    \vertex (mid) at (1,0);
    \vertex (o) at (-0.4,0);
    \vertex (i) at (1.4, 0);
    \vertex (u1) at (1.2, 0.4);
    \vertex (d1) at (1.2, -0.4);
    \diagram* {
      (origin) -- [plain, edge node={node[pos=0.9,auto,swap]{\scalebox{0.7}{$\kvec$}}}] (o),
      (origin) -- [anti fermion, arrow size = 1pt] (mid),
      (mid) -- [plain] (i),
      (origin) -- [anti fermion, arrow size = 1pt] (u1),
      (origin) -- [anti fermion, arrow size = 1pt] (d1),
      (mid) -- [anti fermion, arrow size = 1pt] (u1),
      (mid) -- [anti fermion, arrow size = 1pt] (d1),
    };
  \end{feynman}
  % Filled black square at the origin
  \filldraw[fill=black, draw=black] (-0.075,-0.075) rectangle (0.075,0.075);
  \filldraw[fill=black, draw=black] (1-0.075,-0.075) rectangle (1+0.075,0.075);
  % Perpendicular tick mark on the outgoing leg, close to origin
  % The outgoing leg goes in the -x direction, so perpendicular is vertical
   \draw (-0.20, -0.06) -- (-0.20, 0.06);
  \node[scale=0.8] at (-0.15, -0.15) {$x$};
  \draw (1-0.20, -0.06) -- (1-0.20, 0.06);
  \node[scale=0.8] at (1-0.20, -0.15) {$x$};
\end{tikzpicture}
+ 
\begin{tikzpicture}[baseline = (origin.base), scale = 1.5]
  \begin{feynman}
    \vertex (origin) at (0,0);
    \vertex (mid) at (1,0);
    \vertex (o) at (-0.4,0);
    \vertex (i) at (1.4, 0);
    \vertex (u1) at (1.2, 0.4);
    \vertex (d1) at (1.2, -0.4);
    \vertex (u2) at (1.5, 0.7);
    \vertex (d2) at (1.5, -0.7);
    \diagram* {
      (origin) -- [plain, edge node={node[pos=0.9,auto,swap]{\scalebox{0.7}{$\kvec$}}}] (o),
      (origin) -- [anti fermion, arrow size = 1pt] (mid),
      (mid) -- [plain] (i),
      (origin) -- [anti fermion, arrow size = 1pt] (u1),
      (origin) -- [anti fermion, arrow size = 1pt] (d1),
      (mid) -- [anti fermion, arrow size = 1pt] (u1),
      (mid) -- [anti fermion, arrow size = 1pt] (d1),
      (origin) -- [anti fermion, arrow size = 1pt] (u2),
      (origin) -- [anti fermion, arrow size = 1pt] (d2),
      (mid) -- [anti fermion, arrow size = 1pt] (u2),
      (mid) -- [anti fermion, arrow size = 1pt] (d2),
    };
  \end{feynman}
  % Filled black square at the origin
  \filldraw[fill=black, draw=black] (-0.075,-0.075) rectangle (0.075,0.075);
  \filldraw[fill=black, draw=black] (1-0.075,-0.075) rectangle (1+0.075,0.075);
  % Perpendicular tick mark on the outgoing leg, close to origin
  % The outgoing leg goes in the -x direction, so perpendicular is vertical
   \draw (-0.20, -0.06) -- (-0.20, 0.06);
  \node[scale=0.8] at (-0.15, -0.15) {$x$};
  \draw (1-0.20, -0.06) -- (1-0.20, 0.06);
  \node[scale=0.8] at (1-0.20, -0.15) {$x$};
\end{tikzpicture}
+
\begin{tikzpicture}[baseline = (origin.base), scale = 1.5]
  \begin{feynman}
    \vertex (origin) at (0,0);
    \vertex (mid) at (1,0);
    \vertex (o) at (-0.4,0);
    \vertex (i) at (1.4, 0);
    \vertex (u1) at (1.2, 0.4);
    \vertex (d1) at (1.2, -0.4);
    \vertex (u2) at (1.5, 0.7);
    \vertex (d2) at (1.5, -0.7);
    \vertex (u3) at (1.8, 1);
    \vertex (d3) at (1.8, -1);
    \diagram* {
      (origin) -- [plain, edge node={node[pos=0.9,auto,swap]{\scalebox{0.7}{$\kvec$}}}] (o),
      (origin) -- [anti fermion, arrow size = 1pt] (mid),
      (mid) -- [plain] (i),
      (origin) -- [anti fermion, arrow size = 1pt] (u1),
      (origin) -- [anti fermion, arrow size = 1pt] (d1),
      (mid) -- [anti fermion, arrow size = 1pt] (u1),
      (mid) -- [anti fermion, arrow size = 1pt] (d1),
      (origin) -- [anti fermion, arrow size = 1pt] (u2),
      (origin) -- [anti fermion, arrow size = 1pt] (d2),
      (mid) -- [anti fermion, arrow size = 1pt] (u2),
      (mid) -- [anti fermion, arrow size = 1pt] (d2),
      (origin) -- [anti fermion, arrow size = 1pt] (u3),
      (origin) -- [anti fermion, arrow size = 1pt] (d3),
      (mid) -- [anti fermion, arrow size = 1pt] (u3),
      (mid) -- [anti fermion, arrow size = 1pt] (d3),
    };
  \end{feynman}
  % Filled black square at the origin
  \filldraw[fill=black, draw=black] (-0.075,-0.075) rectangle (0.075,0.075);
  \filldraw[fill=black, draw=black] (1-0.075,-0.075) rectangle (1+0.075,0.075);
  % Perpendicular tick mark on the outgoing leg, close to origin
  % The outgoing leg goes in the -x direction, so perpendicular is vertical
   \draw (-0.20, -0.06) -- (-0.20, 0.06);
  \node[scale=0.8] at (-0.15, -0.15) {$x$};
  \draw (1-0.20, -0.06) -- (1-0.20, 0.06);
  \node[scale=0.8] at (1-0.20, -0.15) {$x$};
\end{tikzpicture}
+\dots \label{eq:even_moth_diagrams}\\
    \normaloddmoths & \defequal 
    \begin{tikzpicture}[baseline = (origin.base), scale = 1.5]
  \begin{feynman}
    \vertex (origin) at (0,0);
    \vertex (mid) at (1,0);
    \vertex (o) at (-0.4,0);
    \vertex (i) at (1.4, 0);
    \vertex (u1) at (1.2, 0.4);
    \diagram* {
      (origin) -- [plain, edge node={node[pos=0.9,auto,swap]{\scalebox{0.7}{$\kvec$}}}] (o),
      (origin) -- [anti fermion, arrow size = 1pt] (mid),
      (mid) -- [plain] (i),
      (origin) -- [anti fermion, arrow size = 1pt] (u1),
      (mid) -- [anti fermion, arrow size = 1pt] (u1),
    };
  \end{feynman}
  % Filled black square at the origin
  \filldraw[fill=black, draw=black] (-0.075,-0.075) rectangle (0.075,0.075);
  \filldraw[fill=black, draw=black] (1-0.075,-0.075) rectangle (1+0.075,0.075);
  % Perpendicular tick mark on the outgoing leg, close to origin
  % The outgoing leg goes in the -x direction, so perpendicular is vertical
   \draw (-0.20, -0.06) -- (-0.20, 0.06);
  \node[scale=0.8] at (-0.15, -0.15) {$y$};
  \draw (1-0.20, -0.06) -- (1-0.20, 0.06);
  \node[scale=0.8] at (1-0.20, -0.15) {$y$};
\end{tikzpicture}
+ 
\begin{tikzpicture}[baseline = (origin.base), scale = 1.5]
  \begin{feynman}
    \vertex (origin) at (0,0);
    \vertex (mid) at (1,0);
    \vertex (o) at (-0.4,0);
    \vertex (i) at (1.4, 0);
    \vertex (u1) at (1.2, 0.4);
    \vertex (u2) at (1.5, 0.7);
    \vertex (d2) at (1.5, -0.7);
    \diagram* {
      (origin) -- [plain, edge node={node[pos=0.9,auto,swap]{\scalebox{0.7}{$\kvec$}}}] (o),
      (origin) -- [anti fermion, arrow size = 1pt] (mid),
      (mid) -- [plain] (i),
      (origin) -- [anti fermion, arrow size = 1pt] (u1),
      (mid) -- [anti fermion, arrow size = 1pt] (u1),
      (origin) -- [anti fermion, arrow size = 1pt] (u2),
      (origin) -- [anti fermion, arrow size = 1pt] (d2),
      (mid) -- [anti fermion, arrow size = 1pt] (u2),
      (mid) -- [anti fermion, arrow size = 1pt] (d2),
    };
  \end{feynman}
  % Filled black square at the origin
  \filldraw[fill=black, draw=black] (-0.075,-0.075) rectangle (0.075,0.075);
  \filldraw[fill=black, draw=black] (1-0.075,-0.075) rectangle (1+0.075,0.075);
  % Perpendicular tick mark on the outgoing leg, close to origin
  % The outgoing leg goes in the -x direction, so perpendicular is vertical
   \draw (-0.20, -0.06) -- (-0.20, 0.06);
  \node[scale=0.8] at (-0.15, -0.15) {$y$};
  \draw (1-0.20, -0.06) -- (1-0.20, 0.06);
  \node[scale=0.8] at (1-0.20, -0.15) {$y$};
\end{tikzpicture}
+
\begin{tikzpicture}[baseline = (origin.base), scale = 1.5]
  \begin{feynman}
    \vertex (origin) at (0,0);
    \vertex (mid) at (1,0);
    \vertex (o) at (-0.4,0);
    \vertex (i) at (1.4, 0);
    \vertex (u1) at (1.2, 0.4);
    \vertex (u2) at (1.5, 0.7);
    \vertex (d2) at (1.5, -0.7);
    \vertex (u3) at (1.8, 1);
    \vertex (d3) at (1.8, -1);
    \diagram* {
      (origin) -- [plain, edge node={node[pos=0.9,auto,swap]{\scalebox{0.7}{$\kvec$}}}] (o),
      (origin) -- [anti fermion, arrow size = 1pt] (mid),
      (mid) -- [plain] (i),
      (origin) -- [anti fermion, arrow size = 1pt] (u1),
      (mid) -- [anti fermion, arrow size = 1pt] (u1),
      (origin) -- [anti fermion, arrow size = 1pt] (u2),
      (origin) -- [anti fermion, arrow size = 1pt] (d2),
      (mid) -- [anti fermion, arrow size = 1pt] (u2),
      (mid) -- [anti fermion, arrow size = 1pt] (d2),
      (origin) -- [anti fermion, arrow size = 1pt] (u3),
      (origin) -- [anti fermion, arrow size = 1pt] (d3),
      (mid) -- [anti fermion, arrow size = 1pt] (u3),
      (mid) -- [anti fermion, arrow size = 1pt] (d3),
    };
  \end{feynman}
  % Filled black square at the origin
  \filldraw[fill=black, draw=black] (-0.075,-0.075) rectangle (0.075,0.075);
  \filldraw[fill=black, draw=black] (1-0.075,-0.075) rectangle (1+0.075,0.075);
  % Perpendicular tick mark on the outgoing leg, close to origin
  % The outgoing leg goes in the -x direction, so perpendicular is vertical
   \draw (-0.20, -0.06) -- (-0.20, 0.06);
  \node[scale=0.8] at (-0.15, -0.15) {$y$};
  \draw (1-0.20, -0.06) -- (1-0.20, 0.06);
  \node[scale=0.8] at (1-0.20, -0.15) {$y$};
\end{tikzpicture}
+\dots \label{eq:odd_moth_diagrams}
    \end{align}
\end{subequations}
where the perturbative vertices are dressed once again by the daisy diagrams in \Eref{daisy_diagrams}. Unlike the previous diagrams, the moth diagrams do depend non-trivially on the external wave-vector, thereby renormalising the diffusivity. 

% XXX CONT HERE 29 June 2026 17.38pm
% Emir to check GP's comments labelled 1 July.        

% What was a bit in the way above was rushed thinking that all it takes is oddness in derivatives. But we need oddity under reflection of the dummy(!!) variable. This won't be the case below any more. Here the arguments will be $\posvec-\posvec'$. Needs climbing down the rabbit hole.

% 30 June 2026 00.01am Have a look at the comment after \Eref{invprop_diagrams}.

Denoting the left-hand side of \Eref{moth_diagrams} as $\diagram{\smallevenmoths}$ and $\diagram{\smalloddmoths}$, respectively, and accounting for the symmetry factors, we can sum them in closed form
\begin{subequations} \elabel{sum_of_moths}
    \begin{align}
        \diagram{\smallevenmoths}(\kvec,\omega) &= \imag k_x \diagram{\smalldaisies}^2\sum_{n=1}^{\infty} \int \ddintx{2}{\pos} \int \dint{t} \exp{-\imag(\kvec \cdot \posvec - \omega t)} (\partial_x G_0(\posvec,t)) \frac{C^{2n}_0(\posvec,t;a)}{(2n)!} \nonumber \\
        &= \imag k_x \diagram{\smalldaisies}^2\int \ddintx{2}{\pos} \int \dint{t} \exp{-\imag(\kvec \cdot \posvec - \omega t)} (\partial_x G_0(\posvec,t)) \left\{ \cosh(C_0(\posvec,t;a)) - 1\right\} \label{eq:sum_even_moths}\\
        \diagram{\smalloddmoths}(\kvec,\omega) & = \imag k_y \diagram{\smalldaisies}^2\sum_{n=1}^{\infty}\int \ddintx{2}{\pos} \int \dint{t}\exp{-\imag(\kvec \cdot \posvec - \omega t)} (\partial_yG_0(\posvec,t)) \frac{C^{2n-1}_0(\posvec,t;a)}{(2n-1)!}\nonumber \\
        &= \imag k_y \diagram{\smalldaisies}^2\int \ddintx{2}{\pos} \int \dint{t} \exp{-\imag(\kvec \cdot \posvec - \omega t)} (\partial_yG_0(\posvec,t))  \sinh(C_0(\posvec,t;a)) \label{eq:sum_odd_moths}
    \end{align}
\end{subequations}
where the prefactors $\imag k_{x,y}$ are due to the differentiation of the internal $\phasetilde(\xvec,t)$ field that connects to the outgoing $\phase(\xvec,t)$ field. According to minimal subtraction \cite{amitRenormalisationGroupAnalysis1980a} we are interested only in the UV-divergent parts of the diagram, corresponding to distances where the correlation function grows logarithmically, \Eref{asymptotic_expansion}. To capture the UV-divergent behaviour of the diagrams we can approximate the hyperbolic functions as
\begin{equation}
    %\cosh\left( C_0(\posvec,t;a)\right) - 1 = \sinh\left( C_0(\posvec,t;a)\right) = \frac{\exp{ C_0(\posvec,t;a)}}{2} + \mathcal{O}(\mass^2 \pos^2)
\cosh(C_0)-1 = \frac{1}{2} \exp{C_0} + \mathcal{O}(C_0^0) 
\qquad\text{and}\qquad
\sinh(C_0) = \frac{1}{2}\exp{C_0} + \mathcal{O}(\exp{-C_0})
\end{equation}
where $C_0(\posvec,0;a)$ diverges for $\posvec=\nullvec$ as the UV-cutoff $a$ vanishes. We therefore rewrite \Erefs{sum_of_moths}
\begin{subequations} \label{eq:moths_UV_divergent}
    \begin{align}
        \diagram{\smallevenmoths}(\kvec,\omega) &= \frac{\imag k_x}{2} \diagram{\smalldaisies}^2\int \ddintx{2}{\pos} \int \dint{t} \exp{-\imag(\kvec \cdot \posvec - \omega t)} (\partial_xG_0(\posvec,t)) \exp{ C_0(\posvec,t;a)} + 
        %\mathcal{O}(\mass^2 \pos^2)
        \text{(UV-convergent terms)}\label{eq:even_moths_UV_divergent}\\
        \diagram{\smalloddmoths}(\kvec,\omega) & =  \frac{\imag k_y}{2} \diagram{\smalldaisies}^2\int \ddintx{2}{\pos} \int \dint{t} \exp{-\imag(\kvec \cdot \posvec - \omega t)} (\partial_yG_0(\posvec,t))  \exp{ C_0(\posvec,t;a)} + 
        %\mathcal{O}(\mass^2 \pos^2) 
        \text{(UV-convergent terms)} \ .
        \label{eq:odd_moths_UV_divergent}
    \end{align}
\end{subequations}
%where the neglected terms $\mathcal{O}(\mass^2 \pos^2)$ yield UV finite expressions. 
The moth diagrams are non-trivial functions of $\kvec$ and $\omega$, but we need only their leading order dependence on $\kvec$ to determine the renormalisation of $\lambda$ or $D$. The integrals over $\pos$ in \Erefs{moths_UV_divergent} vanish once $\kvec$ is set to $\zerovec$ in the integrand as it becomes odd under the parity transformation, $\posvec \to -\posvec$, due to the single derivative acting on the bare propagator, implying that the moth diagrams do not renormalise $\lambda$. To capture the correction to the diffusivity $D$, we expand the exponential, $\exp{-\imag \posvec \cdot \kvec}$, to first order in $\kvec$ and set $\omega = 0$
\begin{subequations} \label{eq:moths_UV_divergent_expanded}
    \begin{align}
        \diagram{\smallevenmoths}(\kvec,\omega) &= \frac{k^2_x}{2} \diagram{\smalldaisies}^2\int \ddintx{2}{\pos} \int \dint{t} x (\partial_xG_0(\posvec,t)) \exp{ C_0(\posvec,t;a)} + \ldots %\mathcal{O}(\mass^2 \pos^2, k_{x,y}^3, \omega)
        \label{eq:even_moths_UV_divergent_expanded}\\
        \diagram{\smalloddmoths}(\kvec,\omega) & =  \frac{k^2_y}{2} \diagram{\smalldaisies}^2\int \ddintx{2}{\pos} \int \dint{t} y (\partial_yG_0(\posvec,t))  \exp{ C_0(\posvec,t;a)} + \ldots
        %\mathcal{O}(\mass^2 \pos^2, k_{x,y}^3, \omega)
        \label{eq:odd_moths_UV_divergent_expanded}
    \end{align}
\end{subequations}
where the terms proportional to $k_x k_y$ vanish in both lines as %contributions to the integrand proportional to $k_xk_y$
they
are odd under the parity transformation $x \to -x$ or $y \to -y$. Since the bare propagator and the correlation function are functions of $\pos^2$, we can change the integration variables in \Eref{odd_moths_UV_divergent_expanded} from $(x,y)$ to $(y,x)$ to find that both moth diagrams renormalise the diffusivity identically. Consequently, we can add them up to find the overall correction to the diffusion term  
\begin{equation} \label{eq:sum_of_both_moths}
    \diagram{\smallevenmoths}(\kvec,\omega) + \diagram{\smalloddmoths}(\kvec,\omega) = \frac{\kvec^2}{2} \diagram{\smalldaisies}^2\int \ddintx{2}{\pos} \int \dint{t} x (\partial_xG_0(\posvec,t)) \exp{ C_0(\posvec,t;a)} + \ldots
    %\mathcal{O}(\mass^2 \pos^2, k_{x,y}^3, \omega). 
\end{equation}
As we are interested in the UV-divergent part of %the integral in
\Eref{sum_of_both_moths} to lowest order in the couplings, and the diagram is already of order $\lambda^2$, we take into account the term of order $\delta^0$ only, \ie can set $\delta = 0$ inside the integrand. Furthermore, using \Erefs{diagramsharmonic}, \eref{corr_func_reg} and \eref{sum_of_daisies_divergent}, we can simplify \Eref{sum_of_both_moths}
\begin{equation}\elabel{sum_of_both_moths_again}
    \diagram{\smallevenmoths}(\kvec,\omega) + \diagram{\smalloddmoths}(\kvec,\omega) = -D\kvec^2 \frac{(\lambda\curlyC \mass a)^2}{32 \pi D} \int \ddintx{2}{\pos} \pos^2 \int_0^{\infty} \frac{\dint{t}}{t^2}\Exp{- \frac{\pos^2 }{4 D t} - Dt \mass^2 + \int_{D t \mass^2}^{\infty} \frac{\dint{v}}{v}\exp{-v - \frac{\mass^2(\pos^2 + a^2)}{4 v}}} + \ldots
\end{equation}
where we have taken the liberty to symmetrise the integral with respect to the spatial integration variable to transform $x^2$ to $\pos^2/2$ under the integral. We focus our attention to the UV-divergent part of \Eref{sum_of_both_moths_again}, \ie its behaviour as $a\to0$. Working along the lines of \cite{amitRenormalisationGroupAnalysis1980a,neudeckerNonlinearCrystalGrowth1982} we change the integration variable from $t$ to $u = (\pos^2 + a^2)/(4Dt)$, to rewrite the integral as
\begin{multline} \label{eq:diffusivity_change_of_variables}
    \diagram{\smallevenmoths}(\kvec,\omega) + \diagram{\smalloddmoths}(\kvec,\omega) = -D\kvec^2 \frac{(\lambda\curlyC \mass a)^2}{8 \pi } \int \ddintx{2}{\pos} \frac{\pos^2}{\pos^2 + a^2}  \\
    \times \int_0^{\infty} \dint{u}\Exp{- u - \frac{\mass^2(\pos^2 + a^2)}{4 u} + u\frac{a^2}{\pos^2 +a^2} + \int_{\frac{\mass^2(\pos^2 + a^2)}{4u}}^{\infty} \frac{\dint{v}}{v}\exp{-v - \frac{\mass^2(\pos^2 + a^2)}{4 v}}} + \ldots \ ,
\end{multline}
which will allow us to expand the rightmost integral over $v$ in small $m^2(\pos^2+a^2)$,
\begin{align} \label{eq:corr_UV_divergent_part}
    \int_{\frac{\mass(\pos^2 + a^2)}{4u}}^{\infty} \frac{\dint{v}}{v}\exp{-v - \frac{\mass^2(\pos^2 + a^2)}{4 v}} &= \int_{0}^{\infty} \frac{\dint{v}}{v}\exp{-v - \frac{\mass^2(\pos^2 + a^2)}{4 v}}- \int_{0}^{\frac{\mass^2(\pos^2 + a^2)}{4u}} \frac{\dint{v}}{v}\exp{-v - \frac{\mass^2(\pos^2 + a^2)}{4 v}} \nonumber \\
    &= 2 K_0(\mass \sqrt{\pos^2 + a^2}) - \sum_{n=0}^{\infty} \frac{(-1)^n}{n!} \int_{0}^{\frac{\mass(\pos^2 + a^2)}{4u}} \dint{v} v^{n-1} \exp{-\frac{\mass(\pos^2 + a^2)}{4v}} \nonumber \\
    &= 2 K_0(\mass \sqrt{\pos^2 + a^2}) - \sum_{n=0}^{\infty} \frac{1}{n!} \left(-\frac{\mass^2(\pos^2+a^2)}{4}\right)^n \Gamma(-n,u)\nonumber \\
    &= -2 \ln(\curlyC \mass \sqrt{\pos^2+a^2}) -\Gamma(0,u) + \mathcal{O}(\mass^2 (\pos^2+a^2)) 
\end{align}
where $\mathcal{O}(\mass^2 (\pos^2+a^2))$ terms contain the UV regular part of the expression and we have used \Eref{asymptotic_expansion} as well as the upper-complete Gamma function $\Gamma(n,z)$ 
\begin{equation}
    \Gamma(n,z)  \defequal \int_{z}^{\infty} \dint{t}t^{n-1}\exp{-t}.
\end{equation}
%To extract the UV-divergent part of the integral, we restrict the spatial integral range to $\pos \leq 1/\mass$ \cite{amitRenormalisationGroupAnalysis1980a}. Furthermore, the integral over $v$ can be written in a power series in $\mass^2 r^2$
Using \Eref{corr_UV_divergent_part} in \Eref{diffusivity_change_of_variables} produces
\begin{multline} \elabel{moths_in_between}
%\label{eq:diffusivity_change_of_variables}
    \diagram{\smallevenmoths}(\kvec,\omega) + \diagram{\smalloddmoths}(\kvec,\omega) = -D\kvec^2 \frac{(\lambda\curlyC \mass a)^2}{8 \pi } \int \ddintx{2}{\pos} \frac{\pos^2}{\pos^2 + a^2}  \\
    \times \int_0^{\infty} \dint{u}
    \Exp{- u - \frac{\mass^2(\pos^2 + a^2)}{4 u} + u\frac{a^2}{\pos^2 +a^2}} 
    \frac{\exp{-\Gamma(0,u)}}{\curlyC^2m^2(\pos^2+a^2)}
    + \ldots \ ,
\end{multline}
where the second factor of $1/(\pos^2+a^2)$ on the far right makes the integral of $\pos^2/(\pos^2+a^2)^2$ over $\pos$ log-divergent in the UV-limit $a\to0$. To extract the UV behaviour, we may therefore take $a\to0$ and $\pos\to0$ in the remainder of the integrand, 
\begin{equation}
    \lim_{\pos\to0} \lim_{a\to0}
    \int_0^{\infty} \dint{u}
    \Exp{- u - \frac{\mass^2(\pos^2 + a^2)}{4 u} + u\frac{a^2}{\pos^2 +a^2}} 
    \exp{-\Gamma(0,u)} = \int_0^{\infty} \dint{u} \Exp{-u-\Gamma(0,u)} =: \alpha \approx 0.62433\ldots
\end{equation}
using \cite{mathematica} to numerically evaluate the integral. From \Eref{moths_in_between} we thus have 
\begin{align}  \label{eq:sum_of_moths_final_step}
    \diagram{\smallevenmoths}(\kvec,\omega) + \diagram{\smalloddmoths}(\kvec,\omega) &= -D\kvec^2 \frac{(\lambda a)^2}{8 \pi } \alpha \int \ddintx{2}{\pos} \frac{\pos^2}{(\pos^2 + a^2)^2}  +
    \ldots
    %+ \mathcal{O}(\mass^2 |\posvec|^2)  
    \\
    &=-D\kvec^2 (\lambda a)^2 \frac{\alpha}{4} \int_{0}^{1/m}\dint{r} \frac{r^3}{(r^2 + a^2)^2} + 
    \ldots
    %\mathcal{O}(\mass^2 |\posvec|^2)
\end{align}
where the IR regularisation of $r<1/m$ has been made explicit. The final integral is readily carried out, 
\begin{equation} \label{eq:sum_of_moth_diagrams_divergent}
    \diagram{\smallevenmoths}(\kvec,\omega) + \diagram{\smalloddmoths}(\kvec,\omega) = Dk^2 \left\{ (\lambda a)^2 \frac{\alpha}{4} \ln(am) \right\} + \mathcal{O}(1).
\end{equation}
where the factor of $a^2$ is to be absorbed in to $\lambda_R$, \Eref{Zfactorsdefn}, rendering the expression dimensionless and finite under $a\to0$, provided $\lambda_R$ remains finite.

% which seems to indicate that it vanishes as $a\to0$, however $\$

% and retaining only the UV-divergent contributions, discarding $-\mass^2\pos^2/4u$ and $u a^2/(\pos^2 + a^2) $ in the exponential, we extract the UV-divergent part of the moth diagrams
% \begin{align}  \label{eq:sum_of_moths_final_step}
%     \diagram{\smallevenmoths}(\kvec,\omega) + \diagram{\smalloddmoths}(\kvec,\omega) &= -D\kvec^2 \frac{(\lambda a)^2}{8 \pi } \int \ddintx{2}{\pos} \frac{\pos^2}{(\pos^2 + a^2)^2} \int_0^{\infty} \dint{u}\exp{- u - \Gamma(0,u)} + \mathcal{O}(\mass^2 |\posvec|^2)  \\
%     &=-D\kvec^2 (\lambda a)^2 \frac{\alpha}{4} \int_{0}^{1/m}\dint{r} \frac{r^3}{(r^2 + a^2)^2} + \mathcal{O}(\mass^2 |\posvec|^2)
% \end{align}
% where we have changed the integration variables from Cartesian to cylindrical polars going from first to second line and defined the constant \cite{mathematica}
% \begin{equation}
%     \alpha \defequal \int_{0}^{\infty}\dint{u}\exp{-u-\Gamma(0,u)} \approx 0.62433\dots.
% \end{equation}
% Evaluating the integral over $r$ in \Eref{sum_of_moths_final_step} in the limit $\mass a \ll 1$, we find
% \begin{equation} \label{eq:sum_of_moth_diagrams_divergent}
%     \diagram{\smallevenmoths}(\kvec,\omega) + \diagram{\smalloddmoths}(\kvec,\omega) = D\kvec^2 \left\{ (\lambda a)^2 \frac{\alpha}{4} \ln(am) \right\} + \mathcal{O}(\mass a).
% \end{equation}
% where the terms in $\mathcal{O}(\mass a)$ are finite as $a \to 0$. 

Using \Erefs{sum_of_daisies_divergent}, \eref{sum_of_sails_divergent} and \eref{sum_of_moth_diagrams_divergent}, to leading order the inverse propagator \Eref{invprop_diagrams} can be written as
\begin{equation} \label{eq:invprop_second_order}
    \invpropphase(\kvec,\omega) = -\imag \omega + D\left(1 - (\lambda a)^2\frac{\alpha} {4}\ln(am)\right)k^2 + D \mass^2 - \imag k_x \curlyC m D\lambda a(1+\delta \ln(\curlyC m a)) + \ldots\ .
\end{equation}
The surprising dependence on $k_x$ but not $k_y$ indicates that the inverse propagator is not invariant under rotations of $\kvec$. The reason for this is the arbitrary choice of measuring angles $\phase(\rvec,t)$ relative to the $x$-axis, as implemented in \Erefs{defn_unit_vectors} resulting ultimately in \Eref{action_pert_expansion}. The $x$-axis thus constitutes a preferred direction and as $\phase$ is \emph{defined} relative to it, the preferred direction featuring should come as no surprise.

%Its invariance under \Eref{sorotation} is obscured by the inverse propagator being the functional derivative of the vertex function at $\phase\equiv0\equiv\phasetilde$, which under \Eref{sorotation} needs to be replaced by $\phase\equiv\psi$.
%\gpcomment{6 July 2026: I am gobsmacked. How is this possible. If you treat the inverse propagator as just that, no evaluation at $\phase=\phasetilde=0$, then there is nothing that stops you from applying 6a. And then there is no reason why 6a doesn't apply --- resulting here in only a change of $k_x$ (rotating $\posvec$), nothing else. HOW???? How can this thing have a preferred direction. Have we ended up with a small angle approximation after all. It doesn't look like it in \Eref{action_pert_expansion}, but maybe we did? Maybe we would have seen what we are seeing if we had done it...  so, what we do amounts to it?}

Evaluating \Eref{invprop_second_order} at the NP, $\mass^2 = \IRreg^2D_R  /D  =\IRreg^2+ \mathcal{O}(u^2)$, \Eref{nice_property}, and imposing the renormalisation conditions, \Eref{ren_conds}, we find the non-trivial $Z$-factors
\begin{subequations} \label{eq:nontrivial_Z_factors}
\begin{align}
    \Zfactor{\lambda} &= 1 + \delta \ln(\curlyC \IRreg a) + \mathcal{O}(u^2) \\
    \Zfactor{D}&= 1 - \lambda^2 a^2 \frac{\alpha}{4}\ln(\IRreg a) + \mathcal{O}(u^3)\label{eq:ZD}
\end{align}
\end{subequations}
where $u$ denotes the expansion parameters $u \in \{\delta, \lambda\}$.

% XXX CONT HERE 6 July 2026

\subsection{Flow Functions}
In this section, we find the flow functions associated with the couplings and derive the results quoted in the main text. The flow functions associated with the couplings using their definitions \Erefs{Zfactorsdefn} and \eref{defn_deltaR}
\begin{subequations} \label{eq:beta_func_defns}
\begin{align}
    \betafunc{\lambda} &\defequal \IRreg \frac{\plaind \lambdaR}{\plaind \IRreg} =  \lambda a \IRreg \frac{\plaind \Zfactor{\lambda}}{\plaind \IRreg} = \lambdaR \left(  \frac{\plaind \logZfactor{\lambda}}{\plaind \ln \IRreg}\right) \label{eq:beta_func_lambda}\\
     \betafunc{\delta} &\defequal \IRreg \frac{\plaind \deltaR}{\plaind \IRreg} =  -\frac{\noise}{4 \pi D_R^2} \IRreg \frac{\plaind \Zfactor{D}}{\plaind \IRreg} = -(1 + \deltaR) \frac{\plaind \logZfactor{D}}{\plaind \ln \IRreg} \label{eq:beta_func_delta}
\end{align}
\end{subequations}
where we have used the fact that $\Zfactor{\noise} = 1$, \Eref{otherexactZ}. Evaluating the log-derivatives and replacing the couplings by their renormalised counterparts, $\lambda = \lambda_R/a + \mathcal{O}(u^2_R)$ and $\delta = \deltaR + \mathcal{O}(u^2_R)$, valid to leading order, yields 
\begin{subequations} \label{eq:beta_funcs}
\begin{align}
    \betafunc{\lambda}  &= \deltaR \lambdaR + \mathcal{O}(g^3_R), \\
    \betafunc{\delta} &= \frac{\alpha}{4}\lambdaR^2 + \mathcal{O}(g^3_R). 
\end{align}
\end{subequations}
\Erefs{beta_funcs} represent the key finding of this section and correspond to the flow function quoted in the main text.

\subsection{Physics at the Gaussian Fixed Point}\label{sec:physicsGFP}
In this subsection, we analyse the physics at the Gaussian fixed point by computing the spin-correlation function
\begin{equation}
    C_{\phi}(\posvec,t) = \av{\cos[\phase(\posvec,t) - \phase(\zerovec, 0)]} = \av{\exp{\imag(\phase(\posvec,t) - \phase(\zerovec,0))}}
\end{equation}
which is the natural observable to compute, given the compact nature of the Goldstone mode and its relation to the order-parameter field $\phivec(\posvec,t)$, \Eref{phaseangledecomp}. 

The spin-correlation function, $C_{\phi}(\posvec,t)$, can be computed by solving its associated Callan-Symanzik equation. We first introduce the renormalised spin-correlation function in terms of its bare counterpart
\begin{equation} \label{eq:ren_spin_corr}
    Z^{-2}(\mu)C_{\phi,R}\left( \posvec,t; D_R(\mu),  \mass^2_R(\mu), \delta_R, \lambda_R;\mu\right) = C_{\phi}\left( \posvec,t; D,  \noise, \mass^2, a, \lambda\right)
\end{equation}
where the $Z$-factor arises from the renormalisation of the exponential operator
\begin{equation}
   \left. \exp{\imag \phase(\posvec,t))}\right|_{R} \equiv Z \exp{\imag \phase(\posvec,t))},
\end{equation}
which absorbs the UV divergence arising in the computation of the average of the operator \cite{zinn-justin_quantum_2002} and encodes its anomalous scaling dimension, \Eref{anomalous_scaling_exp}. To leading order in the couplings, $Z$ is determined by requiring the renormalised operator $\av{\left. \exp{\imag \phase(\posvec,t)} \right|_{R}}$ to remain UV-finite as $am \to 0$ and thus $\av{\exp{\imag \phase(\posvec,t)}}_0\to0$. This gives
\begin{subequations}
\label{eq:anomalous_z_factor}
\begin{align}     Z^{-1}  = \av{\exp{\imag \phase(\posvec,t)}}_0 + \mathcal{O}(\lambda^2) = \exp{-\frac{1}{2} \av{\phase^2(\posvec,t)}_0} + \mathcal{O}(\lambda^2) = \exp{-\frac{1}{2} C(\zerovec, 0;a)}+ \mathcal{O}(\lambda^2)  &= \curlyC m a(1 + \delta \ln(\curlyC m a))  + \mathcal{O}(u^2)\\
    &=  \curlyC \mu a(1 + \delta \ln(\curlyC \mu a)) + \mathcal{O}(u^2)
\end{align}
\end{subequations}
where the average is taken with respect to the harmonic part of the action and $u$ collectively denotes the coupling constants $u \in \{\lambda, \delta\}$. 

Since the right-hand side of \Eref{ren_spin_corr} is independent of the inverse length-scale $\mu$, differentiating with respect to $\mu$ yields the Callan-Symanzik equation. Before writing it down, we introduce the Wilson $\gamma$-functions
\begin{subequations} \elabel{gammafuncs}
\begin{align}
    \gamma_{\mass^2} &= \mu\frac{\partial \ln(\mass_R^2/\mass)}{\partial \mu} = -2 + \mu\frac{\partial \ln Z_{\mass^2}}{\partial \mu} = -2 +\frac{\alpha}{4}\lambdaR^2, \\
    \gamma_{D} &=  \mu\frac{\partial \ln D_R}{\partial \mu} =  \mu\frac{\partial \ln Z_D}{\partial \mu} = -\frac{\alpha}{4}\lambdaR^2, \elabel{gammaDfunc} \\
    \zeta &= \mu\frac{\partial \ln Z}{\partial \mu}  = -1 - \delta_R. \elabel{zetafunc}
\end{align}
\end{subequations}
where the partial derivatives are evaluated using the expressions of the $Z$-factors \Erefs{otherexactZ}, \eref{ZD}, and \eref{anomalous_z_factor}, as well as \Erefs{Zfactorsdefn_propagators} and \Eref{defn_deltaR}. Further, we have used $\lambda = \lambda_R/a + \mathcal{O}(u^2_R)$, \Eref{Zfactorsdefn}, and $\delta = \deltaR + \mathcal{O}(u^2_R)$, 
\Eref{defn_deltaR}. The Callan-Symanzik equation then reads
\begin{equation} \label{eq:CS_eqn_spin_corr}
    \left( \mu\frac{\partial}{\partial \mu} + \beta_{\delta}\frac{\partial}{\partial \delta_R}  + \beta_{\lambda} \frac{\partial}{\partial \lambda_R}  + \gamma_{m^2}m^2 \frac{\partial}{\partial \mass^2} + \gamma_D D_R\frac{\partial}{\partial D_R} -2\zeta\right) C_{\phi,R}\left( \posvec,t; D_R(\mu),  \mass^2_R(\mu), \delta_R, \lambda_R;\mu\right) = 0.
\end{equation}
It can be solved through the method of characteristics, producing
\begin{equation}
    C_{\phi,R}(\posvec,t) = \exp{-2\int_1^{\ell} \frac{\dint{l'}}{l'} \zeta(l')} C_{\phi,R}\left(\posvec,t; D_R(\mu\ell), m^2_R(\mu\ell), \delta_R(\mu\ell), \lambda_R(\mu\ell); \mu \ell\right)
\end{equation}
%XXX CONT HERE 8 July 2026 14.34pm.
where the couplings $\lambda_R(\mu\ell)$ and $\delta_R(\mu\ell)$ solve the RG flow equations, \Erefs{beta_funcs}. As we are interested in the logarithmic corrections in the approach to the Gaussian fixed point, we do not replace them by their fixed point values $\lambda^{*}_R=0$ and $\delta^{*}_R=0$, as is standard \cite{bellac_quantum_1992,tauber_critical_2014}. At $\ell=1$ the two couplings need to be on the separatrix namely $\delta_R(\mu)  = \sqrt{\alpha}\lambda_R(\mu)/2$, \Fref{phase_diagram}, while their flow satisfies \Erefs{beta_funcs}. 

The couplings $D_R(\mu\ell)$ and $m^2_R(\mu\ell)$ similarly flow, 
\begin{subequations}
    \begin{align}
        \ell \frac{d D_R(\mu\ell)}{ d\ell} &= \gamma_{D}(\ell) D_R(\mu\ell) \label{eq:flowofD}\\
        \ell \frac{d \mass_R^2(\ell)}{ d\ell} &= \gamma_{\mass^2}(\ell) \mass_R^2(\mu\ell)
    \end{align}
\end{subequations}
where, again, we do not simply replace $\gamma_{D}(\ell)$ and $\gamma_{\mass^2}(\ell)$ by their fixed point values $\gamma_D^*=\gamma_{D}(0)$ and $\gamma_{m^2}^*=\gamma_{m^2}(0)$. As above, initial conditions are the renormalised couplings at $\ell=1$. 

By dimensional analysis, \sref{DimAna}, the spin-correlation function takes the form
\begin{equation} \label{eq:formal_soln_CS_equation}
    C_{\phi,R}(\posvec,t) = \Exp{-2\int_1^{\ell} \frac{\dint{l'}}{l'} \zeta(l')} C_{\phi,R}\left(\mu \ell|\posvec|, D_R(\mu\ell) \mu^2\ell^2 t; m^2_R(\mu\ell), \delta_R(\mu\ell), \lambda_R(\mu\ell);1\right).
\end{equation}
As $\ell \to 0$, the RG flow approaches the Gaussian fixed point, \Eref{Gaussian_critical_point}.
%, and $\mass_R^2(\mu\ell) \to 0$; 
The logarithmic scaling form is then obtained by evaluating \Eref{formal_soln_CS_equation} in this regime. Solving for the couplings $\lambda_R(\mu\ell)$ and $\delta_R(\mu\ell)$ in this limit, we find, \Eref{asymptotic_solutions},
\begin{equation} \label{eq:asymptotic_couplings}
    \delta_R(\mu\ell) = \frac{\sqrt{\alpha}}{2} \lambda_R(\mu\ell) = -\frac{1}{\ln(\ell)} + \mathcal{O}\left(\frac{1}{\ln^2(\ell)}\right).
\end{equation}
Using this asymptotic solution, we evaluate 
$\int_{1}^{\ell}\dint{\ell'} \zeta(\ell')/\ell'$ in \Eref{formal_soln_CS_equation} using \Eref{zetafunc} 
%\begin{subequations} \label{eq:asymptotic_quantities}
\begin{equation} \label{eq:asymptotic_int_zeta}
    \int_{1}^{\ell}\frac{\dint{\ell'}}{\ell'} \zeta(\ell') \approx -\ln(\ell) + \ln|\ln(\ell)|.
\end{equation}
%\end{subequations}
and solve $D_R(\mu\ell)$ of \Eref{flowofD} using \Eref{gammaDfunc} for small $\ell\ll1$
%, valid in the limit $\ell \ll 1$, finding
%\begin{subequations} \label{eq:asymptotic_quantities}
\begin{equation} \label{eq:asymptotic_D}
    D_R(\mu\ell) \approx D^{*} \Exp{\frac{1}{\ln \ell}}
\end{equation}
%     , \\
%     \int_{1}^{\ell}\frac{\dint{\ell'}}{\ell'} \zeta(\ell') &\approx -\ln(\ell) + \ln|\ln(\ell)|.
% \end{align}
% \end{subequations}

Substituting these asymptotic expressions, \Erefs{asymptotic_int_zeta} and \eref{asymptotic_D}, into \Eref{formal_soln_CS_equation} and taking $\mass^2_R(\ell) \to 0$, we find to leading order
\begin{equation}
    C_{\phi,R}(\posvec,t) = \frac{\ell^2}{\ln^2\ell} C_{\phi, R}\left( \mu \ell |\posvec|, \frac{D^{*}\mu^2 \ell^2 t}{\exp{-1/\ln(\ell)}}; 0, 0, 0;1\right).
\end{equation}
Finally, setting $\ell = 1/(\mu|\posvec|)$, we recover the scaling form presented in the main text
\begin{equation}
    C_{\phi,R}(\posvec,t) = \frac{1}{\pos^2 \ln^2(|\posvec|)} C_{\phi,R}\left(1, \frac{D^{*}t}{\posvec^ 2 \exp{1/\ln(|\posvec|)}}; 0, 0, 0;1\right).
\end{equation}

\bibliography{Novel_PT_VC}% Produces the bibliography via BibTeX.